\documentclass[11pt,a4paper]{article}
\usepackage{jheppub}
\usepackage{multirow}
\usepackage{enumitem}

\newcommand\refr[1]      {ref.\,\cite{#1}}
\newcommand\refrs[1]    {refs.\,\cite{#1}}

\newcommand\Eqn[1]     {Eq.\,(\ref{#1})}
\newcommand\Eqns[2]    {Eqs.\,(\ref{#1}) and~(\ref{#2})}
\newcommand\Eqnss[2]   {Eqs.\,(\ref{#1})--(\ref{#2})}
\newcommand\eqn[1]     {eq.\,(\ref{#1})}
\newcommand\eqns[2]    {eqs.\,(\ref{#1}) and~(\ref{#2})}
\newcommand\eqnss[2]   {eqs.\,(\ref{#1})--(\ref{#2})}

\newcommand\Sect[1]    {Section~{\ref{#1}}}

\newcommand\sect[1]    {section~{\ref{#1}}}
\newcommand\sects[2]   {sections~\ref{#1} and~\ref{#2}}

\newcommand\appx[1]     {appendix~\ref{#1}}

\newcommand\tab[1]     {table~\ref{#1}}

\def\beq{\begin{equation}}
\def\eeq{\end{equation}}
\def\bsp#1\esp{\begin{split}#1\end{split}}
\def\bal#1\eal{\begin{align}#1\end{align}}
\newcommand\nt         {\notag}

\newcommand\tsig[1]    {\sigma^{\mathrm{#1}}}
\newcommand\dsig[1]    {\rd\sigma^{{\rm #1}}}
\newcommand\dsiga[2]   {\rd\sigma^{{\rm #1,A}_{\scriptscriptstyle #2}}}

\newcommand\la         {\langle}
\newcommand\ra         {\rangle}
\newcommand\bra[3]     {\la {\cal M}_{#1}^{#2}#3|}
\newcommand\ket[3]     {|{\cal M}_{#1}^{#2}#3\ra}
\newcommand\SME[3]     {|{\cal M}_{#1}^{(#2)}{(#3)}|^2}
\newcommand\M[2]       {\ensuremath{|{\cal{M}}_{#1}^{#2}|^2}}

\newcommand{\rd}       {{\mathrm{d}}}
\newcommand{\PS}[1]    {\rd\phi_{#1}}
\newcommand{\mom}[1]   {\{p\}^{#1}}
\newcommand{\momt}[1]   {\{\ti{p}\}^{#1}}
\newcommand{\momh}[1]   {\{\ha{p}\}^{#1}}
\newcommand{\momb}[1]   {\{\bar{p}\}^{#1}}
\newcommand{\cmap}[1]   {\stackrel{{\mathsf C}_{#1}}{\longrightarrow}}
\newcommand{\smap}[1]   {\stackrel{{\mathsf S}_{#1}}{\longrightarrow}}

\newcommand\tzz[2]     {z_{#1,#2}}
\newcommand\kT[1]      {k_{\perp,#1}}
\newcommand\kTt[1]     {\tilde{k}_{\perp,#1}}
\newcommand{\Y}[2]     {Y_{\wti{#1}\wti{#2},Q}}
\newcommand{\Yt}[2]     {Y_{\wti{#1}\wti{#2},Q}}
\newcommand{\Yh}[2]     {Y_{\ha{#1}\ha{#2},Q}}

\newcommand{\cC}[2]    {{\cal C}_{#1}^{#2}}
\newcommand{\cS}[2]    {{\cal S}_{#1}^{#2}}
\newcommand{\cCS}[3]   {{\cal C}_{#1}^{~}{\cal S}_{#2}^{#3}}
\newcommand{\cSCS}[2]  {{\cal C}\kern-2pt{\cal S}_{#1}^{#2}}

\newcommand{\rC}       {{\mathrm C}}
\newcommand{\rS}       {{\mathrm S}}
\newcommand{\rSCS}     {{\rC}\kern-2pt{\rS}}
\newcommand{\IcC}[2]   {{\rC}_{#1}^{#2}}
\newcommand{\IcS}[2]   {{\rS}_{#1}^{#2}}
\newcommand{\IcCS}[3]  {{\rC}_{#1}^{~}{\rS}_{#2}^{#3}}
\newcommand{\IcSCS}[2] {\rC\kern-2pt\rS_{#1}^{#2}}
\newcommand{\bI}       {\bom{I}}

\newcommand{\cI}       {{\cal I}}
\newcommand{\cJ}       {{\cal J}}
\newcommand{\cK}       {{\cal K}}
\newcommand{\cJm}      {{\cal J}^{{\rm (1m)}}}
\newcommand{\cKm}      {{\cal K}^{{\rm (1m)}}}
\newcommand{\cM}       {{\cal M}}

\newcommand{\CF}       {C_{\mathrm{F}}}
\newcommand{\CA}       {C_{\mathrm{A}}}
\newcommand{\TR}       {T_{\mathrm{R}}}
\newcommand{\Nc}       {N_{\mathrm{c}}}
\newcommand{\Nf}       {\ensuremath{n_{\mathrm{f}}}}

\newcommand{\bT}       {\bom{T}}
\newcommand\qb         {{\bar q}}
\newcommand{\br}[1]	{\{#1\}}
\newcommand{\Sym}[1]	{S_{\br{#1}}}
\newcommand{\fb}		{{\bar f}}

\newcommand\bom[1]     {{\mbox{\boldmath $#1$}}}
\newcommand\lbb		{\Big(}
\newcommand\rbb		{\Big)}
\newcommand{\wti}[1]   {\widetilde{\,#1\,}}
\newcommand{\wha}[1]   {\widehat{#1}}

\newcommand{\ti}[1]   {\widetilde{\,#1\,}}
\newcommand{\ha}[1]   {\widehat{\,#1\,}}

\newcommand\al	{\alpha}
\newcommand\be	{\beta}
\newcommand\delf[2]	{\delta_{#1,#2}}
\newcommand{\ep}       {\epsilon}
\newcommand{\vth}     {\vartheta}
\newcommand{\vphi}     {\varphi}

\newcommand\B	{{\mathcal B}_0}
\newcommand\Bf[1]		{{\mathcal B}_0\!\left(#1\right)\!}
\newcommand\LL	{l}
\newcommand\KK	{k}
\newcommand\hr	{\ha{r}}

\newcommand{\cA}       {{\cal A}}
\newcommand{\bcA}[1]   {{\bom{\cal A}}_{#1}}

\newcommand\e          {{\mathrm e}}
\newcommand\Oe[1]      {\ensuremath{\mathrm O(\ep^{#1})}}
\newcommand{\pole}[2]  {\frac{#1}{\epsilon^{#2}}}

\newcommand\ldot       {\!\cdot\!}
\newcommand\msbar      {\ensuremath{{\overline {\rm MS}}}}
\newcommand\as  	       {\ensuremath{\alpha_{\mathrm{s}}}}

\newcommand{\calS}     {{\cal S}}
\def\hP			      {\hat{P}}

\title{
A subtraction scheme for computing QCD jet cross sections at NNLO: 
integrating the iterated singly-unresolved subtraction terms
}

\author[a]{Paolo Bolzoni,}
\author[b]{G\'abor Somogyi}
\author[c]{and Zolt\'an Tr\'ocs\'anyi}

\affiliation[a]{
II.~Institut f\"ur Theoretische Physik, Universit\"at Hamburg\\
Luruper Chaussee 149, D-22761 Hamburg, Germany
}
\affiliation[b]{
DESY\\
Platanenalle 6, D-15738 Zeuthen, Germany
}
\affiliation[c]{
University of Debrecen and Institute of Nuclear Research of the 
Hungarian Academy of Sciences\\ 
H-4001 Debrecen P.O.Box 51, Hungary
}

\emailAdd{paolo.bolzoni@desy.de}
\emailAdd{gabor.somogyi@desy.de}
\emailAdd{z.trocsanyi@atomki.hu}

\abstract{
We perform the integration of all iterated singly-unresolved subtraction
terms, as defined in \refr{Somogyi:2006da}, over the two-particle
factorized phase space. We also sum over the unresolved parton flavours.
The final result can be written as a convolution (in colour space) of the
Born cross section and an insertion operator. We spell out the insertion
operator in terms of 24 basic integrals that are defined explicitly. We
compute the coefficients of the Laurent expansion of these integrals
in two different ways, with the method of Mellin-Barnes representations 
and sector decomposition. Finally, we present the Laurent expansion 
of the full insertion operator for the specific examples of electron-positron
annihilation into two and three jets.
}


\arxivnumber{1011.1909}

\proceeding{DESY 10-191 \\ SFB/CPP-10-106}

\begin{document}
\maketitle
\flushbottom


\section{Introduction}
\label{sec:intro}

The most frequently occurring final states in high energy particle  
collisions contain hadronic jets. 
Because of their large production cross sections, 
jet observables can be measured to a high statistical accuracy, and are 
therefore ideal for precision studies. Examples include: the measurement of 
the strong coupling, $\as$, from jet rates and event shapes in 
electron-positron annihilation; the determination of the gluon parton 
distribution function (and also $\as$) in deep inelastic lepton-hadron
scattering into two plus one jets; the measurement of parton distributions 
in hadron-hadron scattering from single jet inclusive production and vector 
boson plus jet production. Often the relevant observables are measured with 
experimental precision of a few per cent or better. Thus, theoretical 
predictions with the same level of accuracy are necessary to fully 
exploit the physics potential of these measurements. This usually requires 
the computation of next-to-next-to-leading order (NNLO) corrections in 
perturbative QCD.

The straightforward calculation of jet cross sections in QCD perturbation 
theory is however 
hampered by the presence of infrared singularities in the intermediate  
stages of the calculation, which must be treated consistently before 
any numerical computation may be performed. At next-to-leading order 
(NLO) accuracy, using a subtraction scheme to handle infrared divergences
is the approach of choice. Exploiting the fact that the kinematic 
singularities of QCD matrix elements are universal, one builds process and 
observable independent counterterms that simultaneously cancel both the 
kinematic singularities in real-emission phase space integrals and the 
explicit $\ep$-poles in one-loop virtual corrections (here the use of 
dimensional regularisation in $d=4 - 2\ep$ dimensions is implied).

At NNLO accuracy, the calculation of fully differential cross sections is a 
challenging problem, and various extensions of the subtraction method at NNLO 
have been proposed, see e.g.~\refrs{Somogyi:2006da,Somogyi:2006db,
Somogyi:2006cz,Somogyi:2005xz,Weinzierl:2003fx,Weinzierl:2003ra,
Frixione:2004is,GehrmannDeRidder:2005cm,Daleo:2006xa,Daleo:2009yj,
Glover:2010im,Czakon:2010td}. 
In very broad terms, when setting up 
any subtraction algorithm, two quite distinct difficulties must be addressed. 
First, one must define subtraction terms that properly regularise the 
real-emission phase space integrals and second, one must combine the 
integrated form of these counterterms with the virtual contributions, so as 
to cancel the infrared divergences of the loop matrix elements. 
In a rigorous mathematical sense, the cancellation of both the kinematic 
singularities in the real-emission pieces and the explicit $\ep$-poles in 
the virtual pieces must be local. On the one hand, this means 
that the subtraction terms and the real-emission contributions must
tend to the same value in $d$ dimensions, in all kinematic limits where 
the latter diverge. 
On the other hand, the cancellation of explicit $\ep$-poles between the 
integrated subtraction terms and the virtual contributions must take place 
point-wise in phase space. This in particular implies that it is possible 
to write the integrated counterterms in such a form that they can be 
explicitly combined with virtual contributions, before phase space 
integration. From a practical point of view, the full locality of the 
subtraction scheme is also important to insure good numerical efficiency 
of the algorithm. 
Finally, the construction should be universal, i.e.\ independent of the 
process and observable being considered. This avoids the need to tediously 
adapt the algorithm to every specific problem.

However, defining universal subtraction terms that are completely local 
in the real-emission phase space is rather delicate, and there is very 
little freedom to define these in such a way, that their integration over 
the unresolved phase spaces becomes convenient. 
One way to address these difficulties is to use counterterms 
that are not fully local, but whose complete analytic integration is 
tractable. For example, the antenna subtraction method 
\cite{GehrmannDeRidder:2005cm,Daleo:2006xa,Daleo:2009yj,Glover:2010im} 
builds the subtraction terms from so-called antenna functions. These are simple 
enough to be integrated analytically, but they do not reproduce azimuthal 
correlations in gluon splitting, and the cancellation of $\ep$ poles in the 
real-virtual contributions is also nonlocal. 
Because of this, actual numerical computations with the antenna scheme, 
such as the calculation of total rates \cite{GehrmannDeRidder:2007jk,
GehrmannDeRidder:2008ug,Weinzierl:2008iv,Weinzierl:2009nz} 
and event shapes \cite{GehrmannDeRidder:2007bj,GehrmannDeRidder:2007hr,
GehrmannDeRidder:2009dp,Weinzierl:2009ms,Weinzierl:2009yz} in electron-positron 
annihilation, require the use of an auxiliary phase space slicing. 
Another option is to develop dedicated subtraction schemes that are 
applicable only to some specific processes, such as the production of 
colourless final states, vector bosons \cite{Catani:2009sm,Catani:2010en} or 
the Higgs boson \cite{Catani:2007vq}, in hadron collisions. 
Then, one may even propose to dispense with the subtraction method altogether, 
and adopt a strategy such as sector decomposition (see \cite{Heinrich:2008si} and 
references therein), where the full Laurent expansions (in $\ep$) of the real-emission 
and virtual pieces are computed directly, and their finite pieces combined. 

Nevertheless, despite the subtleties, it is possible to define completely 
local counterterms for real radiation, by first carefully matching the various 
QCD factorisation formulae for unresolved emission \cite{Somogyi:2005xz,
Nagy:2007mn}, and then 
extending the expressions obtained over the full phase space \cite{Somogyi:2006cz,
Somogyi:2006da,Somogyi:2006db}. Recall that in the scheme of 
\refrs{Somogyi:2005xz,Somogyi:2006cz,Somogyi:2006da,Somogyi:2006db}, 
the NNLO correction to a generic $m$-jet cross section with no coloured particles 
in the initial state (work towards an extension to hadron-initiated processes is 
presented in \refr{Somogyi:2009ri}),
\beq
\tsig{NNLO} =
	\int_{m+2}\!\dsig{RR}_{m+2} J_{m+2}
	+ \int_{m+1}\!\dsig{RV}_{m+1} J_{m+1}
	+ \int_m\!\dsig{VV}_m J_m\,,
\label{eq:sigmaNNLO}
\eeq
is rewritten as a sum of finite integrals
\beq
\tsig{NNLO} =
	\int_{m+2}\!\dsig{NNLO}_{m+2}
	+ \int_{m+1}\!\dsig{NNLO}_{m+1}
	+ \int_m\!\dsig{NNLO}_m\,,
\label{eq:sigmaNNLOfin}
\eeq
where
\bal
\dsig{NNLO}_{m+2} &=
	\Big\{\dsig{RR}_{m+2} J_{m+2} - \dsiga{RR}{2}_{m+2} J_{m}
	-\Big[\dsiga{RR}{1}_{m+2} J_{m+1} - \dsiga{RR}{12}_{m+2} J_{m}\Big]
	\Big\}_{\ep=0}\,,
\label{eq:sigmaNNLOm+2}
\\[2mm]
\dsig{NNLO}_{m+1} &=
	\Big\{\Big[\dsig{RV}_{m+1} + \int_1\dsiga{RR}{1}_{m+2}\Big] J_{m+1}
	-\Big[\dsiga{RV}{1}_{m+1} + \Big(\int_1\dsiga{RR}{1}_{m+2}\Big)
	\strut^{{\rm A}_{\scriptscriptstyle 1}}
\Big] J_{m} \Big\}_{\ep=0}\,,
\label{eq:sigmaNNLOm+1}
\\[2mm]
\intertext{and}
\dsig{NNLO}_{m} &=
	\Big\{\dsig{VV}_m + \int_2\Big[\dsiga{RR}{2}_{m+2} - \dsiga{RR}{12}_{m+2}\Big]
	+\int_1\Big[\dsiga{RV}{1}_{m+1} + \Big(\int_1\dsiga{RR}{1}_{m+2}\Big)
	\strut^{{\rm A}_{\scriptscriptstyle 1}}\Big]\Big\}_{\ep=0} J_{m}\,,
\label{eq:sigmaNNLOm}
\eal
are each integrable in four dimensions by construction. 
In \eqnss{eq:sigmaNNLO}{eq:sigmaNNLOm} $J_n$ ($n=m+2$, $m+1$ and $m$) 
denotes the jet function which defines the infrared safe observable being 
calculated. All approximate cross sections appearing in 
\eqnss{eq:sigmaNNLOm+2}{eq:sigmaNNLOm} above have been defined 
explicitly in \refrs{Somogyi:2006cz,Somogyi:2006da,Somogyi:2006db}. 
To finish the definition of the scheme, we must compute once and for all
the one- and two-particle integrals appearing in 
\eqns{eq:sigmaNNLOm+1}{eq:sigmaNNLOm}. In previous publications, we 
have shown that it is possible to adapt and employ well-known techniques of 
loop integration, such as integration-by-parts identities and solving of 
differential equations \cite{Kotikov:1991pm,Remiddi:1997ny}, the method 
of Mellin--Barnes (MB) representations with harmonic summation 
\cite{Smirnov:1999gc,Tausk:1999vh,Smirnov:2004ym,Czakon:2005rk,Gluza:2007rt} 
and also sector decomposition \cite{Heinrich:2008si}, to compute the integrals that 
arise, analytically and numerically. Indeed, all one-particle integrals, denoted formally 
by $\int_1$ above, have been evaluated with these methods \cite{Somogyi:2008fc,
Aglietti:2008fe,Bolzoni:2009ye}. 
The actual computation of integrated subtraction terms leads to a large 
number of rather elaborate phase space integrals, however these can to 
be computed once and for all.

In this paper, we compute the integral of the iterated singly-unresolved 
subtraction term, $\int_2 \dsiga{RR}{12}_{m+2}$, over the phase space of the 
unresolved partons. We find that the integrated approximate cross section can 
be written as the product of the Born cross section for the production of $m$ 
partons, times a new insertion operator (in colour space), $\bI^{(0)}_{12}$.
We use the method of MB representations, as developed in this 
context in \refr{Bolzoni:2009ye}, to compute the integrals appearing in the various 
building blocks of the insertion operator. In several cases we find multi-dimensional 
MB integrals that are very difficult to compute fully analytically, and hence complete 
analytic expressions cannot be obtained at present. Nevertheless, in these 
cases direct numerical integration of the appropriate MB representations 
provides a fast and reliable way to obtain final results. We stress that for 
phenomenological applications, this is all that is required, hence, we make no 
severe effort to compute analytic expressions beyond those that are trivial to derive. 
As a numerical check, all integrals are evaluated using sector decomposition as well. 
Thus, each integral in this paper is obtained by two independent computations. 

Since the paper is quite long and rather technical, readers mainly interested 
in understanding the general structure of the results or in some applications 
are advised to first read sections~\ref{sec:notation}, \ref{sec:IntRRA12} and 
\ref{sec:results}. \Sect{sec:notation} sets the notation, and in \sect{sec:IntRRA12} 
we present the final expression for the integrated iterated singly-unresolved 
approximate cross section in \eqn{eq:IntRRA12-fin} and the new insertion 
operator $\bI^{(0)}_{12}$ in \eqn{eq:I12}. These two equations are the main 
results of this paper. In \sect{sec:results} we discuss some examples, 
specialising the general formulae to the case of two- and three-jet production 
processes. The explicit definitions of the integrated counterterms are then 
presented in \sects{sec:flavsummedICTs}{sec:IntsCTs}. 
The technical details of computing the integrated subtraction terms are given
in appendices.




\section{Notation}
\label{sec:notation}

%
%

\subsection{Matrix elements}
\label{sec:ME}

We consider processes with coloured particles (partons) in the final
state, while the initial-state particles are colourless (typically
electron-positron annihilation into hadrons).  Any number of additional
non-coloured final-state particles are allowed, too, but they will be
suppressed in the notation. Resolved partons in the final state are labeled 
with letters chosen form the middle of the alphabet, $i,\, j,\, k,\, l,\,\dots$, 
while letters chosen form the end of the alphabet,  $r,\,s,\,t,\dots$, denote 
unresolved final-state partons.

We adopt the colour- and spin-state notation of \refr{Catani:1996vz}. In
this notation the amplitude for a scattering process involving the
final-state momenta $\mom{}$, $\ket{}{}{(\mom{})}$, is an abstract
vector in colour and spin space, and its normalisation is fixed such
that the squared amplitude summed over colours and spins is
\beq
\label{eq:M2}
|\cM|^2 = \bra{}{}{}\ket{}{}{}\:.
\eeq
This matrix element has the following expansion in the number of loops:
\beq
\ket{}{}{} = \ket{}{(0)}{} + \dots\,,
\label{eq:FormalLoopExpansion}
\eeq
where $\ket{}{(0)}{}$ denotes the tree-level contribution and the dots
stand for higher-loop contributions, which are not used in this paper. 

Colour interactions at QCD vertices are represented by associating
colour charges $\bT_i$ with the emission of a gluon from each
parton $i$.  In the colour-state notation, each vector $\ket{}{}{}$ is
a colour-singlet state, and colour conservation implies
\beq
\biggl(\sum_j \bT_j \biggr) \,\ket{}{}{} = 0\,,
\label{eq:colourcons}
\eeq
where the sum over $j$ extends over all the final state partons 
of the state vector $\ket{}{}{}$ (recall that we are considering 
processes where the initial state is colourless), and the equation is valid 
order by order in the loop expansion of \eqn{eq:FormalLoopExpansion}. 

Using the colour-state notation, we define the two-parton colour-correlated
squared tree amplitudes as
\beq
|\cM^{(0)}_{(i,k)}|^2 \equiv
\bra{}{(0)}{} \,\bT_i \ldot \bT_k \, \ket{}{(0)}{}
\label{eq:colam2}
\eeq
and similarly the four-parton colour-correlated squared
tree amplitudes, 
\beq
|\cM^{(0)}_{(i,k),(j,l)}|^2 \equiv
\bra{}{(0)}{} \{\bT_i \ldot \bT_k, \bT_j \ldot \bT_l\} \ket{}{(0)}{}\,.
\label{eq:colam4}
\eeq
We will also use the following $\otimes$ product notation to indicate 
the insertion of colour charge operators between $\bra{}{(0)}{}$ and $\ket{}{(0)}{}$:
\beq
\bsp
|\cM^{(0)}|^2 \otimes \bT_i \ldot \bT_k &\equiv 
	\bra{}{(0)}{} \,\bT_i \ldot \bT_k \, \ket{}{(0)}{}\,,
\\[2mm]
|\cM^{(0)}|^2 \otimes \{\bT_i \ldot \bT_k, \bT_j \ldot \bT_l\} &\equiv 
	\bra{}{(0)}{} \{\bT_i \ldot \bT_k, \bT_j \ldot \bT_l\} \ket{}{(0)}{}\,.
\esp
\label{eq:otimes-def}
\eeq
The colour-charge algebra for the product 
$\sum_{n} (\bT_i)^n (\bT_k)^n \equiv \bT_i \ldot \bT_k$ is:
\beq
\bT_i \ldot \bT_k =\bT_k \ldot \bT_i \quad  {\rm if}
\quad i \neq k; \qquad \bT_i^2= C_{f_i}\:.
\label{eq:colalg}
\eeq
Here $C_{f_i}$ is the quadratic Casimir operator in the representation of
particle $i$ and we have $C_q \equiv \CF= \TR(\Nc^2-1)/\Nc=
(\Nc^2-1)/(2\Nc)$ in the fundamental and $C_g \equiv \CA=2\,\TR
\Nc=\Nc$ in the adjoint representation, i.e.~we are using the customary
normalisation $\TR=1/2$.  

We also use squared colour-charges with multiple indices, such as
$\bT_{ir}^2\equiv C_{f_{ir}}$ and $\bT_{irs}^2\equiv C_{f_{irs}}$.
In such cases the multiple index denotes a single parton with flavour
obtained using the flavour summation rules: odd/even number of quarks
plus any number of gluons gives a quark/gluon, or explicitly for the
relevant cases at NNLO:
\beq
\bsp
& q + g = q\,,\quad q + \qb = g\,,\quad g + g = g\,,\quad
\\[2mm]
& q + g + g = q\,,\quad q + q + \qb = q\,,\quad
g + q + \qb = g\,,\quad g + g + g = g\,.
\label{eq:flavoursummation}
\esp
\eeq
%

%
%

\subsection{Cross sections}
\label{sec:xsecs}

In this paper we shall need to use only cross sections of 
producing $n$ partons at tree level with $n=m$, the Born cross 
section, and $n=m+2$, the so-called doubly-real correction. 
We have
\beq
\dsig{(0)}_{n}(\mom{}) =
{\cal N}\;\sum_{\{n\}}\PS{n}{(\mom{})}\frac{1}{S_{\{n\}}}
\SME{n}{0}{\mom{}}\,,
\label{eq:dsig0n}
\eeq
where ${\cal N}$ includes all QCD-independent factors and
$\PS{n}{(\mom{})}$ is the $d$-dimensional phase space for
$n$ outgoing particles with momenta
$\mom{} \equiv \{p_1,\dots,p_{n}\}$ and total incoming momentum 
$Q$,
\beq
\PS{n}(p_1,\ldots,p_n;Q) = 
\prod_{i=1}^{n}\frac{\rd^d p_i}{(2\pi)^{d-1}}\,\delta_+(p_i^2)\,
(2\pi)^d \delta^{(d)}\bigg(Q-\sum_{i=1}^{n}p_i\bigg)\,.
\label{eq:PSn}
\eeq
The symbol $\sum_{\{n\}}$ denotes summation over the different 
subprocesses and $S_{\{n\}}$ is the Bose symmetry factor for identical
particles in the final state. Then the Born cross section and the 
doubly-real correction are simply
\beq
\dsig{B}_{m}(\mom{}) \equiv \dsig{(0)}_{m}(\mom{})\qquad\mbox{and}\qquad
\dsig{RR}_{m+2}(\mom{}) \equiv \dsig{(0)}_{m+2}(\mom{})\,.
\eeq

The final result will also contain the
phase space factor due to the integral over the $(d-3)$-dimensional
solid angle, which is included in the definition of the running
coupling in the \msbar\ renormalisation scheme,\footnote{In the \msbar\
renormalisation scheme as often employed in the literature, the 
definition of the running coupling includes the factor 
$S_\ep=(4\pi)^\ep \e^{-\ep\gamma_E}$. The two definitions lead to the 
same expressions in a computation at the NLO accuracy. At NNLO these 
lead to slightly different bookkeeping of the IR and UV poles at intermediate 
steps of the computation, but the physical cross section of infrared safe
observables is the same.  Our definition leads to somewhat simpler
expressions at the NNLO level.}
\beq
S_\ep = \int\!\frac{\rd^{(d-3)}\Omega}{(2\pi)^{d-3}} =
\frac{(4\pi)^\ep}{\Gamma(1-\ep)}\,.
\label{eq:Seps}
\eeq
%

%
%

\subsection{Momentum mappings and phase space factorisation}
\label{sec:psfact}

The iterated subtraction terms are written in terms of momenta obtained 
via various combinations of the basic collinear and soft mappings of 
\refr{Somogyi:2006da}:
\beq
\mom{}_{m+2} \stackrel{{\mathsf X}_{R}}{\longrightarrow}
\momh{(R)}_{m+1} \stackrel{{\mathsf Y}_{\ha{S}}}{\longrightarrow}
\momt{(\ha{S},R)}_{m}\,,
\eeq
where both $\stackrel{{\mathsf X}_{R}}{\longrightarrow}$ and 
$\stackrel{{\mathsf Y}_{\ha{S}}}{\longrightarrow}$ may label either a 
collinear or soft mapping. (In general, both $R$ and $\ha{S}$ are 
multiple indices.) As the above notation suggests, the final set of $m$ 
momenta are denoted by tildes, while hats indicate the intermediate 
set of $m+1$ momenta. In {\em kinematic} expressions where only 
the label of a momentum is displayed (we shall discuss several examples 
below), the tilde and/or hat is inherited by the label, and we write 
for instance $\ha{i}$, $\wha{ir}$ and $\wti{irs}$, where the latter two 
label {\em single momenta}. However, since these mappings affect only 
the momenta, but not the colour and flavour (apart from the flavour summation 
rules of \eqn{eq:flavoursummation}), we shall omit the hat and tilde from 
flavour and colour indices.

We also use labels such as $(ir)$ to denote 
a {\em single momentum} that is simply the sum of two momenta,
$p_{(ir)} \equiv p_i + p_r$.

Importantly, both the collinear and soft momentum mappings lead to 
an exact factorisation of phase space as follows:
\bal
& \mom{}_{n+1} \cmap{ir} \momb{(ir)}_{n} &:\quad\;\; &
\PS{n+1}(\mom{}_{n+1};Q) = \PS{n}(\momb{(ir)}_{n},Q)
[\rd p_{1,n}^{(ir)}(p_r,\bar{p}_{ir};Q)]\,,
\\[2mm]
& \mom{}_{n+1} \smap{r} \momb{(r)}_{n} &:\quad\;\; &
\PS{n+1}(\mom{}_{n+1};Q) = \PS{n}(\momb{(r)}_{n},Q)
[\rd p_{1,n}^{(r)}(p_r;Q)]\,,
\eal
where here and below the bar denotes either a hat, when $n=m+1$, or 
a tilde, when $n=m$. The one-particle factorized phase spaces can be 
written in the following form. For the collinear mapping we have
\beq
[\rd p_{1,n}^{(ir)}(p_r,\bar{p}_{ir};Q)] =
     \rd \al_{ir} (1-\al_{ir})^{2(n-1)(1-\ep)-1}\frac{s_{\overline{ir}Q}}{2\pi}
     \PS{2}(p_i,p_r;p_{(ir)})\Theta(\al_{ir})\Theta(1-\al_{ir})\,,
\label{eq:dpir_n}
\eeq
where the $p_{(ir)}$ appearing on the right hand side is understood to be written 
in terms of mapped momenta, that is 
$p_{(ir)}^\mu = (1-\al_{ir})\bar{p}_{ir}^\mu + \al_{ir}Q^\mu$. For the soft mapping 
we find
\beq
[\rd p_{1,n}^{(r)}(p_r;Q)] =
     \rd y_{rQ} (1-y_{rQ})^{(n-1)(1-\ep)-1} \frac{Q^2}{2\pi}
     \PS{2}(p_r,K;Q)\Theta(y_{rQ})\Theta(1-y_{rQ})\,,
\label{eq:dpr_n}
\eeq
where the momentum $K$ is massive and $K^2 = (1-y_{rQ})Q^2$. As the
notation above indicates, $\al_{ir}$ and $y_{rQ}$ will become integration 
variables, hence their precise definitions are not important and will not be 
recalled here. (See \refr{Somogyi:2006da} for details.)

%
%

\subsection{Kinematic variables}
\label{sec:kinematic}

Three types of kinematic variables are used to write the iterated subtraction 
terms. The precise definitions of these were given in \refr{Somogyi:2006da}. 
Here we recall only those formulae that are needed for defining every 
expression precisely in the present paper.
\begin{itemize}
\item Two-particle invariants, such as 
\beq
s_{ir} = 2 p_i\ldot p_r\,,\quad 
s_{\wha{kt}\ha{r}} = 2 \ha{p}_{kt}\ldot \ha{p}_r\,,\qquad
\mbox{or}\qquad s_{iQ} = 2 p_i\ldot Q\quad \mbox{and}\quad
s_{i k_{\perp}} = 2 p_i\ldot k_{\perp}\,.
\eeq
The two-particle invariants scaled with $Q^2$ are denoted by
$y_{ij} = s_{ij}/Q^2$.
\item Momentum fractions $z_{i,r}$ and $z_{\ha{k},\ha{t}}$ for the 
splittings $\ha{p}_{ir} \to p_i + p_r$ and $\ti{p}_{kt} \to \ha{p}_k + \ha{p}_t$,
\beq
z_{i,r} = \frac{y_{iQ}}{y_{iQ}+y_{rQ}}\qquad\mbox{and}\qquad
z_{\ha{k},\ha{t}} = \frac{y_{\ha{k}Q}}{y_{\ha{k}Q}+y_{\ha{t}Q}}\,,
\eeq
with $z_{r,i} = 1-z_{i,r}$ and $z_{\ha{t},\ha{k}} = 1-z_{\ha{k},\ha{t}}$.
Momentum fractions for three-particle splittings are denoted by
\beq
z_{k,tr} = \frac{y_{kQ}}{y_{kQ}+y_{tQ}+y_{rQ}}\,,
\eeq
with the expressions for $z_{r,kt}$ and $z_{t,rk}$ obtained by cyclic permutation. 
In the following, all momentum fractions will be integrated out, and so they will be 
expressed using the integration variables.
\end{itemize}
We also use extensively the uncontracted and contracted eikonal factors:
\beq
\calS_{jl}^{\mu\nu}(r) = \frac{p_j^\mu p_l^\nu}{p_j\ldot p_r\, p_r \ldot p_l}
\,,\qquad
\calS_{jl}(r) = g_{\mu\nu}\calS_{jl}^{\mu\nu}(r) = \frac{2s_{jl}}{s_{jr} s_{lr}}
\,.
\label{eq:cSdef}
\eeq

As mentioned above, the sum of two momenta is often abbreviated with the 
two indices in parenthesis, e.g.\ $p_i^\mu + p_r^\mu = p_{(ir)}^\mu$, which is 
used systematically in other occurrences, such as
\beq
s_{(ir)k} =  s_{ik}+s_{rk}
\qquad\mbox{and}\qquad
\calS_{(ir)l}(t) =
\frac{2s_{(ir)l}}{s_{(ir)t} s_{lt}}
= \frac{s_{il} + s_{rl}}{(s_{it} + s_{rt}) s_{lt}}\,.
\label{eq:softcolleikonal}
\eeq

Finally, we express the integrated iterated subtraction terms as functions of the 
following (combinations of) invariants: 
\beq
x_i = y_{iQ}\qquad\mbox{and}\qquad 
Y_{ij,Q} = \frac{y_{ij}}{y_{iQ}\,y_{jQ}}\,.
\label{eq:xi_YijQdef}
\eeq
In the centre-of-momentum frame (i.e.\ the rest frame of $Q^\mu$), we find that 
$x_i = 2E_i/\sqrt{s}$ ($s=Q^2$) is simply the scaled energy of parton $i$, while 
$Y_{ij,Q} = (1-\cos\chi_{ij})/2$, where $\chi_{ij}$ is the angle between momenta 
$p_i^\mu$ and $p_j^\mu$.




\newpage


\section{Integrating the iterated singly-unresolved approximate cross section}
\label{sec:RRA12}

%
%

\subsection{The integrated approximate cross section and insertion operator}
\label{sec:IntRRA12}

We begin by recalling that the doubly-real emission cross section 
is defined precisely as in \eqn{eq:dsig0n}, with $n=m+2$. Then the 
iterated singly-unresolved approximate cross section times the jet 
function reads \cite{Somogyi:2006da}
\beq
\bsp
\dsiga{RR}{12}_{m+2} \odot J_m &=
	{\cal N} \sum_{\{m+2\}} \PS{m+2}(\mom{})\frac{1}{S_{\{m+2\}}}
	\sum_t \bigg[\sum_{k\ne t} \frac{1}{2}\cC{kt}{(0,0)}\cA_{2}
	\SME{m+2}{0}{\mom{}}
\\[2mm] &
	+\bigg(\cS{t}{(0,0)}\cA_{2}\SME{m+2}{0}{\mom{}}
	-\sum_{k\ne t}\cCS{kt}{t}{(0,0)}\cA_{2}\SME{m+2}{0}{\mom{}}\bigg)
	\bigg] \odot J_m(\momt{})
\,,
\esp
\label{eq:dsigRRA12Jm}
\eeq
where the notation $\odot\, J_m$ means that the jet function
multiplying the different terms in the sum depends on different sets of
momenta. Explicitly, the three terms in \eqn{eq:dsigRRA12Jm} are given 
by
\beq
\bsp
&
\cC{kt}{(0,0)}\cA_{2}\SME{m+2}{0}{\mom{}} \odot J_m(\momt{}) \equiv
\\[2mm] &\qquad \equiv
	\sum_{r\ne k,t}\bigg[\cC{kt}{}\cC{ktr}{(0,0)} J_m(\momt{(\wha{kt}\ha{r},kt)})
	+ \cC{kt}{}\cSCS{kt;r}{(0,0)} J_m(\momt{(\ha{r},kt)})
\\[2mm] &\qquad\qquad\qquad
	-\cC{kt}{}\cC{ktr}{}\cSCS{kt;r}{(0,0)} J_m(\momt{(\ha{r},kt)})
	-\cC{kt}{}\cC{rkt}{}\cS{kt}{(0,0)} J_m(\momt{(\wha{kt},kt)})
\\[2mm] &\qquad\qquad\qquad
	+\sum_{i\ne r,k,t}\bigg(
	\frac{1}{2}\cC{kt}{}\cC{ir;kt}{(0,0)} J_m(\momt{(\ha{i}\ha{r},kt)})
	-\cC{kt}{}\cC{ir;kt}{}\cSCS{kt;r}{(0,0)}  J_m(\momt{\ha{r},kt)})\bigg)\bigg]
\\[2mm] &\qquad
	+\cC{kt}{}\cS{kt}{(0,0)} J_m(\momt{(\wha{kt},kt)})\,,
\label{eq:CktA2} 
\esp
\eeq
\beq
\bsp
&
\cS{t}{(0,0)}\cA_{2}\SME{m+2}{0}{\mom{}} \odot J_m(\momt{}) \equiv
\\[2mm] &\qquad \equiv
	\sum_{r\ne t}\bigg\{\sum_{i\ne r,t}\bigg[\frac{1}{2}\bigg(
	\cS{t}{}\cC{irt}{(0,0)} J_m(\momt{(\ha{i}\ha{r},t)})
	+\cS{t}{}\cSCS{ir;t}{(0,0)} J_m(\momt{(\ha{i}\ha{r},t)})
\\[2mm] &\qquad\qquad\qquad\qquad
	-\cS{t}{}\cC{irt}{}\cSCS{ir;t}{(0,0)} J_m(\momt{(\ha{i}\ha{r},t)})\bigg)
	-\cS{t}{}\cC{irt}{}\cS{rt}{(0,0)} J_m(\momt{(\ha{r},t)})
\\[2mm] &\qquad\qquad\qquad\qquad
	-\cS{t}{}\cSCS{ir;t}{}\cS{rt}{(0,0)} J_m(\momt{(\ha{r},t)})
	+\cS{t}{}\cC{irt}{}\cSCS{ir;t}{}\cS{rt}{(0,0)} J_m(\momt{(\ha{r},t)})\bigg]
\\[2mm] &\qquad\qquad\qquad
	+\cS{t}{}\cS{rt}{(0,0)} J_m(\momt{(\ha{r},t)})\bigg\}
\label{eq:StA2} 
\esp
\eeq
and
\beq
\bsp
&
\cCS{kt}{t}{(0,0)}\cA_{2}\SME{m+2}{0}{\mom{}} \odot J_m(\momt{}) \equiv
\\[2mm] &\qquad \equiv
	\sum_{r\ne k,t}\bigg[
	\cCS{kt}{t}{}\cC{krt}{(0,0)} J_m(\momt{(\ha{k}\ha{r},t)})
\\[2mm] &\qquad\qquad
	+\sum_{i\ne r,k,t}\bigg(
	\frac{1}{2}\cCS{kt}{t}{}\cSCS{ir;t}{(0,0)} J_m(\momt{(\ha{i}\ha{r},t)})
	-\cCS{kt}{t}{}\cSCS{ir;t}{}\cS{rt}{(0,0)} J_m(\momt{(\ha{r},t)})\bigg)
\\[2mm] &\qquad\qquad
	-\cCS{kt}{t}{}\cC{krt}{}\cS{rt}{(0,0)} J_m(\momt{(\ha{r},t)})
	-\cCS{kt}{t}{}\cC{rkt}{}\cS{kt}{(0,0)} J_m(\momt{(\ha{k},t)})
\\[2mm] &\qquad\qquad
	+\cCS{kt}{t}{}\cS{rt}{(0,0)} J_m(\momt{(\ha{r},t)})\bigg]
	+\cCS{kt}{t}{}\cS{kt}{(0,0)} J_m(\momt{(\ha{k},t)})\,.
\label{eq:CktStA2}
\esp
\eeq

All momentum mappings in \eqnss{eq:CktA2}{eq:CktStA2} lead to the
factorisation of the original $m+2$ parton phase space into an $m$ parton
phase space times two one-parton phase space measures, as discussed  
originally in \refr{Somogyi:2006da}, and recalled in \sect{sec:psfact} above. 
Symbolically, we may write 
\beq
\PS{m+2}(\mom{}) = \PS{m}(\momt{}) [\rd p_{1,m}] [\rd p_{1,m+1}]\,.
\eeq
The jet function does not depend on the variables of the factorized
one-parton measures, $[\rd p_{1,m}] [\rd p_{1,m+1}]$, so we can compute
the integral of \eqn{eq:dsigRRA12Jm} over the phase space of the two
unresolved partons, independently of the jet function $J_m$, that we
shall omit in the following. The result of the integration is a long
expression of kinematics-dependent functions --- each corresponding to a
specific iteration of unresolved limits of the squared matrix elements
--- times colour factors,
\beq
\bsp
&
\int_2\dsiga{RR}{12}_{m+2} =
	{\cal N} \sum_{\{m+2\}} \PS{m}(\momt{}) \frac{1}{S_{\{m+2\}}}
	\left[\frac{\as}{2\pi}S_\ep \left(\frac{\mu^2}{Q^2}\right)^\ep\right]^2
\label{eq:I2dsigRRA12Jm-1}
\\[2mm] &\qquad\times
\sum_{t}\bigg\{
\sum_{k\ne t} \frac{1}{2}\bigg[
\sum_{r\ne k,t}\bigg(
	[\IcC{kt}{}\IcC{ktr}{(0)}]_{f_k f_t f_r} (\bT_{ktr}^2)^2
	+ \sum_{j,l} 
	[\IcC{kt}{}\IcSCS{kt;r}{(0)}]_{f_k f_t}^{(j,l)} \bT_{kt}^2 \bT_j \bT_l
\\[2mm] &\qquad\qquad\qquad\qquad\qquad\qquad
	-[\IcC{kt}{}\IcC{ktr}{}\IcSCS{kt;r}{(0)}]_{f_k f_t} (\bT_{kt}^2)^2
	-[\IcC{kt}{}\IcC{rkt}{}\IcS{kt}{(0)}]_{f_r f_k f_t} (\bT_{r}^2)^2\bigg)
\\[2mm] &\qquad\qquad\qquad\quad
	+\sum_{r\ne k,t}\sum_{i\ne r,k,t}\bigg(
	\frac{1}{2}[\IcC{kt}{}\IcC{ir;kt}{(0)}]_{f_k f_t;f_i f_r} \bT_{kt}^2 \bT_{ir}^2
	-[\IcC{kt}{}\IcC{ir;kt}{}\IcSCS{kt;r}{(0)}]_{f_k f_t} \bT_{kt}^2 \bT_{i}^2 \bigg)
\\[2mm] &\qquad\qquad\qquad\quad
	+\sum_{j,l} [\IcC{kt}{}\IcS{kt}{(0)}]_{f_k f_t}^{(j,l)} \CA \bT_j \bT_l\bigg]
\\[2mm] &\qquad\qquad
+\sum_{r\ne t}\sum_{i\ne r,t}\bigg[\frac{1}{2}\bigg(
	[\IcS{t}{}\IcC{irt}{(0)}]_{f_i f_r} (\bT_{ir}^2)^2
	+\sum_{j,l} 
	[\IcS{t}{}\IcSCS{ir;t}{(0)}]_{f_i f_r}^{(j,l)} \bT_{ir}^2 \bT_j \bT_l
\\[2mm] &\qquad\qquad\qquad\qquad\qquad
	-[\IcS{t}{}\IcC{irt}{}\IcSCS{ir;t}{(0)}]_{f_i f_r} (\bT_{ir}^2)^2\bigg)
	-[\IcS{t}{}\IcC{irt}{}\IcS{rt}{(0)}]_{f_i} (\bT_i^2)^2
\\[2mm]&\qquad\qquad\qquad\qquad
	-\sum_{j,l} 
	[\IcS{t}{}\IcSCS{ir;t}{}\IcS{rt}{(0)}]^{(j,l)} \bT_i^2 \bT_j \bT_l
	+[\IcS{t}{}\IcC{irt}{}\IcSCS{ir;t}{}\IcS{rt}{(0)}] (\bT_i^2)^2\bigg]
\\[2mm]&\qquad\qquad\qquad
	+\sum_{r\ne t}\bigg[
		\sum_{i,k,j,l}
		[\IcS{t}{}\IcS{rt}{(0)}]^{(i,k)(j,l)} \{\bT_i \bT_k , \bT_j \bT_l\}
		+\sum_{i,k} 
		[\IcS{t}{}\IcS{rt}{(0)}]^{(i,k)} \CA \bT_i \bT_k\bigg]
\\[2mm] &\qquad\qquad
-\sum_{k\ne t}\bigg[\sum_{r\ne k,t}\bigg(
	[\IcCS{kt}{t}{}\IcC{krt}{(0)}]_{f_k f_t} (\bT_{krt}^2)^2
	-[\IcCS{kt}{t}{}\IcC{krt}{}\IcS{rt}{(0)}] (\bT_k^2)^2
\\[2mm] &\qquad\qquad\qquad\qquad\quad
	-[\IcCS{kt}{t}{}\IcC{rkt}{}\IcS{kt}{(0)}] (\bT_r^2)^2
	+\sum_{j,l} 
	[\IcCS{kt}{t}{}\IcS{rt}{(0)}]^{(j,l)} \bT_k^2 \bT_j \bT_l\bigg)
\\[2mm] &\qquad\qquad\qquad\quad
	+\sum_{r\ne k,t}\sum_{i\ne r,k,t}\bigg(
	\frac{1}{2}[\IcCS{kt}{t}{}\IcSCS{ir;t}{(0)}]_{f_i f_r} \bT_{ir}^2 \bT_{k}^2
	-[\IcCS{kt}{t}{}\IcSCS{ir;t}{}\IcS{rt}{(0)}] \bT_i^2 \bT_k^2\bigg)
\\[2mm] &\qquad\qquad\qquad\quad
	+\sum_{j,l} 
	[\IcCS{kt}{t}{}\IcS{kt}{(0)}]^{(j,l)} \CA \bT_j \bT_l
\bigg]\bigg\} \otimes \M{m}{(0)}\,,
\esp
\eeq
where the operation $\otimes$ means insertion of the colour charges
between $\bra{m}{(0)}{}$ and $\ket{m}{(0)}{}$, see \eqn{eq:otimes-def}.  
Three types of colour connections appear in \eqn{eq:I2dsigRRA12Jm-1}, 
and the functions on the right hand side --- the ``non flavour summed'' 
(see below) integrated counterterms --- take three different forms accordingly:
\bal
\int_2 {\cal X}_1^{(0,0)} &=
\left[\frac{\as}{2\pi}S_\ep \left(\frac{\mu^2}{Q^2}\right)^\ep\right]^2
[X_1^{(0)}]_{f_i\dots} \bT^2_x \bT^2_y \M{m}{(0)}\,,
\label{eq:intX}
\\[2mm]
\int_2 {\cal X}_2^{(0,0)} &=
\left[\frac{\as}{2\pi}S_\ep \left(\frac{\mu^2}{Q^2}\right)^\ep\right]^2
[X_2^{(0)}]_{f_i\dots} ^{(j,l)} \bT^2_x \M{m,(j,l)}{(0)}\,,
\label{eq:intXjl}
\\[2mm]
\int_2 {\cal X}_3^{(0,0)} &=
\left[\frac{\as}{2\pi}S_\ep \left(\frac{\mu^2}{Q^2}\right)^\ep\right]^2
[X_3^{(0)}]^{(i,k)(j,l)} \M{m,(i,k)(j,l)}{(0)}\,,
\label{eq:intXikjl}
\eal
where e.g.\ $[X_1^{(0)}]_{f_i\dots}$ represents a function that results in the 
integration of the counterterm ${\cal X}_1^{(0,0)}$. The quadratic Casimir 
operators that appear in \eqns{eq:intX}{eq:intXjl} are factored out to make the
integrals $[X_1^{(0)}]_{f_i\dots} $ and $[X_2^{(0)}]_{f_i\dots} ^{(j,l)}$ (together with
$[X_3^{(0)}]^{(i,k)(j,l)}$) dimensionless in colour-space. As the notation implies, 
$[X_1^{(0)}]_{f_i\dots}$ and $[X_2^{(0)}]_{f_i\dots}$ may also carry flavour 
dependence. Incidentally, we note that for every integrated counterterm in 
\eqn{eq:I2dsigRRA12Jm-1}, we consider everything inside the square brackets 
to be simply part of the function's name. In particular, the lower indices 
{\em inside square brackets} loose their meaning. Nevertheless, we choose to 
keep these in order to exhibit from which counterterm each function derives. 

\Eqn{eq:I2dsigRRA12Jm-1} is not yet in the form of an $m$-parton
contribution times a factor. In order to obtain such a form, we must
still perform summation over unresolved flavours, rewriting the symmetry 
factor of an $m+2$-parton configuration to the symmetry factor of an $m$ 
parton configuration. The complete details of this counting are not very 
difficult, but rather long, and are given in \appx{sec:flavoursum}.
As a result of these manipulations, we obtain functions --- the flavour summed 
integrated counterterms --- denoted by 
$\big(X^{(0)}\big)_{f_i\dots} ^{(j,l)\dots}$, which are specific sums of the 
non flavour summed integrated subtraction terms, symbolically
\beq
\Big(X^{(0)}\Big)_{f_i\dots} ^{(j,l)\dots} = \sum \,[X^{(0)}]_{f_k\dots} ^{(j,l)\dots}\,. 
\eeq
It is important to realise that the {\em flavour indices} on the left and right hand 
sides of the above equation need not match up. Indeed, the non flavour summed 
functions on the right hand side carry dependence on unresolved flavours, while 
the flavour summed functions on the left do not, by definition.
Similar change in notation was introduced in the dipole subtraction scheme 
\cite{Catani:1996vz}, where we find the definitions (see eqs.~(7.21) and (7.22))
\bal
{\cal V}_i(\ep) &\equiv {\cal V}_{qg}(\ep)\,, 
& \mbox{if}\;\; i=q,\qb\,,
\\[2mm]
{\cal V}_i(\ep) &\equiv \frac{1}{2}{\cal V}_{gg}(\ep) + \Nf {\cal V}_{q\qb}(\ep)\,, 
& \mbox{if}\;\; i=g\,,
\eal
where the functions ${\cal V}_i(\ep)$ on the left hand side represent 
flavour summed integrated counterterms, while the ${\cal V}_{ij}(\ep)$ functions 
on the right hand side are not flavour summed.
In the present paper, due to the extra complications of an NNLO subtraction scheme, 
we believe that it is helpful to make a sharper notational distinction between flavour 
summed and non flavour summed integrated counterterms.

After summation over unresolved flavours in \eqn{eq:I2dsigRRA12Jm-1}, we find 
that the final result can be written in the form
\beq
\int_2\dsiga{RR}{12}_{m+2} = \dsig{B}_{m} \otimes
\bI^{(0)}_{12}(\mom{}_m;\ep)\,,
\label{eq:IntRRA12-fin}
\eeq
where the insertion operator (in colour space) has three contributions according
to the possible colour structures in \eqnss{eq:intX}{eq:intXikjl}:
\beq
\bsp
\bI^{(0)}_{12} = 
\left[\frac{\as}{2\pi}S_\ep \left(\frac{\mu^2}{Q^2}\right)^\ep\right]^2
&\bigg\{
\sum_{i}
\bigg[\rC^{(0)}_{12,f_i}\,C_{f_i} + \sum_{k\ne i}\rC^{(0)}_{12,f_i f_k}\,C_{f_k}\bigg]
C_{f_i}
\\[2mm]&
+\sum_{j,l\ne j}
\bigg[\rS^{(0),(j,l)}_{12} \CA +\sum_{i} \rSCS^{(0),(j,l)}_{12,f_i} C_{f_i}\bigg]
\bT_j \bT_l
\\[2mm]&
+\sum_{i,k\ne i} \sum_{j,l\ne j}
\rS^{(0),(i,k)(j,l)}_{12} \{\bT_i \bT_k , \bT_j \bT_l\}
\bigg\}\,,
\label{eq:I12}
\esp
\eeq
with $f_i$ denoting flavours, and $C_q = \CF \equiv \bT_q^2$,  $C_g = \CA
\equiv \bT_g^2$ as in \eqn{eq:colalg}. 

\Eqns{eq:IntRRA12-fin}{eq:I12} 
are the main results of this paper. 

In the following, we shall define and compute each term appearing in 
\eqn{eq:I12}. First, in terms of flavour summed integrated counterterms 
discussed above, we get:
\bal
\rC^{(0)}_{12,f_i} & = 
\Big(\IcC{kt}{}\IcC{ktr}{(0)}\Big)_{f_i}
-\Big(\IcC{kt}{}\IcC{ktr}{}\IcSCS{kt;r}{(0)}\Big)_{f_i}
-\Big(\IcC{kt}{}\IcC{rkt}{}\IcS{kt}{(0)}\Big)_{f_i}
\nt \\[2mm] &
+\Big(\IcS{t}{}\IcC{irt}{(0)}\Big)_{f_i}
-\Big(\IcS{t}{}\IcC{irt}{}\IcSCS{ir;t}{(0)}\Big)_{f_i}
-\Big(\IcS{t}{}\IcC{irt}{}\IcS{rt}{(0)}\Big)_{f_i}
\label{eq:Ifi}
\\[2mm]\nt  &
+\Big(\IcS{t}{}\IcC{irt}{}\IcSCS{ir;t}{}\IcS{rt}{(0)}\Big)_{f_i}
-\Big(\IcCS{kt}{t}{}\IcC{krt}{(0)}\Big)_{f_i}
+\Big(\IcCS{kt}{t}{}\IcC{krt}{}\IcS{rt}{(0)}\Big)_{f_i}
\,,
\\[2mm]
\rC^{(0)}_{12,f_if_k} &=
\Big(\IcC{kt}{}\IcC{ir;kt}{(0)}\Big)_{f_if_k}
-\Big(\IcC{kt}{}\IcC{ir;kt}{}\IcSCS{kt;r}{(0)}\Big)_{f_if_k}
\nt \\[2mm] &
-\Big(\IcCS{kt}{t}{}\IcSCS{ir;t}{(0)}\Big)_{f_if_k}
+\Big(\IcCS{kt}{t}{}\IcSCS{ir;t}{}\IcS{rt}{(0)}\Big)_{f_if_k}
\label{eq:Ififk}
\,,
\\[2mm]
\rS^{(0),(j,l)}_{12} &=
\Big(\IcC{kt}{}\IcS{kt}{(0)}\Big)^{(j,l)}
+\Big(\IcS{t}{}\IcS{rt}{(0)}\Big)^{(j,l)}
-\Big(\IcCS{kt}{t}{}\IcS{kt}{(0)}\Big)^{(j,l)}
\label{eq:Ijl}
\,,
\\[2mm]
\rSCS^{(0),(j,l)}_{12,f_i} &=
\Big(\IcC{kt}{}\IcSCS{kt;r}{(0)}\Big)^{(j,l)}_{f_i}
+\Big(\IcS{t}{}\IcSCS{ir;t}{(0)}\Big)^{(j,l)}_{f_i}
-\Big(\IcS{t}{}\IcSCS{ir;t}{}\IcS{rt}{(0)}\Big)^{(j,l)}_{f_i}
\nt \\[2mm] &
-\Big(\IcCS{kt}{t}{}\IcS{rt}{(0)}\Big)^{(j,l)}_{f_i}
\label{eq:Ifijl}
\,,
\\[2mm]
\rS^{(0),(i,k)(j,l)}_{12} &=
\Big(\IcS{t}{}\IcS{rt}{(0)}\Big)^{(i,k)(j,l)}\,.
\label{eq:Iikjl}
\eal
On the right hand sides of these equations the flavour dependent functions
are the flavour summed integrated counterterms discussed above. They depend 
on the kinematics through variables of the type $x_i$ and $Y_{ij,Q}$. The latter 
dependence derives from integrating an eikonal factor which is always multiplied 
with a colour-connected squared matrix element. In order to make the results more 
transparent, we hid the arguments of the functions, but kept the relation to the 
colour-connected matrix elements, shown as upper indices.

%
%

\subsection{Flavour summed integrated counterterms}
\label{sec:flavsummedICTs}

Here we list the flavour summed integrated counterterms appearing on the 
right hand sides of \eqnss{eq:Ifi}{eq:Iikjl}, written in terms of the integrated 
subtraction terms.


\paragraph{Collinear-type terms:}
\begin{enumerate}[leftmargin=*]
\item Collinear-triple collinear:
\beq
\bsp
\Big(\IcC{kt}{}\IcC{ktr}{(0)}\Big)_q &= 
  [\IcC{kt}{}\IcC{ktr}{(0)}]_{qgg}
+ \frac12 [\IcC{kt}{}\IcC{ktr}{(0)}]_{ggq} 
+ \Nf [\IcC{kt}{}\IcC{ktr}{(0)}]_{q'\qb' q}
\label{eq:C-TC}
\,,
\\[2mm]
\Big(\IcC{kt}{}\IcC{ktr}{(0)}\Big)_g &= 
\frac12 [\IcC{kt}{}\IcC{ktr}{(0)}]_{ggg}
+ \Nf [\IcC{kt}{}\IcC{ktr}{(0)}]_{q\qb g}
+ 2 \Nf [\IcC{kt}{}\IcC{ktr}{(0)}]_{g q\qb}
\,.
\esp
\eeq
\item Collinear-double collinear:
\beq
\bsp
\Big(\IcC{kt}{}\IcC{ir;kt}{(0)}\Big)_{qq} &= 
[\IcC{kt}{}\IcC{ir;kt}{(0)}]_{qg;qg}
\,,
\\[2mm]
\Big(\IcC{kt}{}\IcC{ir;kt}{(0)}\Big)_{qg} &= 
\frac12 [\IcC{kt}{}\IcC{ir;kt}{(0)}]_{qg;gg}
+ \Nf [\IcC{kt}{}\IcC{ir;kt}{(0)}]_{qg;q\qb}
\,,
\\[2mm]
\Big(\IcC{kt}{}\IcC{ir;kt}{(0)}\Big)_{gq} &= 
\frac12 [\IcC{kt}{}\IcC{ir;kt}{(0)}]_{gg;qg}
+ \Nf [\IcC{kt}{}\IcC{ir;kt}{(0)}]_{q\qb;qg}
\,,
\\[2mm]
\Big(\IcC{kt}{}\IcC{ir;kt}{(0)}\Big)_{gg} &= 
\frac14 [\IcC{kt}{}\IcC{ir;kt}{(0)}]_{gg;gg}
+ \Nf^2 [\IcC{kt}{}\IcC{ir;kt}{(0)}]_{q\qb;q\qb}
\\[2mm] &
+ \frac12 \Nf
\Big([\IcC{kt}{}\IcC{ir;kt}{(0)}]_{q\qb;gg}
 +   [\IcC{kt}{}\IcC{ir;kt}{(0)}]_{gg;q\qb}\Big)
\,.
\esp
\eeq
\item Collinear-soft-collinear:
\beq
\bsp
\Big(\IcC{kt}{}\IcSCS{kt;r}{(0)}\Big)^{(j,l)}_q &= 
[\IcC{kt}{}\IcSCS{kt;r}{(0)}]_{qg}^{(j,l)}
\,,
\\[2mm]
\Big(\IcC{kt}{}\IcSCS{kt;r}{(0)}\Big)^{(j,l)}_g &= 
\frac12 [\IcC{kt}{}\IcSCS{kt;r}{(0)}]_{gg}^{(j,l)}
+ \Nf [\IcC{kt}{}\IcSCS{kt;r}{(0)}]_{q\qb}^{(j,l)}
\,.
\esp
\eeq
\item Collinear-triple collinear-soft-collinear:
\beq
\bsp
\Big(\IcC{kt}{}\IcC{ktr}{}\IcSCS{kt;r}{(0)}\Big)_q &=
[\IcC{kt}{}\IcC{ktr}{}\IcSCS{kt;r}{(0)}]_{qg}
\,,
\\[2mm]
\Big(\IcC{kt}{}\IcC{ktr}{}\IcSCS{kt;r}{(0)}\Big)_g &=
\frac12 [\IcC{kt}{}\IcC{ktr}{}\IcSCS{kt;r}{(0)}]_{gg}
+ \Nf [\IcC{kt}{}\IcC{ktr}{}\IcSCS{kt;r}{(0)}]_{q\qb}
\,.
\esp
\eeq
\item Collinear-double collinear-soft-collinear:
\beq
\bsp
\Big(\IcC{kt}{}\IcC{ir;kt}{}\IcSCS{kt;r}{(0)}\Big)_{qf} &=
[\IcC{kt}{}\IcC{ir;kt}{}\IcSCS{kt;r}{(0)}]_{qg}
\,,
\\[2mm]
\Big(\IcC{kt}{}\IcC{ir;kt}{}\IcSCS{kt;r}{(0)}\Big)_{gf} &=
\frac12 [\IcC{kt}{}\IcC{ir;kt}{}\IcSCS{kt;r}{(0)}]_{gg}
+ \Nf [\IcC{kt}{}\IcC{ir;kt}{}\IcSCS{kt;r}{(0)}]_{q\qb}
\,,
\esp
\eeq
i.e.~it is independent of the flavour $f$.
\item Collinear-double soft:
\beq
\Big(\IcC{kt}{}\IcS{kt}{(0)}\Big)^{(j,l)} =
\frac12 [\IcC{kt}{}\IcS{kt}{(0)}]_{gg}^{(j,l)}
+ \Nf [\IcC{kt}{}\IcS{kt}{(0)}]_{q\qb}^{(j,l)}
\,.
\label{eq:C2S2jl}
\eeq
\item Collinear-triple collinear-double soft:
\beq
\Big(\IcC{kt}{}\IcC{rkt}{}\IcS{kt}{(0)}\Big)_f =
\frac12 [\IcC{kt}{}\IcC{rkt}{}\IcS{kt}{(0)}]_{fgg}
+ \Nf [\IcC{kt}{}\IcC{rkt}{}\IcS{kt}{(0)}]_{fq\qb}
\,.
\eeq
\end{enumerate}


\paragraph{Soft-type terms:}
\begin{enumerate}[leftmargin=*]
\item Soft-triple collinear:
\beq
\bsp
\Big(\IcS{t}{}\IcC{irt}{(0)}\Big)_q &=
[\IcS{t}{}\IcC{irt}{(0)}]_{qg}
\,,
\\[2mm]
\Big(\IcS{t}{}\IcC{irt}{(0)}\Big)_g &=
\frac12 [\IcS{t}{}\IcC{irt}{(0)}]_{gg}
+ \Nf [\IcS{t}{}\IcC{irt}{(0)}]_{q\qb}
\,.
\esp
\eeq
\item Soft-soft-collinear:
\beq
\bsp
\Big(\IcS{t}{}\IcSCS{ir;t}{(0)}\Big)^{(j,l)}_q &=
[\IcS{t}{}\IcSCS{ir;t}{(0)}]_{qg}^{(j,l)}
\,,
\\[2mm]
\Big(\IcS{t}{}\IcSCS{ir;t}{(0)}\Big)^{(j,l)}_g &=
\frac12 [\IcS{t}{}\IcSCS{ir;t}{(0)}]_{gg}^{(j,l)}
+ \Nf [\IcS{t}{}\IcSCS{ir;t}{(0)}]_{q\qb}^{(j,l)}
\,.
\esp
\eeq
\item Soft-triple collinear-soft-collinear:
\beq
\bsp
\Big(\IcS{t}{}\IcC{irt}{}\IcSCS{ir;t}{(0)}\Big)_q &=
[\IcS{t}{}\IcC{irt}{}\IcSCS{ir;t}{(0)}]_{qg}
\,,
\\[2mm]
\Big(\IcS{t}{}\IcC{irt}{}\IcSCS{ir;t}{(0)}\Big)_g &=
\frac12 [\IcS{t}{}\IcC{irt}{}\IcSCS{ir;t}{(0)}]_{gg}
+ \Nf [\IcS{t}{}\IcC{irt}{}\IcSCS{ir;t}{(0)}]_{q\qb}
\,.
\esp
\eeq
\item Soft-triple collinear-double soft:
\beq
\Big(\IcS{t}{}\IcC{irt}{}\IcS{rt}{(0)}\Big)_f =
[\IcS{t}{}\IcC{irt}{}\IcS{rt}{(0)}]_f
\,.
\eeq
\item Soft-soft-collinear-double soft:
\beq
\Big(\IcS{t}{}\IcSCS{ir;t}{}\IcS{rt}{(0)}\Big)^{(j,l)}_f =
[\IcS{t}{}\IcSCS{ir;t}{}\IcS{rt}{(0)}]^{(j,l)}
\,,
\eeq
i.e.~it is independent of the flavour $f$.
\item Soft-triple collinear-soft-collinear-double soft
\beq
\Big(\IcS{t}{}\IcC{irt}{}\IcSCS{ir;t}{}\IcS{rt}{(0)}\Big)_f =
[\IcS{t}{}\IcC{irt}{}\IcSCS{ir;t}{}\IcS{rt}{(0)}]
\,,
\eeq
i.e.~it is independent of the flavour $f$.
\item Soft-double soft:
\beq
\bsp
\Big(\IcS{t}{}\IcS{rt}{(0)}\Big)^{(i,k)(j,l)} =
[\IcS{t}{}\IcS{rt}{(0)}]^{(i,k)(j,l)}
\,,
\\[2mm]
\Big(\IcS{t}{}\IcS{rt}{(0)}\Big)^{(j,l)} =
[\IcS{t}{}\IcS{rt}{(0)}]^{(j,l)}
\,.
\esp
\eeq
\end{enumerate}


\paragraph{Soft-collinear-type terms:}
\begin{enumerate}[leftmargin=*]
\item Soft-collinear-triple collinear:
\beq
\bsp
\Big(\IcCS{kt}{t}{}\IcC{krt}{(0)}\Big)_q &=
  [\IcCS{kt}{t}{}\IcC{krt}{(0)}]_{qg}
+ [\IcCS{kt}{t}{}\IcC{krt}{(0)}]_{gq}
\,,
\\[2mm]
\Big(\IcCS{kt}{t}{}\IcC{krt}{(0)}\Big)_g &=
[\IcCS{kt}{t}{}\IcC{krt}{(0)}]_{gg}
+ 2 \Nf [\IcCS{kt}{t}{}\IcC{krt}{(0)}]_{q\qb}
\,.
\esp
\eeq
\item Soft-collinear-soft-collinear:
\beq
\bsp
\Big(\IcCS{kt}{t}{}\IcSCS{ir;t}{(0)}\Big)_{f q} &=
[\IcCS{kt}{t}{}\IcSCS{ir;t}{(0)}]_{qg}
\,,
\\[2mm]
\Big(\IcCS{kt}{t}{}\IcSCS{ir;t}{(0)}\Big)_{f g} &=
\frac12 [\IcCS{kt}{t}{}\IcSCS{ir;t}{(0)}]_{gg}
+ \Nf [\IcCS{kt}{t}{}\IcSCS{ir;t}{(0)}]_{q\qb}
\,,
\esp
\eeq
i.e.~it is independent of the flavour $f$.
\item Soft-collinear-triple collinear-double soft:
\beq
\bsp
\Big(\IcCS{kt}{t}{}\IcC{krt}{}\IcS{rt}{(0)}\Big)_f &=
  [\IcCS{kt}{t}{}\IcC{krt}{}\IcS{rt}{(0)}]
+ [\IcCS{kt}{t}{}\IcC{rkt}{}\IcS{kt}{(0)}]
\,,
\esp
\eeq
i.e.~it is independent of the flavour $f$.
\item Flavour-dependent soft-collinear-double soft:
\beq
\bsp
\Big(\IcCS{kt}{t}{}\IcS{rt}{(0)}\Big)^{(j,l)}_f =
[\IcCS{kt}{t}{}\IcS{rt}{(0)}]^{(j,l)}
\,.
\esp
\eeq
Actually, as shown here, and also seen in the precise definition of
integrated flavour-dependent soft-collinear-double soft subtraction in
\eqn{eq:IcCktIcStIcSkt}, the integral itself does not depend on the
flavour. Distinguishing the flavour dependence serves book-keeping
purposes: the flavour-dependent subtraction contributes to
$\rSCS^{(0),(j,l)}_{12,f_i}$ in \eqn{eq:Ifijl}, while the flavour-independent
one in \eqn{eq:IcCSIcSjl} contributes to $\rS^{(0),(j,l)}_{12}$ in
\eqn{eq:Ijl}.
\item Soft-collinear-soft-collinear-double-soft:
\beq
\Big(\IcCS{kt}{t}{}\IcSCS{ir;t}{}\IcS{rt}{(0)}\Big)_{f_1 f_2} =
[\IcCS{kt}{t}{}\IcSCS{ir;t}{}\IcS{rt}{(0)}]
\,,
\label{eq:SC-SC-DS}
\eeq
i.e.~it is independent of both flavours $f_1$, $f_2$.
\item Flavour-independent soft-collinear-double soft:
\beq
\bsp
\Big(\IcCS{kt}{t}{}\IcS{kt}{(0)}\Big)^{(j,l)} =
[\IcCS{kt}{t}{}\IcS{kt}{(0)}]^{(j,l)}
\,.
\label{eq:IcCSIcSjl}
\esp
\eeq
\end{enumerate}

%
%

\subsection{Analytic expressions to $\Oe{-2}$}

In the next section we compute the functions on the right hand sides of
\eqnss{eq:C-TC}{eq:IcCSIcSjl} in terms of basic integrals that are
calculated in appendices. Expanding these integrals in $\ep$ we obtain
the Laurent expansions of the functions on the left hand sides of 
\eqnss{eq:C-TC}{eq:IcCSIcSjl}. 
Analytic expressions for the expansion coefficients have been obtained to 
$\Oe{-2}$ accuracy in all cases, and we present these below.
However, in the case of the single and double poles 
as well as the finite terms, we encountered several instances where obtaining 
complete analytic expressions was not feasible. This being the case, we made no 
severe effort to derive analytic expressions beyond those presented here. 
Higher order coefficients in the Laurent expansions will be given numerically 
in the form of a computer code elsewhere. In the following results we 
use $d_0' = D_0' + d_1' \ep$ (see \appx{sec:modsubterms} and especially 
\eqn{eq:d0s}) and the abbreviations
\beq
\B(\Nf) \equiv
\frac{\beta_0(\Nf)}{\CA} =
\frac{11\CA - 4\TR\Nf}{3\CA}
\,,
\label{eq:beta0}
\eeq
\beq
\Sigma(z,N) = \ln z - \sum_{k=1}^N \frac{1 - (1 - z)^k}{k}
\,.
\label{eq:Sigma}
\eeq
%



\paragraph{Collinear-type terms:}
\begin{enumerate}[leftmargin=*]
\item Collinear-triple collinear:
\beq
\bsp
\lbb\IcC{kt}{}\IcC{ktr}{(0)}\rbb_{f_i}(x_i) &= 
{\CA + 2 C_{f_i} \over 2 C_{f_i}} \bigg[
  \pole14
- \pole13
\bigg(4\ln{x_i} - 3\,\delf{f_i}{q} - \frac12 \Bf{\Nf} \,\delf{f_i}{g}\bigg)\bigg]
\\[2mm]&
+ \pole13 {\CA \over C_{f_i}} \Bf{{\CF \over C_{f_i}}\Nf}
  \bigg(\frac14\, \delf{f_i}{q} + \delf{f_i}{g} \bigg)
+ \Oe{-2}
\,.
\esp
\label{eq:seriesbegin}
\eeq
\item Collinear-double collinear:
\beq
\bsp
\lbb\IcC{kt}{}\IcC{ir;kt}{(0)}\rbb_{f_if_k}(x_i,x_k) &= 
  \pole14 - \pole13 \bigg[2 \bigg(\ln{x_i} + \ln{x_k}\bigg)
- 3\,\delf{f_i}{q}\,\delf{f_k}{q} - \Bf{\Nf}\,\delf{f_i}{g}\,\delf{f_k}{g}
\\[2mm]&\qquad\qquad
- \frac12 \bigg(3 + \Bf{\Nf}\bigg)
  \bigg(\delf{f_i}{q}\,\delf{f_k}{g} + \delf{f_i}{g}\,\delf{f_k}{q}\bigg)
\bigg]
+ \Oe{-2}
\,.
\esp
\eeq
\item Collinear-soft-collinear:
\beq
\bsp
\lbb\IcC{kt}{}\IcSCS{kt;r}{(0)}\rbb^{(j,l)}_{f_i}(x_i,Y_{jl,Q}) &= 
- \pole14 + \pole23 \bigg(\ln{x_i} + \Sigma(y_0, D'_0-1)\bigg)
\\[2mm]&
+ \pole13 \bigg[\ln{Y_{jl,Q}}
- \frac12 \bigg(3\,\delf{f_i}{q} + \Bf{\Nf}\,\delf{f_i}{g} \bigg)
\bigg]
+ \Oe{-2}
\,.
\esp
\eeq
When e.g.\ $i=j$, the functions
$\lbb\IcC{kt}{}\IcSCS{kt;r}{(0)}\rbb^{(i,l)}_{f_i}(x_i,Y_{il,Q})$
and
$\lbb\IcC{kt}{}\IcSCS{kt;r}{(0)}\rbb^{(j,l)}_{f_i}(x_i,Y_{jl,Q})$ 
are different (where $i \ne j,l$ is understood) but up to this order, 
the functional dependence on the variables is the same.
\item Collinear-triple collinear-soft-collinear:
\beq
\bsp
\lbb\IcC{kt}{}\IcC{ktr}{}\IcSCS{kt;r}{(0)}\rbb_{f_i}(x_i) &=
  \pole14
- \pole23 \bigg(\ln{x_i} + \Sigma(y_0, D'_0-1)\bigg)
\\[2mm]&
+ \pole13
  \frac12 \bigg(3\,\delf{f_i}{q} + \Bf{\Nf}\,\delf{f_i}{g} \bigg)
+ \Oe{-2}
\,.
\esp
\eeq
\item Collinear-double collinear-soft-collinear:
\beq
\bsp
\lbb\IcC{kt}{}\IcC{ir;kt}{}\IcSCS{kt;r}{(0)}\rbb_{f_if_k}(x_i,Y_{ik,Q}) &=
  \pole14
- \pole23 \bigg(\ln{x_i} + \Sigma(y_0, D'_0-1)\bigg)
\\[2mm]&
+ \pole13
  \frac12 \bigg(3\,\delf{f_i}{q} + \Bf{\Nf}\,\delf{f_i}{g} \bigg)
+ \Oe{-2}
\,.
\esp
\eeq
As implied by the notation, this function depends on both $x_i$ and 
$Y_{ik,Q}$. However, the dependence on the latter vanishes up to this 
order.
\item Collinear-double soft:
\beq
\lbb\IcC{kt}{}\IcS{kt}{(0)}\rbb^{(j,l)}(Y_{jl,Q}) =
\frac12\bigg[
- \pole14 + \pole13 \bigg(\ln{Y_{jl,Q}} + 4 \Sigma(y_0, D'_0-1)
+ \frac34 - \frac34 \Bf{\Nf}\bigg)\bigg]
+ \Oe{-2}
\,.
\eeq
\item Collinear-triple collinear-double soft:
\beq
\lbb\IcC{kt}{}\IcC{rkt}{}\IcS{kt}{(0)}\rbb_{f_i} =
{\CA \over 2C_{f_i}} \bigg[
  \pole14
- \pole13 \bigg(4 \Sigma(y_0, D'_0-1) - \frac12\,\Bf{\Nf} \bigg)
\bigg]
+ \Oe{-2}
\,.
\eeq
This function is independent of the kinematics.
\end{enumerate}


\paragraph{Soft-type terms:}
\begin{enumerate}[leftmargin=*]
\item Soft-triple collinear:
\beq
\bsp
\lbb\IcS{t}{}\IcC{irt}{(0)}\rbb_{f_i}(x_i) &=
{\CA + 2 C_{f_i} \over 2 C_{f_i}} \bigg[
  \pole14
- \pole23 \bigg(\ln{x_i} + \Sigma(y_0, D'_0)\bigg)
\\[2mm]&
+ \pole13
  \frac12 \bigg(3\,\delf{f_i}{q}
+ \frac13 \Bf{\Nf}\,\delf{f_i}{g}
+ \frac23 \Bf{{\CF \over \CA}\Nf} \delf{f_i}{g}\bigg)
\bigg]
+ \Oe{-2}
\,.
\esp
\eeq
\item Soft-soft-collinear:
\beq
\bsp
\lbb\IcS{t}{}\IcSCS{ir;t}{(0)}\rbb^{(j,l)}_{f_i}(x_i,Y_{jl,Q}) &=
- \pole14 + \pole23 \bigg(\ln{x_i} + \Sigma(y_0, D'_0)\bigg)
\\[2mm]&
+ \pole13 \bigg[\ln{Y_{jl,Q}}
- \frac12 \bigg(3\,\delf{f_i}{q} + \Bf{\Nf}\,\delf{f_i}{g} \bigg)
\bigg]
+ \Oe{-2}
\,,
\\[2mm]
\lbb\IcS{t}{}\IcSCS{ir;t}{(0)}\rbb^{(i,l)}_{f_i}(x_i,Y_{il,Q}) &=
\frac56 \bigg[
- \pole14 + \pole23 \bigg(\ln{x_i} + \Sigma(y_0, D'_0)\bigg)\bigg]
\\[2mm]&
+ \pole13 \bigg[\ln{Y_{il,Q}}
- \frac38 \bigg(3\,\delf{f_i}{q} + \Bf{\Nf}\,\delf{f_i}{g} \bigg)
\bigg]
+ \Oe{-2}
\,.
\esp
\eeq
\item Soft-triple collinear-soft-collinear:
\beq
\bsp
\lbb\IcS{t}{}\IcC{irt}{}\IcSCS{ir;t}{(0)}\rbb_{f_i}(x_i) &=
\frac23 \bigg[
  \pole14 - \pole23 \bigg(\ln{x_i} + \Sigma(y_0, D'_0)\bigg)\bigg]
\\[2mm]&
+ \pole13
  \frac14 \bigg(3\,\delf{f_i}{q} + \Bf{\Nf}\,\delf{f_i}{g} \bigg)
+ \Oe{-2}
\,.
\esp
\eeq
\item Soft-triple collinear-double soft:
\beq
\lbb\IcS{t}{}\IcC{irt}{}\IcS{rt}{(0)}\rbb_{f_i} =
{\CA + 2 C_{f_i} \over 2 C_{f_i}} \bigg[
  \pole14
- \pole23 \bigg(\Sigma(y_0, D'_0) + \Sigma(y_0, D'_0-1)\bigg)
\bigg]
+ \Oe{-2}
\,.
\eeq
This function is independent of the kinematics.
\item Soft-soft-collinear-double soft:
\beq
\bsp
\lbb\IcS{t}{}\IcSCS{ir;t}{}\IcS{rt}{(0)}\rbb^{(j,l)}_{f_i}(x_i,Y_{jl,Q}) &=
- \pole14 + \pole23 \bigg(\Sigma(y_0, D'_0) + \Sigma(y_0, D'_0-1)\bigg)
\\[2mm] &
+ \pole13 \ln{Y_{jl,Q}}
+ \Oe{-2}
\,,
\\[2mm] 
\lbb\IcS{t}{}\IcSCS{ir;t}{}\IcS{rt}{(0)}\rbb^{(i,l)}_{f_i}(x_i,Y_{il,Q}) &=
\frac16\bigg[- \pole54
+ \pole13 \bigg(\ln{x_i} + 10 \Sigma(y_0, D'_0) + 9 \Sigma(y_0, D'_0-1)\bigg)
\bigg]
\\[2mm] &
+ \pole13 \ln{Y_{il,Q}}
+ \Oe{-2}
\,.
\esp
\eeq
\item Soft-triple collinear-soft-collinear-double soft
\beq
\lbb\IcS{t}{}\IcC{irt}{}\IcSCS{ir;t}{}\IcS{rt}{(0)}\rbb_{f_i}(x_i) =
\frac13\bigg[\pole24
- \pole13 \bigg(\ln{x_i} + 4 \Sigma(y_0, D'_0) + 3 \Sigma(y_0, D'_0-1)\bigg)
\bigg]
+ \Oe{-2}
\,.
\eeq
\item Soft-double soft:
\beq
\bsp
\lbb\IcS{t}{}\IcS{rt}{(0)}\rbb^{(i,k)(j,k)}(Y_{ik,Q},Y_{ij,Q},Y_{jk,Q}) &=
\frac12\bigg[\pole14 - \pole13 \bigg(\ln{Y_{ik,Q}} + \ln{Y_{jk,Q}}\bigg) \bigg]
\\[2mm]&
- \pole13 \bigg(\Sigma(y_0, D'_0) + \Sigma(y_0, D'_0-1)\bigg)
+ \Oe{-2}
\,,
\\[2mm]
\lbb\IcS{t}{}\IcS{rt}{(0)}\rbb^{(i,k)}(Y_{ik,Q}) &=
- \frac12\bigg(\pole14 - \pole23 \ln{Y_{ik,Q}} \bigg)
\\[2mm]&
+ \pole13 \bigg(\Sigma(y_0, D'_0) + \Sigma(y_0, D'_0-1)\bigg)
+ \Oe{-2}
\,.
\esp
\eeq
The expansion for $\lbb\IcS{t}{}\IcS{rt}{(0)}\rbb^{(i,k)(j,k)}$ is
valid for the restricted kinematics discussed around \eqn{eq:frameCktCSktr-3jet}.
\end{enumerate}


\paragraph{Soft-collinear-type terms:}
\begin{enumerate}[leftmargin=*]
\item Soft-collinear-triple collinear:
\beq
\bsp
\lbb\IcCS{kt}{t}{}\IcC{krt}{(0)}\rbb_{f_i}(x_i) =
{\CA + C_{f_i} \over C_{f_i}} &\bigg[
  \pole14
- \pole23 \bigg(\ln{x_i} + \Sigma(y_0, D'_0)\bigg)
\\[2mm]&
+ \pole13 \frac12 \bigg(
  3\,\delf{f_i}{q} + \Bf{{\CF \over \CA}\Nf}\,\delf{f_i}{g}
\bigg)
\bigg]
+ \Oe{-2}
\,.
\esp
\eeq
\item Soft-collinear-soft-collinear:
\beq
\bsp
\lbb\IcCS{kt}{t}{}\IcSCS{ir;t}{(0)}\rbb_{f_if_k}(x_k) &=
  \pole14
- \pole23 \bigg(\ln{x_k} + \Sigma(y_0, D'_0)\bigg)
\\[2mm]&
+ \pole13 \frac12 \bigg( 3\,\delf{f_k}{q} + \Bf{\Nf}\,\delf{f_k}{g} \bigg)
+ \Oe{-2}
\,.
\esp
\eeq
\item Soft-collinear-triple collinear-double soft:
\beq
\lbb\IcCS{kt}{t}{}\IcC{krt}{}\IcS{rt}{(0)}\rbb_{f_i} =
  \pole24
- \pole43 \bigg(\Sigma(y_0, D'_0) + \Sigma(y_0, D'_0-1)\bigg)
+ \Oe{-2}
\,.
\eeq
This function is independent of the kinematics.
\item Flavour-dependent soft-collinear-double soft:
\beq
\bsp
\lbb\IcCS{kt}{t}{}\IcS{rt}{(0)}\rbb^{(j,l)}_{f_i}(Y_{jl,Q}) &=
- \pole14
+ \pole13 \bigg(\ln{Y_{jl,Q}} + 2 \Sigma(y_0, D'_0) + 2 \Sigma(y_0, D'_0-1)\bigg)
+ \Oe{-2}
\,.
\esp
\eeq
\item Soft-collinear-soft-collinear-double-soft:
\beq
\lbb\IcCS{kt}{t}{}\IcSCS{ir;t}{}\IcS{rt}{(0)}\rbb_{f_if_k} =
  \pole14
- \pole23 \bigg(\Sigma(y_0, D'_0) + \Sigma(y_0, D'_0-1)\bigg)
+ \Oe{-2}
\,.
\eeq
This function is independent of the kinematics.
\item Flavour-independent soft-collinear-double soft:
\beq
\lbb\IcCS{kt}{t}{}\IcS{kt}{(0)}\rbb^{(j,l)}(Y_{jl,Q}) =
- \pole14
+ \pole13 \bigg(\ln{Y_{jl,Q}} + 2 \Sigma(y_0, D'_0) + 2 \Sigma(y_0, D'_0-1)\bigg)
+ \Oe{-2}
\,.
\label{eq:seriesend}
\eeq
\end{enumerate}

Substituting the expansions in \eqnss{eq:seriesbegin}{eq:seriesend} into 
\eqnss{eq:Ifi}{eq:Iikjl}, we obtain the following explicit expressions for the 
kinematics dependent functions appearing on the right hand side of 
\eqn{eq:I12}.
For $\rC^{(0)}_{12,f_i}$ we find
\beq
\bsp
\rC^{(0)}_{12,f_i}(x_i) &= 
{C_{f_i} - \CA \over C_{f_i}}
\bigg[\pole14 - \pole23 \bigg(\Sigma(y_0, D'_0) + \Sigma(y_0, D'_0-1)\bigg)\bigg]
\\[2mm] &
+ {C_{f_i} + \CA \over C_{f_i}}
  \pole13\bigg[\Sigma(y_0, D'_0-1) - \ln{x_i}
+ \frac14 \bigg(3\,\delf{f_i}{q} + \Bf{{\CF \Nf \over \CA}}\,\delf{f_i}{g} \bigg)\bigg]
\\[2mm] &
+ \Oe{-2}
\,.
\esp
\label{eq:serIfi}
\eeq
The function $\rC^{(0)}_{12,f_if_k}$ simply vanishes up to this order in the 
$\ep$-expansion
\beq
\rC^{(0)}_{12,f_if_k}(x_i,x_k,Y_{ik,Q}) = \Oe{-2}
\,.
\label{eq:serIfifk}
\eeq
The two-parton colour-correlated soft function, $\rS^{(0),(j,l)}_{12}$, is
\beq
\bsp
\rS^{(0),(j,l)}_{12}(Y_{jl,Q}) &=
  \pole13 \bigg(\frac12 \ln{Y_{jl,Q}} - \Sigma(y_0, D'_0) + \Sigma(y_0, D'_0-1)
+ \frac38 - \frac38 \Bf{\Nf}\bigg)
\\[2mm] &
+ \Oe{-2}
\,.
\esp
\label{eq:serIjl}
\eeq
For $\rSCS^{(0),(j,l)}_{12,f_i}$ we obtain  (the $Y_{jl,Q}$ dependence 
vanishes up to this order)
\beq
\bsp
\rSCS^{(0),(j,l)}_{12,f_i}(x_i,Y_{jl,Q}) &=
  \pole13 \bigg[4 \ln{x_i} - 2 \Sigma(y_0, D'_0) - 2 \Sigma(y_0, D'_0-1)
\\[2mm] &\qquad\quad
- \bigg(3\,\delf{f_i}{q} + \Bf{\Nf}\,\delf{f_i}{g} \bigg)\bigg]
+ \Oe{-2}
\,,
\esp
\label{eq:serIfijl}
\eeq
if $i$ is distinct from both $j$ and $l$, while for e.g.\ $i=j$ we have
\beq
\bsp
\rSCS^{(0),(i,l)}_{12,f_i}(x_i,Y_{il,Q}) &=
  \pole13
  \bigg[\frac72 \ln{x_i} - 2 \Sigma(y_0, D'_0) - \frac32 \Sigma(y_0, D'_0-1)
\\[2mm] &\qquad\quad
- \frac78 \bigg(3\,\delf{f_i}{q} + \Bf{\Nf}\,\delf{f_i}{g} \bigg)\bigg]
+ \Oe{-2}
\,.
\esp
\label{eq:serIfiil}
\eeq
Finally, the four-parton colour-correlated soft function, $\rS^{(0),(i,k)(i,k)}_{12}$, reads
\beq
\bsp
\rS^{(0),(i,k)(i,k)}_{12}(Y_{ik,Q}) &=
  \pole{2^{-1}}4
- \pole13\bigg(\ln{Y_{ik,Q}} + \Sigma(y_0, D'_0) + \Sigma(y_0, D'_0-1)\bigg)
\\[2mm]&
+ \Oe{-2}
\,,
\esp
\label{eq:serIikik}
\eeq
if only two indices, say $i$ and $k$ are distinct, while
\beq
\bsp
\rS^{(0),(i,k)(j,k)}_{12}(Y_{ik,Q},Y_{ij,Q},Y_{jk,Q}) &=
  \pole{2^{-1}}4
- \pole13\bigg(\frac12 \ln{Y_{ik,Q}} + \frac12 \ln{Y_{jk,Q}}
\\[2mm] &\qquad\quad
+ \Sigma(y_0, D'_0) + \Sigma(y_0, D'_0-1)\bigg)
+ \Oe{-2}
\,,
\esp
\label{eq:serIikjk}
\eeq
for three distinct indices. Furthermore, the above expression is valid in 
the case of restricted kinematics of three hard partons.




\section{Integrated counterterms}
\label{sec:IntsCTs}

In this section we list the explicit definitions of the functions that
appear on the right hand side of an equation among
\eqnss{eq:C-TC}{eq:IcCSIcSjl}.

%
%

\subsection{Integrated collinear-type counterterms}
\label{sec:IntsCktCTs}

\begin{enumerate}[leftmargin=*]
\item Collinear-triple collinear:
\beq
\bsp
[\IcC{kt}{}\IcC{ktr}{(0)}]_{f_k f_t f_r} =
\left(\frac{16\pi^2}{S_\ep}Q^{2\ep}\right)^2
\int_2 &[\rd p^{(\wha{kt}\ha{r})}_{1,m}] [\rd p^{(kt)}_{1,m+1}]
\frac{1}{s_{kt}} \frac{1}{s_{\wha{kt}\ha{r}}} 
\\[2mm] &\times
\frac{1}{(\bT_{ktr}^2)^2} 
P^{\mathrm{s.o.} (0)}_{f_k f_t f_r}
\Big(\tzz{t}{k},\tzz{\ha{r}}{\wha{kt}},
     y_{\wha{kt}\ha{r}}/y_{\wha{kt}Q}^2;\ep\Big)
\\[2mm] &\times
f(\al_0,\al_{kt},d(m,\ep)) f(\al_0,\al_{\wha{kt}\ha{r}},d(m,\ep))
\,,
\label{eq:IcCktIcCktr}
\esp
\eeq
where $P^{\mathrm{s.o.} (0)}_{f_k f_t f_r}$ are the strongly-ordered 
three-parton splitting kernels averaged over the spin states of the 
parent parton (see \appx{sec:Intspindep} and especially \eqn{eq:Psounified}).
The subtraction terms contain the spin-dependent kernels, that together
with the corresponding kinematic variables can be found in
\refr{Somogyi:2006da}. In \appx{sec:Intspindep} we prove that the
integrals of the spin-dependent kernels give the same contribution as
those of the spin-averaged ones, therefore, we can use the latter 
when integrating the subtraction terms.  

The $f(\al_0,\al,d)$ functions, defined in \eqn{eq:f_def},
represent simple modifications to the original subtraction scheme of
\refr{Somogyi:2006da}. As discussed in \appx{sec:modsubterms} in detail, 
these modifications do not destroy the cancellation of singularities, but
serve improved numerical control, efficiency and stability, and result
in simpler, $m$ independent, integrated counterterms. The rest of the counterterms 
are modified similarly by appropriate $f$ functions, and below we shall simply 
include these factors without further comment.

The integrated counterterm is computed in \appx{sec:IntsCktCktr}. 
In terms of the functions $\cI^{(i)}_{\cC{}{}}$ ($i=1$, 2 and 3) of \eqnss{eq:IC1}{eq:IC3} 
we find
\bal
\bsp
[\IcC{kt}{}\IcC{ktr}{(0)}]_{f_k f_t f_r} &=
\frac{C_{f_{kt}}}{C_{f_{ktr}}}
\bigg[
\sum_{i=1}^2 \sum_{j=-1}^2 \sum_{l=-1}^2
d^{(0)}_{f_{kt} f_r,i} c^{(0)}_{f_k f_t,j} c^{(0)}_{f_{kt} f_r,l}
\cI^{(i)}_{\cC{}{}}(x_{\wti{ktr}};\ep,\al_{0},d_0;j,l)
\\[2mm]
&- \delta_{f_{kt} g} b^{(0)}_{f_k f_t} b^{(0)}_{f_{kt} f_r}
\Big(\cI^{(3)}_{\cC{}{}}(x_{\wti{ktr}};\ep,\al_{0},d_0,1)
    -\cI^{(3)}_{\cC{}{}}(x_{\wti{ktr}};\ep,\al_{0},d_0,2)\Big)
\bigg]
\,.
\label{eq:IcCktIcCktrresult}
\esp
\eal
The various coefficients read:
\bal
b^{(0)}_{qg} &= b^{(0)}_{gq} = 2 \,, &
b^{(0)}_{gg} &= 2 \,, &
b^{(0)}_{q\qb} &= -\frac{2}{1-\ep} \frac{\TR}{\CA} \,,
\label{eq:bkt}
\eal
\bal
c^{(0)}_{qg,-1}&=2\,, & 
c^{(0)}_{qg,0}&=-2\,, & 
c^{(0)}_{qg,1}&=1-\ep\,, & 
c^{(0)}_{qg,2}&=0\,,
\nt\\[2mm]
c^{(0)}_{q\qb,-1}&=0\,, & 
c^{(0)}_{q\qb,0}&=\frac{\TR}{\CA}\,, & 
c^{(0)}_{q\qb,1}&=b^{(0)}_{q\qb}\,, & 
c^{(0)}_{q\qb,2}&=-b^{(0)}_{q\qb}\,,
\label{eq:ckt}
\\[2mm]
c^{(0)}_{gg,-1}&=4\,, & 
c^{(0)}_{gg,0}&=-4\,, & 
c^{(0)}_{gg,1}&=b^{(0)}_{gg}\,, & 
c^{(0)}_{gg,2}&=-b^{(0)}_{gg}\,,
\nt
\eal
with $c^{(0)}_{gq,j} = c^{(0)}_{qg,j}$, and finally
\bal
d^{(0)}_{qg,2} &= d^{(0)}_{gq,1} = 1\,, &
d^{(0)}_{gq,2} &= d^{(0)}_{qg,1} = 0\,, &
d^{(0)}_{gg,1} &= d^{(0)}_{gg,2} = 
d^{(0)}_{q\qb,1} = d^{(0)}_{q\qb,2} = \frac12\,.
\label{eq:diktr}
\eal
\item Collinear-double collinear:
\beq
\bsp
[\IcC{kt}{}\IcC{ir;kt}{(0)}]_{f_k f_t;f_i f_r} &=
\left(\frac{16\pi^2}{S_\ep}Q^{2\ep}\right)^2
\\[2mm] &\times
\int_2 [\rd p^{(\ha{i}\ha{r})}_{1,m}] [\rd p^{(kt)}_{1,m+1}]
\frac{1}{s_{kt}} \frac{1}{\bT_{kt}^2} P^{(0)}_{f_k f_t}(\tzz{t}{k};\ep)
\frac{1}{s_{\ha{i}\ha{r}}}
\frac{1}{\bT_{ir}^2}
P^{(0)}_{f_{\ha{i}} f_{\ha{r}}}(\tzz{\ha{r}}{\ha{i}};\ep)
\\[2mm] &\qquad\times
f(\al_0,\al_{kt},d(m,\ep)) f(\al_0,\al_{\ha{i}\ha{r}},d(m,\ep))
\,.
\esp
\eeq
The integrated counterterm is computed in \appx{sec:IntsCktCirkt}. In
terms of the function $\cI^{(4)}_{\cC{}{}}$ of \eqn{eq:IC4} we find
\beq
[\IcC{kt}{}\IcC{ir;kt}{(0)}]_{f_k f_t;f_i f_r} =
\sum_{j=-1}^2 \sum_{l=-1}^2 c^{(0)}_{f_k f_t,j}
c^{(0)}_{f_if_r,l}
\cI^{(4)}_{\cC{}{}}(x_{\wti{kt}},x_{\wti{ir}};\ep,\al_{0},d_0;j,l)\,,
\label{eq:CktCirktresult}
\eeq
with coefficients given in \eqn{eq:ckt}.
\item Collinear-soft-collinear:
\beq
\bsp
[\IcC{kt}{}\IcSCS{kt;r}{(0)}]_{f_k f_t}^{(j,l)} =
	-\left(\frac{16\pi^2}{S_\ep}Q^{2\ep}\right)^2
	\int_2 &[\rd p^{(\ha{r})}_{1,m}] [\rd p^{(kt)}_{1,m+1}]
	\frac{1}{2} \calS_{\ha{j}\ha{l}}(\ha{r}) \frac{1}{s_{kt}}
	\frac{1}{\bT_{kt}^2} P^{(0)}_{f_k f_t}(\tzz{t}{k};\ep)
\\[2mm] &\times
f(\al_0,\al_{kt},d(m,\ep)) f(y_0,y_{\ha{r}Q},d'(m,\ep))
\,,
\label{eq:IcCktIcSCSktr}
\esp
\eeq
where $\calS_{\ha{j}\ha{l}}(\ha{r})$ is the eikonal factor defined in
\eqn{eq:cSdef} and $P^{(0)}_{f_k f_t}$ are the spin-averaged two-parton
Altarelli--Parisi splitting kernels (see \eqnss{eq:P0gg}{eq:P0qg}).  

The integrated counterterm is computed in \appx{sec:IntsCktCSktr}. In
terms of the function $\cI^{(5)}_{\cC{}{}}$ of \eqn{eq:IC5} we find
\beq
[\IcC{kt}{}\IcSCS{kt;r}{(0)}]_{f_k f_t}^{(j,l)} =
\sum_{i=-1}^2 c^{(0)}_{f_k f_t,i}
\cI^{(5)}_{\cC{}{}}
(x_{\wti{kt}},\Y{j}{l},\Y{j}{kt},\Y{l}{kt};\ep,\al_0,d_0,y_0,d'_0;i)\,,
\label{eq:IcCktIcSCSktrresult}
\eeq
with coefficients given in \eqn{eq:ckt}.
The integral $\cI^{(5)}_{\cC{}{}}$ with full kinematic dependence, as
written above, first appears in computing NNLO corrections to four-jet
production.
\item Collinear-triple collinear-soft-collinear:
\beq
\bsp
[\IcC{kt}{}\IcC{ktr}{}\IcSCS{kt;r}{(0)}]_{f_k f_t} =
\left(\frac{16\pi^2}{S_\ep}Q^{2\ep}\right)^2
\int_2 &[\rd p^{(\ha{r})}_{1,m}] [\rd p^{(kt)}_{1,m+1}]
\frac{2}{s_{\wha{kt}\ha{r}}}
\frac{\tzz{\wha{kt}}{\ha{r}}}{\tzz{\ha{r}}{\wha{kt}}}
\frac{1}{s_{kt}} \frac{1}{\bT_{kt}^2} P^{(0)}_{f_k f_t}(\tzz{t}{k};\ep)
\\[2mm] &\times
f(\al_0,\al_{kt},d(m,\ep)) f(y_0,y_{\ha{r}Q},d'(m,\ep))
\,.
\label{eq:IcCktIcCktrIcSCSktr}
\esp
\eeq
This integrated counterterm is also computed in
\appx{sec:IntsCktCSktr}. In terms of the function $\cI^{(6)}_{\cC{}{}}$
of \eqn{eq:IC6} we find
\beq
[\IcC{kt}{}\IcC{ktr}{}\IcSCS{kt;r}{(0)}]_{f_k f_t} =
\sum_{j=-1}^2 c^{(0)}_{f_k f_t,j}
\cI^{(6)}_{\cC{}{}}(x_{\wti{kt}Q},0;\ep,\al_0,d_0,y_0,d'_0;j)
\label{eq:IcCktIcCktrIcSCSktrresult}
\,,
\eeq
with coefficients given in \eqn{eq:ckt}.
\item Collinear-double collinear-soft-collinear:
\beq
\bsp
[\IcC{kt}{}\IcC{ir;kt}{}\IcSCS{kt;r}{(0)}]_{f_kf_t} =
\left(\frac{16\pi^2}{S_\ep}Q^{2\ep}\right)^2
\int_2 &[\rd p^{(\ha{r})}_{1,m}] [\rd p^{(kt)}_{1,m+1}]
\frac{2}{s_{\ha{i}\ha{r}}} \frac{\tzz{\ha{i}}{\ha{r}}}{\tzz{\ha{r}}{\ha{i}}}
\frac{1}{s_{kt}} \frac{1}{\bT_{kt}^2} P^{(0)}_{f_k f_t}(\tzz{t}{k};\ep)
\\[2mm] &\times
 f(\al_0,\al_{kt},d(m,\ep)) f(y_0,y_{\ha{r}Q},d'(m,\ep))
\,.
\label{eq:IcCktIcCirktIcSCSktr}
\esp
\eeq
This integrated counterterm is also computed in
\appx{sec:IntsCktCSktr}. In terms of the function $\cI^{(6)}_{\cC{}{}}$
of \eqn{eq:IC6} we find
\beq
[\IcC{kt}{}\IcC{ir;kt}{}\IcSCS{kt;r}{(0)}]_{f_kf_t} =
\sum_{j=-1}^2 c^{(0)}_{f_k f_t,j}
\cI^{(6)}_{\cC{}{}}(x_{\wti{kt}Q},\Y{i}{kt};\ep,\al_0,d_0,y_0,d_0';j)
\,,
\label{eq:IcCktIcCirktIcSCSktrresult}
\eeq 
with coefficients given in \eqn{eq:ckt}.
\item Collinear-double soft:
\beq
\bsp
[\IcC{kt}{}\IcS{kt}{(0)}]_{f_kf_t}^{(j,l)} =
\left(\frac{16\pi^2}{S_\ep}Q^{2\ep}\right)^2
\int_2 &[\rd p^{(\wha{kt})}_{1,m}] [\rd p^{(kt)}_{1,m+1}]
\\[2mm] &\times
\frac{1}{2} \calS^{\mu\nu}_{\ha{j}\ha{l}}(\wha{kt})
\frac{1}{s_{kt}} \frac{1}{\CA}
\la\mu| \hP^{(0)}_{f_k f_t}(\tzz{k}{t},\tzz{t}{k},\kT{k,t};\ep) |\nu\ra\
\\[2mm] &\times
 f(\al_0,\al_{kt},d(m,\ep)) f(y_0,y_{\wha{kt}Q},d'(m,\ep))
\,,
\label{eq:IcCktIcS}
\esp
\eeq
where $\calS^{\mu\nu}_{\ha{j}\ha{l}}(\wha{kt})$ is defined in
\eqn{eq:cSdef} and $\la\mu|\hP^{(0)}_{f_k f_t} |\nu\ra$ is the
spin-dependent Altarelli--Parisi splitting kernel for gluon
splitting (see \eqn{eq:P0munu}). 

The integrated counterterm is computed
in \appx{sec:IntsCktSkt}.  In terms of the functions
$\cI^{(i)}_{\cC{}{}}$ ($i=7$, 8 and 9) of \eqnss{eq:IC7}{eq:IC9} we
find
\beq
\bsp
[\IcC{kt}{}\IcS{kt}{(0)}]_{f_kf_t}^{(j,l)} &=
\sum_{i=-1}^2 c^{(0)}_{f_k f_t,i}
\cI^{(7)}_{\cC{}{}}(\Y{j}{l};\ep,\al_{0},d_0;i)
\\[2mm] &
+ b^{(0)}_{f_k f_t}
\sum_{i=8}^{9}
\Big(
\cI^{(i)}_{\cC{}{}}(\ep,\al_{0},d_0;1)-\cI^{(i)}_{\cC{}{}}(\ep,\al_{0},d_0;2)
\Big)
\,,
\label{eq:IcCktIcSresult}
\esp
\eeq
with $b^{(0)}_{f_k f_t}$ and $c^{(0)}_{f_k f_t,k}$ given in
\eqns{eq:bkt}{eq:ckt}, respectively.
\item Collinear-triple collinear-double soft:
\beq
\bsp
[\IcC{kt}{}\IcC{rkt}{}\IcS{kt}{(0)}]_{f_rf_kf_t} &=
\left(\frac{16\pi^2}{S_\ep}Q^{2\ep}\right)^2
\int_2 [\rd p^{(\wha{kt})}_{1,m}] [\rd p^{(kt)}_{1,m+1}]
\frac{2}{s_{kt}s_{\wha{kt}\ha{r}}}
\\[2mm] &\qquad\qquad\qquad\times
\frac{\bT_{kt}^2}{\bT_r^2} 
\left(
\frac{\tzz{\ha{r}}{\wha{kt}}}{\tzz{\wha{kt}}{\ha{r}}}
 \frac{1}{\bT_{kt}^2}P^{(0)}_{f_kf_t}(\tzz{t}{k};\ep)
-\tzz{k}{t}\tzz{t}{k} b^{(0)}_{f_k f_t} \frac{y_{\wha{kt}\ha{r}}}{y_{\wha{kt}Q}^2}
\right)
\\[2mm] &\qquad\qquad\qquad\times
 f(\al_0,\al_{kt},d(m,\ep)) f(y_0,y_{\wha{kt}Q},d'(m,\ep))
\,,
\label{eq:CktCrktSkt}
\esp
\eeq
with the flavour-dependent constants $b^{(0)}_{f_k f_t}$ given in 
\eqn{eq:bkt}.  Of course, only $f_k + f_t = g$ gives a non-vanishing 
result (see \eqn{eq:C2S2jl}). In obtaining the form (\ref{eq:CktCrktSkt}) 
of the integrated counterterm, we exploited that the integrals of the two
expressions in \eqn{eq:replace} are equal.

The integrated counterterm is computed in \appx{sec:IntsCktSkt}. In
terms of the functions $\cI^{(8)}_{\cC{}{}}$ and $\cI^{(9)}_{\cC{}{}}$ of
\eqns{eq:IC8}{eq:IC9} we find
\beq
\bsp
[\IcC{kt}{}\IcC{rkt}{}\IcS{kt}{(0)}]_{f_rf_kf_t} &=
\frac{\CA}{C_{f_r}}\Bigg[
\sum_{j=-1}^2 c^{(0)}_{f_k f_t,j} \cI^{(8)}_{\cC{}{}}(\ep,\al_{0},d_0;j)
\\[2mm] &\qquad\quad
+ b^{(0)}_{f_k f_t}
\Big(
\cI^{(9)}_{\cC{}{}}(\ep,\al_{0},d_0;1)-\cI^{(9)}_{\cC{}{}}(\ep,\al_{0},d_0;2)
\Big)
\Bigg]
\,,
\label{eq:CktCrktSktresult}
\esp
\eeq
with coefficients given in \eqns{eq:bkt}{eq:ckt}.
\end{enumerate}

%
%

\subsection{Integrated soft-type counterterms}
\label{sec:IntsStCTs}

\begin{enumerate}[leftmargin=*]
\item Soft-triple collinear:
\beq
\bsp
[\IcS{t}{}\IcC{irt}{(0)}]_{f_if_r} &=
\left(\frac{16\pi^2}{S_\ep}Q^{2\ep}\right)^2
\int_2 [\rd p^{(\ha{i}\ha{r})}_{1,m}] [\rd p^{(t)}_{1,m+1}]
\frac{1}{\bT_{ir}^2}P^{(\mathrm{S})}_{f_i f_r g}
(\tzz{i}{rt},\tzz{r}{it},\tzz{t}{ir},s_{ir},s_{it},s_{rt};\ep)
\\[2mm] &\qquad\qquad\qquad\times
\frac{1}{s_{\ha{i}\ha{r}}} \frac{1}{\bT_{ir}^2}
P^{(0)}_{f_i f_r}(\tzz{\ha{r}}{\ha{i}};\ep)
f(y_0,y_{tQ},d'(m,\ep)) f(\al_0,\al_{\ha{i}\ha{r}},d(m,\ep))
\,,
\label{eq:IcStIcCirt}
\esp
\eeq
where the functions $P^{(\mathrm{S})}_{f_i f_r f_t}$ are the soft limits
of the triple-collinear splitting functions, introduced in
\refr{Somogyi:2005xz} (see \eqn{eq:PSirt}), and we used $\bT_{irt}^2 =
\bT_{ir}^2$ because the soft parton $t$ can only be gluon, $f_t = g$.

This integrated counterterm is computed in \appx{sec:StCirt-type}.
In terms of the functions $\cI^{(1)}_{\cS{}{}}$ and $\cI^{(2)}_{\cS{}{}}$
of \eqns{eq:IS1}{eq:IS2} we find
\bal
\bsp
[\IcS{t}{}\IcC{irt}{(0)}]_{f_if_r} =
\sum_{k=-1}^2 \frac{c^{(0)}_{f_i f_r,k}}{C_{f_{ir}}}
&\Big[
(C_{f_{i}} + C_{f_{r}} - C_{f_{ir}})
\,\cI^{(2)}_{\cS{}{}}(x_{\wti{ir}};\ep,\al_0,d_0,y_0,d'_0;k)
\\[2mm] &
+ 2 C_{f_{ir}}
\cI^{(1)}_{\cS{}{}}(x_{\wti{ir}};\ep,\al_0,d_0,y_0,d'_0;k)
\Big]
\,,
\label{eq:IcStIcCirtresult}
\esp
\eal
with coefficients given in \eqn{eq:ckt}.
\item Soft-soft-collinear:
\beq
\bsp
[\IcS{t}{}\IcSCS{ir;t}{(0)}]_{f_if_r}^{(j,l)} =
-\left(\frac{16\pi^2}{S_\ep}Q^{2\ep}\right)^2
\int_2 &[\rd p^{(\ha{i}\ha{r})}_{1,m}] [\rd p^{(t)}_{1,m+1}]
\frac{1}{2} \calS_{jl}(t)
\frac{1}{s_{\ha{i}\ha{r}}} \frac{1}{\bT_{ir}^2}
P^{(0)}_{f_i f_r}(\tzz{\ha{r}}{\ha{i}};\ep)
\\[2mm] &\times
f(y_0,y_{tQ},d'(m,\ep)) f(\al_0,\al_{\ha{i}\ha{r}},d(m,\ep))
\,.
\label{eq:IcStIcCSirt}
\esp
\eeq
This integrated counterterm is computed in \appx{sec:StCirt-type}.
In terms of the functions $\cI^{(3)}_{\cS{}{}}$ and $\cI^{(4)}_{\cS{}{}}$
of \eqns{eq:IS3}{eq:IS4} we find
\bal
\bsp
[\IcS{t}{}\IcSCS{ir;t}{(0)}]_{f_if_r}^{(j,l)} =
\sum_{k=-1}^2 c^{(0)}_{f_i f_r,k}
&\Big[
(1 - \delta_{j(ir)} - \delta_{l(ir)})
\,\cI^{(3)}_{\cS{}{}}(x_{\wti{ir}},\Y{j}{l};\ep,\al_0,d_0,y_0,d'_0;k)
\\[2mm] &
+ (\delta_{j(ir)} + \delta_{l(ir)})
\cI^{(4)}_{\cS{}{}}(x_{\wti{ir}},\Y{j}{l};\ep,\al_0,d_0,y_0,d'_0;k)
\Big]
\,,
\label{eq:IcStIcCSirtresult}
\esp
\eal
with coefficients given in \eqn{eq:ckt}. Note that $j$ and $l$ are always distinct.
\item Soft-triple collinear-soft-collinear:
\beq
\bsp
[\IcS{t}{}\IcC{irt}{}\IcSCS{ir;t}{(0)}]_{f_i f_r} =
\left(\frac{16\pi^2}{S_\ep}Q^{2\ep}\right)^2
\int_2 &[\rd p^{(\ha{i}\ha{r})}_{1,m}] [\rd p^{(t)}_{1,m+1}]
\frac{2}{s_{(ir)t}}\,\frac{1-\tzz{t}{ir}}{\tzz{t}{ir}}
\frac{1}{s_{\ha{i}\ha{r}}} \frac{1}{\bT_{ir}^2}
P^{(0)}_{f_i f_r}(\tzz{\ha{r}}{\ha{i}};\ep)
\\[2mm] &\times
f(y_0,y_{tQ},d'(m,\ep)) f(\al_0,\al_{\ha{i}\ha{r}},d(m,\ep))
\,.
\label{eq:IcStIcCirtIcCSirt}
\esp
\eeq
This integrated counterterm is computed in \appx{sec:StCirt-type}.
In terms of the function $\cI^{(5)}_{\cS{}{}}$ of \eqn{eq:IS5} we find
\bal
\bsp
[\IcS{t}{}\IcC{irt}{}\IcSCS{ir;t}{(0)}]_{f_i f_r} =
\sum_{k=-1}^2 c^{(0)}_{f_i f_r,k}
\cI^{(5)}_{\cS{}{}}(x_{\wti{ir}};\ep,\al_0,d_0,y_0,d'_0;k)
\,,
\label{eq:IcStIcCirtIcCSirtresult}
\esp
\eal
with coefficients given in \eqn{eq:ckt}.
\item Soft-triple collinear-double soft:%
\footnote{We note a harmless misprint in the definition of the
subtraction term $\cS{t}{}\cC{irt}{}\cS{rt}{(0,0)}$ in eq.~(7.38) of
\refr{Somogyi:2006da}, where in the last term of the square bracket 
$\frac{\tzz{i}{t}}{\tzz{t}{i}}$ was used as
compared to $\frac{\tzz{i}{rt}}{\tzz{t}{ir}}$ here. Our definition of the
momentum fractions gives the same for these ratios:
$\frac{\tzz{i}{rt}}{\tzz{t}{ir}}=\frac{\tzz{i}{t}}{\tzz{t}{i}}
=\frac{y_{iQ}}{y_{tQ}}$.}
\beq
\bsp
[\IcS{t}{}\IcC{irt}{}\IcS{rt}{(0)}]_{f_i} &=
\left(\frac{16\pi^2}{S_\ep}Q^{2\ep}\right)^2
\\[2mm] &\times
\int_2 [\rd p^{(\ha{r})}_{1,m}] [\rd p^{(t)}_{1,m+1}]
\Bigg[\frac{\CA}{\bT_i^2}
\left(\frac{s_{ir}}{s_{it} s_{rt}}
+ \frac{1}{s_{rt}} \frac{\tzz{r}{it}}{\tzz{t}{ir}}
- \frac{1}{s_{it}} \frac{\tzz{i}{rt}}{\tzz{t}{ir}}\right)
+ \frac{2}{s_{it}} \frac{\tzz{i}{rt}}{\tzz{t}{ir}} \Bigg]
\\[2mm] &\qquad\times
\frac{2}{s_{\ha{i}\ha{r}}}
\frac{\tzz{\ha{i}}{\ha{r}}}{\tzz{\ha{r}}{\ha{i}}}
f(y_0,y_{tQ},d'(m,\ep)) f(y_0,y_{\ha{r}Q},d'(m,\ep))
\,.
\label{eq:IcStIcCirtIcSrt}
\esp
\eeq
The integrated counterterm is computed in \appx{sec:StCirtSrt-type}.
In terms of the functions $\cI^{(6)}_{\cS{}{}}$ and $\cI^{(7)}_{\cS{}{}}$ 
of \eqns{eq:IS6}{eq:IS7} we find
\bal
\bsp
[\IcS{t}{}\IcC{irt}{}\IcS{rt}{(0)}]_{f_i} =
2 \cI^{(6)}_{\cS{}{}}(\ep,y_0,d'_0)
+ \frac{\CA}{C_{f_i}} \cI^{(7)}_{\cS{}{}}(\ep,y_0,d'_0)
\,.
\label{eq:IcStIcCirtIcSrtresult}
\esp
\eal
\item Soft-soft-collinear-double soft:
\beq
\bsp
[\IcS{t}{}\IcSCS{ir;t}{}\IcS{rt}{(0)}]^{(j,l)} =
-\left(\frac{16\pi^2}{S_\ep}Q^{2\ep}\right)^2
\int_2 &[\rd p^{(\ha{r})}_{1,m}] [\rd p^{(t)}_{1,m+1}]
\frac{2}{s_{\ha{i}\ha{r}}} \frac{\tzz{\ha{i}}{\ha{r}}}{\tzz{\ha{r}}{\ha{i}}}
\frac{1}{2}\calS_{jl}(t)
\\[2mm] &\times
f(y_0,y_{tQ},d'(m,\ep)) f(y_0,y_{\ha{r}Q},d'(m,\ep))
\label{eq:IcStIcSCSirtIcSrt}
\,.
\esp
\eeq
The integrated counterterm is computed in \appx{sec:StCirtSrt-type}.
In terms of the functions $\cI^{(8)}_{\cS{}{}}$ and $\cI^{(9)}_{\cS{}{}}$ 
of \eqns{eq:IS8}{eq:IS9} we find
\bal
\bsp
[\IcS{t}{}\IcSCS{ir;t}{}\IcS{rt}{(0)}]^{(j,l)} &=
(1 - \delta_{j(ir)} - \delta_{l(ir)})
\,\cI^{(8)}_{\cS{}{}}(\Yt{j}{l};\ep,y_0,d'_0)
\\[2mm] &\qquad
+ (\delta_{j(ir)} + \delta_{l(ir)})
\cI^{(9)}_{\cS{}{}}(x_{\wti{i}},\Yt{j}{l};\ep,y_0,d'_0)
\,.
\label{eq:IcStIcSCSirtIcSrtresult}
\esp
\eal
Note that $j$ and $l$ are always distinct.
\item Soft-triple collinear-soft-collinear-double soft:
\beq
\bsp
[\IcS{t}{}\IcC{irt}{}\IcSCS{ir;t}{}\IcS{rt}{(0)}] =
\left(\frac{16\pi^2}{S_\ep}Q^{2\ep}\right)^2
\int_2 &[\rd p^{(\ha{r})}_{1,m}] [\rd p^{(t)}_{1,m+1}]
\frac{2}{s_{\ha{i}\ha{r}}} \frac{\tzz{\ha{i}}{\ha{r}}}{\tzz{\ha{r}}{\ha{i}}}
\frac{2}{s_{(ir)t}}\,\frac{1-\tzz{t}{ir}}{\tzz{t}{ir}}
\\[2mm] &\times
f(y_0,y_{tQ},d'(m,\ep)) f(y_0,y_{\ha{r}Q},d'(m,\ep))
\,.
\label{eq:IcStIcCirtIcSCSirtIcSrt}
\esp
\eeq
This integrated counterterm is computed in \appx{sec:StCirtSrt-type}.
In terms of the function $\cI^{(10)}_{\cS{}{}}$
of \eqn{eq:IS10} we find
\bal
\bsp
[\IcS{t}{}\IcC{irt}{}\IcSCS{ir;t}{}\IcS{rt}{(0)}] &=
\cI^{(10)}_{\cS{}{}}(x_{\wti{i}};\ep,y_0,d'_0)
\,.
\label{eq:IcStIcCirtIcSCSirtIcSrtresult}
\esp
\eal
\item Soft-double soft:
\bal
[\IcS{t}{}\IcS{rt}{(0)}]^{(i,k)(j,l)} =
\left(\frac{16\pi^2}{S_\ep}Q^{2\ep}\right)^2
\int_2 &[\rd p^{(\ha{r})}_{1,m}] [\rd p^{(t)}_{1,m+1}]
\frac{1}{8} \calS_{\ha{i}\ha{k}}(\ha{r}) \calS_{jl}(t)
\nt\\[2mm] &\times
f(y_0,y_{tQ},d'(m,\ep)) f(y_0,y_{\ha{r}Q},d'(m,\ep))
\label{eq:IcStIcSrtab}
\,,
\\[2mm]
[\IcS{t}{}\IcS{rt}{(0)}]^{(i,k)} =
-\left(\frac{16\pi^2}{S_\ep}Q^{2\ep}\right)^2
\int_2 &[\rd p^{(\ha{r})}_{1,m}] [\rd p^{(t)}_{1,m+1}]
\frac{1}{4} \calS_{\ha{i}\ha{k}}(\ha{r}) 
\Big(\calS_{ir}(t) + \calS_{kr}(t) - \calS_{ik}(t)\Big)
\nt\\[2mm] &\times
f(y_0,y_{tQ},d'(m,\ep)) f(y_0,y_{\ha{r}Q},d'(m,\ep))
\,.
\label{eq:IcStIcSrtnab}
\eal
In this paper we do not discuss the case when $i$, $j$, $k$, and $l$ are
all distinct, which first appears in computing NNLO corrections to 
four-jet production.
For the specific cases of two and three hard partons in the final state, 
we compute the integrated counterterms in \appx{sec:StCirtSrt-type}.
In terms of the functions $\cI^{(11)}_{\cS{}{}}$ and $\cI^{(12)}_{\cS{}{}}$
of \eqns{eq:IS11}{eq:IS12} we find
\bal
[\IcS{t}{}\IcS{rt}{(0)}]^{(i,k)(j,k)} &=
\frac12 \cI^{(11)}_{\cS{}{};ik,jk}(\Yt{i}{k},\Yt{i}{j},\Yt{j}{k};\ep,y_0,d'_0)
\label{eq:IcStIcSrtabresult}
\,,
\intertext{ and}
[\IcS{t}{}\IcS{rt}{(0)}]^{(i,k)} &=
2 \cI^{(12)}_{\cS{}{};ik}(\Yt{i}{k};\ep,y_0,d'_0)
+ \cI^{(11)}_{\cS{}{};ik,ik}(\Yt{i}{k};\ep,y_0,d'_0)
\label{eq:IcStIcSrtnabresult}
\,.
\eal
\end{enumerate}

%
%

\subsection{Integrated soft-collinear-type counterterms}
\label{sec:IntsCktStCTs}

\begin{enumerate}[leftmargin=*]
\item Soft-collinear-triple collinear:
\beq
\bsp
[\IcC{kt}{}\IcS{t}{}\IcC{krt}{(0)}]_{f_k f_t} =
\left(\frac{16\pi^2}{S_\ep}Q^{2\ep}\right)^2
\int_2 &[\rd p^{(\ha{k}\ha{r})}_{1,m}] [\rd p^{(t)}_{1,m+1}]
\frac{2}{s_{kt}} \frac{\tzz{k}{t}}{\tzz{t}{k}} \frac{1}{s_{\ha{k}\ha{r}}}
\frac{\bT_{kt}^2}{\bT_{krt}^2}
\frac{1}{\bT_{kr}^2} P^{(0)}_{f_k f_r}(\tzz{\ha{r}}{\ha{k}};\ep)
\\[2mm] &\times
f(y_0,y_{tQ},d'(m,\ep)) f(\al_0,\al_{\ha{k}\ha{r}},d(m,\ep))
\label{eq:IcCktIcStIcCkrt}
\,,
\esp
\eeq
where we used $\bT_{krt}^2 = \bT_{kr}^2$ because the soft parton $t$
can only be gluon, $f_t = g$. For the same reason 
$\bT_{kt}^2/\bT_{krt}^2 = \bT_{k}^2/\bT_{kr}^2$, to be used in 
\eqn{eq:softcollresults}.
\item Soft-collinear-soft-collinear:%
\footnote{We note a misprint in the definition of the subtraction term 
$\cC{kt}{}\cS{t}{}\cSCS{ir;t}{(0,0)}$ in eq.~(7.46) of \refr{Somogyi:2006da}.
The quadratic Casimir has to be changed from $\bT_i^2$ to $\bT_k^2$ (see
eq.~(7.22) of \refr{Somogyi:2005xz}).}
\beq
\bsp
[\IcC{kt}{}\IcS{t}{}\IcSCS{ir;t}{(0)}]_{f_i f_r} =
\left(\frac{16\pi^2}{S_\ep}Q^{2\ep}\right)^2
\int_2 &[\rd p^{(\ha{i}\ha{r})}_{1,m}] [\rd p^{(t)}_{1,m+1}]
\frac{2}{s_{kt}} \frac{\tzz{k}{t}}{\tzz{t}{k}} \frac{1}{s_{\ha{i}\ha{r}}}
\frac{1}{\bT_{ir}^2} P^{(0)}_{f_i f_r}(\tzz{\ha{r}}{\ha{i}};\ep)
\\[2mm] &\times
f(y_0,y_{tQ},d'(m,\ep)) f(\al_0,\al_{\ha{i}\ha{r}},d(m,\ep))
\label{eq:IcCktIcStIcSCSirt}
\,.
\esp
\eeq
\item Soft-collinear-triple collinear-double soft:
\beq
\bsp
[\IcC{kt}{}\IcS{t}{}\IcC{krt}{}\IcS{rt}{(0)}] =
\left(\frac{16\pi^2}{S_\ep}Q^{2\ep}\right)^2
\int_2 &[\rd p^{(\ha{r})}_{1,m}] [\rd p^{(t)}_{1,m+1}]
\frac{2}{s_{\ha{k}\ha{r}}} \frac{\tzz{\ha{k}}{\ha{r}}}{\tzz{\ha{r}}{\ha{k}}} 
\frac{2}{s_{kt}} \frac{\tzz{k}{t}}{\tzz{t}{k}}
\\[2mm] &\times
f(y_0,y_{tQ},d'(m,\ep)) f(y_0,y_{\ha{r}Q},d'(m,\ep))
\,,
\\[2mm]
[\IcC{kt}{}\IcS{t}{}\IcC{rkt}{}\IcS{kt}{(0)}] =
\left(\frac{16\pi^2}{S_\ep}Q^{2\ep}\right)^2
\int_2 &[\rd p^{(\ha{k})}_{1,m}] [\rd p^{(t)}_{1,m+1}]
\frac{2}{s_{\ha{k}\ha{r}}} \frac{\tzz{\ha{r}}{\ha{k}}}{\tzz{\ha{k}}{\ha{r}}} 
\frac{2}{s_{kt}} \frac{\tzz{k}{t}}{\tzz{t}{k}}
\\[2mm] &\times
f(y_0,y_{tQ},d'(m,\ep)) f(y_0,y_{\ha{k}Q},d'(m,\ep))
\label{eq:IcCktIcStIcCkrtIcSrt}
\,.
\esp
\eeq
The integrals over $[\rd p^{(\ha{k})}_{1,m}]$ (or $[\rd p^{(\ha{r})}_{1,m}]$) 
and $[\rd p^{(t)}_{1,m+1}]$ factorize in the two equations above. Therefore, 
$[\IcC{kt}{}\IcS{t}{}\IcC{krt}{}\IcS{rt}{(0)}] =
[\IcC{kt}{}\IcS{t}{}\IcC{rkt}{}\IcS{kt}{(0)}]$ (as seen by the simple exchange of 
indices $k\leftrightarrow r$) and the distinction of these functions is purely 
formal, and serves bookkeeping purposes only.
\item Soft-collinear-double soft:
\beq
\bsp
[\IcC{kt}{}\IcS{t}{}\IcS{rt}{(0)}]^{(j,l)} =
-\left(\frac{16\pi^2}{S_\ep}Q^{2\ep}\right)^2
\int_2 &[\rd p^{(\ha{r})}_{1,m}] [\rd p^{(t)}_{1,m+1}]
\frac{1}{2} \calS_{\ha{j}\ha{l}}(\ha{r})
\frac{2}{s_{kt}} \frac{\tzz{k}{t}}{\tzz{t}{k}}
\\[2mm] &\times
f(y_0,y_{tQ},d'(m,\ep)) f(y_0,y_{\ha{r}Q},d'(m,\ep))
\,,
\\[2mm]
[\IcC{kt}{}\IcS{t}{}\IcS{kt}{(0)}]^{(j,l)} =
-\left(\frac{16\pi^2}{S_\ep}Q^{2\ep}\right)^2
\int_2 &[\rd p^{(\ha{k})}_{1,m}] [\rd p^{(t)}_{1,m+1}]
\frac{1}{2} \calS_{\ha{j}\ha{l}}(\ha{k})
\frac{2}{s_{kt}} \frac{\tzz{k}{t}}{\tzz{t}{k}}
\\[2mm] &\times
f(y_0,y_{tQ},d'(m,\ep)) f(y_0,y_{\ha{k}Q},d'(m,\ep))
\,,
\label{eq:IcCktIcStIcSkt}
\esp
\eeq
where the first equation above defines the flavour-dependent 
soft-collinear-double soft function, while the second equation gives the
flavour-independent one. Again, the integrals over 
$[\rd p^{(\ha{k})}_{1,m}]$ (or $[\rd p^{(\ha{r})}_{1,m}]$) 
and $[\rd p^{(t)}_{1,m+1}]$ factorize in both equations. Therefore, 
$[\IcC{kt}{}\IcS{t}{}\IcS{rt}{(0)}]^{(j,l)} =
[\IcC{kt}{}\IcS{t}{}\IcS{kt}{(0)}]^{(j,l)}$ (as seen by the simple exchange of 
indices $k\leftrightarrow r$) and the distinction of these functions is formal, 
and serves bookkeeping purposes.
\item Soft-collinear-soft-collinear-double soft:
\beq
\bsp
[\IcC{kt}{}\IcS{t}{}\IcSCS{ir;t}{}\IcS{rt}{(0)}] =
\left(\frac{16\pi^2}{S_\ep}Q^{2\ep}\right)^2
\int_2 &[\rd p^{(\ha{r})}_{1,m}] [\rd p^{(t)}_{1,m+1}]
\frac{2}{s_{\ha{i}\ha{r}}} \frac{\tzz{\ha{i}}{\ha{r}}}{\tzz{\ha{r}}{\ha{i}}}
\frac{2}{s_{kt}} \frac{\tzz{k}{t}}{\tzz{t}{k}}
\\[2mm] &\times
f(y_0,y_{tQ},d'(m,\ep)) f(y_0,y_{\ha{r}Q},d'(m,\ep))
\label{eq:IcCktIcStIcSCSirtIcSrt}
\,.
\esp
\eeq
\end{enumerate}

The soft-collinear-type integrated counterterms are computed in 
\appx{sec:IntsCktStA2}.
The calculation of the integrals in
\eqnss{eq:IcCktIcStIcCkrt}{eq:IcCktIcStIcSCSirtIcSrt} is fairly
straightforward because the integrals over the soft phase space measures
decouple in all cases. Using the functions $\cI^{(i)}_{\cSCS{}{}}$ ($i =
1$, 2, 3) computed in \eqnss{eq:ISCS1}{eq:ISCS3}, we find:
\beq
\bsp
&
[\IcC{kt}{}\IcS{t}{}\IcC{krt}{(0)}]_{f_k f_t} =
\frac{C_{f_{k}}}{C_{f_{kr}}}
\sum_{j=-1}^2 c^{(0)}_{f_k f_r,j}
\cI^{(1)}_{\cSCS{}{}}(y_{\wti{kr}Q};\ep,\al_0,d_0,y_0,d'_0;j)
\\[2mm]&
[\IcC{kt}{}\IcS{t}{}\IcSCS{ir;t}{(0)}]_{f_i f_r} =
\sum_{j=-1}^2 c^{(0)}_{f_i f_r,j}
\cI^{(1)}_{\cSCS{}{}}(y_{\wti{kr}Q};\ep,\al_0,d_0,y_0,d'_0;j)
\,,
\\[2mm]&
[\IcC{kt}{}\IcS{t}{}\IcC{krt}{}\IcS{rt}{(0)}] =
[\IcC{kt}{}\IcS{t}{}\IcC{rkt}{}\IcS{kt}{(0)}] =
\cI^{(3)}_{\cSCS{}{}}(\ep,y_0,d'_0)
\,,
\\[2mm]&
[\IcC{kt}{}\IcS{t}{}\IcS{rt}{(0)}]^{(j,l)} =
[\IcC{kt}{}\IcS{t}{}\IcS{kt}{(0)}]^{(j,l)} =
\cI^{(2)}_{\cSCS{}{}}(\Yt{j}{l};\ep,y_0,d'_0)
\,,
\\[2mm]&
[\IcC{kt}{}\IcS{t}{}\IcSCS{ir;t}{}\IcS{rt}{(0)}] =
\cI^{(3)}_{\cSCS{}{}}(\ep,y_0,d'_0)
\,.
\label{eq:softcollresults}
\esp
\eeq
with coefficients given in \eqn{eq:ckt}.

This ends the definition of the integrals of the iterated subtraction
terms, that can be used to construct the insertion operator
$\bI^{(0)}_{12}$ as given by \eqnss{eq:I12}{eq:SC-SC-DS}. The computation
of these integrals is presented in the appendices.




\section{Insertion operator for two- and three-jet production}
\label{sec:results}

In this section, we present illustrative numerical results for the insertion 
operator $\bI^{(0)}_{12}$
of \eqn{eq:I12}, at various specific 
phase space points, for processes with at most three hard partons in the 
final state.

While it is true that the most general collinear and/or soft configuration 
at NNLO accuracy involves four hard partons, if these are all in the final 
state, which is the subject of this paper, then the two-loop amplitudes 
needed in the doubly-virtual correction have at least six massless (or 
four massless and one massive) external legs. Results for such amplitudes 
are not foreseen in the near future, therefore, we restrict our discussion to 
computing NNLO corrections to two- and three-jet quantities. 

In an explicit computation of a jet cross section at NNLO, we require the 
expansion coefficients (up to and including the finite part) of the Laurent 
series in $\ep$ of the insertion operator. In general, these expansion 
coefficients are functions of various kinematic variables (and also parameters 
such as $\al_0$, $y_0$, $d_0$ and $d'_0$) which depend  on the particular 
phase space point. One may either attempt to compute these functions 
analytically, or numerically. The former is important as a matter of principle 
only. For practical purposes (from the point of phenomenology) the latter is
sufficient. Indeed, the higher order expansion coefficients  (starting form 
$\Oe{-2}$) of the results we will present were obtained numerically.

%
%

\subsection{Two-jet production}
\label{sec:2jet-prod}

Let us consider the process $e^+e^- \to 2$ jets. The corresponding
squared matrix element at tree level is $|\cM^{(0)}_2(1_q,2_\qb)|^2$,
i.e.~the quark carries label 1 and the antiquark label 2. Both the colour
algebra and kinematics are trivial. Colour conservation implies 
\beq
\bT_1\,\bT_2 = -\CF
\eeq
and $C_{f_1} = C_{f_2} = \CF$. Hence, the insertion operator is a scalar 
in colour space. On the other hand, momentum conservation requires that 
the two final state momenta are back-to-back, i.e.\ in a properly oriented 
frame we have
\beq
\renewcommand{\arraystretch}{1.5}
\begin{tabular}{lcc @{\big(} l @{\;\;,\;\;} l @{\;\;,\;\;} l @{\;\;,\;\;} l @{\big)\,,}}
$p_1^\mu$ &=& $\sqrt{s}\,$ & $\frac12$ & 0 & 0 & $\frac12$\\
$p_2^\mu$ &=& $\sqrt{s}\,$ & $\frac12$ & 0 & 0 & $-\frac12$\\
\end{tabular}
\label{eq:2jetPSpoint}
\eeq
which implies
\beq
y_{12} = x_1 = x_2 = Y_{12,Q} = 1
\,.
\eeq

The insertion operator \eqn{eq:I12} becomes
\beq
\bsp
\bI^{(0)}_{12}(p_1, p_2;\ep)  =
\left[\frac{\as}{2\pi}S_\ep \left(\frac{\mu^2}{Q^2}\right)^\ep\right]^2
&\bigg[
2 \CF^2\left(\rC^{(0)}_{12,q} + \rC^{(0)}_{12,q q}
- 2 \rSCS^{(0),(1,2)}_{12,q} + 4 \rS^{(0),(1,2)(1,2)}_{12}\right)
\\[2mm]&
- 2 \CF \CA \rS^{(0),(1,2)}_{12}
\bigg]\,,
\label{eq:I122jet}
\esp
\eeq
with all arguments being equal to 1. Substituting the Laurent expansions
of the kinematic functions, we obtain
\beq
\bI^{(0)}_{12}(p_1, p_2;\ep) =
\left[\frac{\as}{2\pi}S_\ep \left(\frac{\mu^2}{Q^2}\right)^\ep\right]^2
\CF^2
\sum_{i=-4}^0
\ep^i
\Big(
  \cI_{12,2j}^{(0,i)}
+ x\,\cI_{12,2j}^{(x,i)}
+ y\Nf\,\cI_{12,2j}^{(y,i)}
\Big)
\,,
\label{eq:serI122jet}
\eeq
where \cite{Nagy:1997md}
\beq
x = \frac{\CA}{\CF}\,,\qquad y = \frac{\TR}{\CF}\,.
\eeq
With this decomposition the Abelian case is obtained by setting
$\CF = 1$, $x =0$, $y = 1$.
We can compute the two leading terms in the $\ep$ expansion analytically:
\beq
\bsp
\bI^{(0)}_{12}(p_1, p_2;\ep) &=
\left[\frac{\as}{2\pi}S_\ep \left(\frac{\mu^2}{Q^2}\right)^\ep\right]^2
\CF^2
\bigg\{(3 - x)\pole24 
+ \frac16 \bigg[72 + 21 x - 6 y\Nf 
\\[2mm] &\qquad
	- 24 (1 - x) \Sigma(y_0, D'_0 - 1) 
	- 12 (2 - 3 x) \Sigma(y_0, D'_0)\bigg]\pole13 + \Oe{-2}\bigg\}\,.
\esp
\eeq
The rest of the expansion coefficients are computed numerically. 
We present the results in \tab{tab:2jet}. To obtain these numbers,  we used the 
specific values  of $\al_0 = y_0 = 1$, $d_0 = d'_0 = 3$.
\begin{table}
\begin{center}
\renewcommand{\arraystretch}{1.5}
\begin{tabular}{|l|c|c|c|c|c|}
\hline\hline
$i$ & $-4$ & $-3$ & $-2$ & $-1$ & $0$ \\
\hline
$\cI_{12,2j}^{(0,i)}$ & 
	$6$ & 
	$\frac{76}{3}$ & 
	$32.10 \pm 0.23$ &
	$-87.90 \pm 0.66$ &
	$-554.5 \pm 1.8$\\
\hline
$\cI_{12,2j}^{(x,i)}$ & 
	$-2$ & 
	$-\frac{27}{2}$ & 
	$-52.40 \pm 0.01$ &
	$-150.7 \pm 0.09$ &
	$-339.5 \pm 0.43$\\
\hline
$\cI_{12,2j}^{(y,i)}$ & 
	$0$ & 
	$-1$ &
	$-6.332 \pm 0.001$ &
	$-17.66 \pm 0.008$ &
	$1.013 \pm 0.069$\\
\hline\hline
\end{tabular}
\caption{\label{tab:2jet} Coefficients of the Laurent expansion of the $\cI_{12,2j}$ 
functions appearing in the insertion operator $\bI_{12}^{(0)}(p_1, p_2;\ep) $ in the 
case of two-jet production. The numbers for $i=-4,-3$ are obtained by evaluating 
the appropriate analytic expressions. We used the parameters $\al_0 = y_0 = 1$, 
$d_0 = d'_0 = 3$.}
\end{center}
\end{table}

Finally, we show the value of the complete insertion operator for the case of QCD 
with $\Nf=5$ light flavours:
\beq
\bI_{12}^{(0)}(p_1, p_2;\ep) = 
	\frac{3}{2\ep^4}
	-\frac{83}{12\ep^3}
	-\frac{97.68 \pm 0.27}{\ep^2}
	-\frac{460.2 \pm 0.87}{\ep}
	-\Big(1317. \pm 2.9\Big) + \Oe{1}\,.
\eeq
In the above equation, the coefficients of $1/\ep^4$ and $1/\ep^3$ were 
computed by evaluating the appropriate analytic expressions. However, 
we have also computed these term with the same numerical algorithms that 
we used for computing the higher order expansion coefficients. It is then 
instructive to compare this numerical result to the analytic one. We find:
\bal
\bI_{12}^{(0)}(p_1, p_2;\ep)\big|_{\rm A} &= 
	\frac{1.5}{\ep^4}
	-\frac{6.917}{\ep^3} + \Oe{-2}\,,
\\[2mm] 
\bI_{12}^{(0)}(p_1, p_2;\ep)\big|_{\rm N} &= 
	\frac{1.498 \pm 0.0014}{\ep^4}
	-\frac{6.932 \pm 0.11}{\ep^3} + \Oe{-2}\,.
\eal
Comparing the exact and numerical results, we see first of all that the two 
match up to the uncertainty of the numerical computation, and second, that 
the error estimate on the next-to-leading pole is very conservative.

%
%

\subsection{Three-jet production}
\label{sec:3jet-prod}

Next, let us consider the process $e^+e^- \to 3$ jets. The corresponding
squared matrix element at tree level is $|\cM^{(0)}_2(1_q,2_\qb,3_g)|^2$,
i.e.~the quark carries label 1, the antiquark label 2 and the gluon
carries label 3. The colour algebra is again trivial. Colour
conservation implies 
\beq
\bT_1\,\bT_2 = \frac{\CA - 2\CF}{2}\,,\qquad
\bT_1\,\bT_3 = \bT_2\,\bT_3 = -\frac{\CA}{2}\,,
\eeq
and $C_{f_1} = C_{f_2} = \CF$, while $C_{f_3} = \CA$. Thus, the insertion 
operator is again a scalar in colour space.
On the other hand, the kinematics is no longer trivial, since the relative 
orientation of the three final state momenta are not fully fixed by momentum 
conservation. (Note that the insertion operator is independent of the overall 
event orientation with respect to the beam.) Since the three-particle phase 
space in $d=4$ dimensions is 5-dimensional, but three of the independent 
variables just correspond to the three Euler angles needed to specify the 
overall orientation, we find that out of the six kinematic
variables
\beq
x_i = y_{iQ}\,,\quad\mbox{and}\quad 
Y_{ij}\equiv Y_{ij,Q} = \frac{y_{ij}}{x_i x_j}\,,\qquad
i,j=1,2,3 \quad\mbox{and}\quad i\ne j\,,
\eeq
only two are independent. Nevertheless, we choose not to fix the independent 
ones below, in order to better exhibit the structure of the insertion operator.

The insertion operator \eqn{eq:I12} becomes
\beq
\bsp
\bI^{(0)}_{12} =
\left[\frac{\as}{2\pi}S_\ep \left(\frac{\mu^2}{Q^2}\right)^\ep\right]^2
&\bigg\{
  \CF^2 \Big(\rC^{(0)}_{12,q}(x_1) + \rC^{(0)}_{12,q}(x_2)\Big)
+ \CA^2 \rC^{(0)}_{12,g}(x_3)
\\[2mm]&
+ \CF^2 \Big(\rC^{(0)}_{12,qq}(x_1,x_2,Y_{12}) + \rC^{(0)}_{12,qq}(x_2,x_1,Y_{12})\Big)
\\[2mm]&
+ \CF \CA \Big(\rC^{(0)}_{12,qg}(x_1,x_3,Y_{13}) + \rC^{(0)}_{12,gq}(x_3,x_1,Y_{13})
\\[2mm]&\qquad\qquad
+ \rC^{(0)}_{12,qg}(x_2,x_3,Y_{23}) + \rC^{(0)}_{12,gq}(x_3,x_2,Y_{23})\Big)
\\[2mm]&
+ (\CA - 2 \CF) \Big[
  \CA \Big(\rSCS^{(0),(1,2)}_g(x_3, Y_{12}) + \rS^{(0),(1,2)}(Y_{12})\Big)
\\[2mm]&\qquad\qquad\qquad
+ \CF \Big(\rSCS^{(0),(1,2)}_q(x_1, Y_{12}) + \rSCS^{(0),(1,2)}_q(x_2, Y_{12})\Big)
\Big]
\\[2mm]&
- \CA \Big[
  \CA \Big(\rSCS^{(0),(1,3)}_g(x_3, Y_{13}) + \rS^{(0),(1,3)}(Y_{13})\Big)
\\[2mm]&\qquad
+ \CF \Big(\rSCS^{(0),(1,3)}_q(x_1, Y_{13}) + \rSCS^{(0),(1,3)}_q(x_2, Y_{13})\Big)
\\[2mm]&\qquad
+ \CA \Big(\rSCS^{(0),(2,3)}_g(x_3, Y_{23}) + \rS^{(0),(2,3)}(Y_{23})\Big)
\\[2mm]&\qquad
+ \CF \Big(\rSCS^{(0),(2,3)}_q(x_1, Y_{23}) + \rSCS^{(0),(2,3)}_q(x_2, Y_{23})\Big)
\Big]
\\[2mm]&
+ 2 \CA (2 \CF - \CA)
\\[2mm]&\quad\times
  \Big(\rS^{(0),(1,2)(1,3)}_{12}(Y_{12},Y_{23},Y_{13})
     + \rS^{(0),(1,2)(2,3)}_{12}(Y_{12},Y_{13},Y_{23})
\\[2mm]&\qquad
     + \rS^{(0),(2,3)(1,2)}_{12}(Y_{23},Y_{13},Y_{12})
     + \rS^{(0),(1,3)(1,2)}_{12}(Y_{13},Y_{23},Y_{12})
\Big)
\\[2mm]&
+ 2 \CA^2 \Big(
  \rS^{(0),(1,3)(2,3)}_{12}(Y_{13},Y_{12},Y_{23})
+ \rS^{(0),(2,3)(1,3)}_{12}(Y_{23},Y_{12},Y_{13})
\Big)
\\[2mm]&
+ \Big(8 \CF^2 - 8 \CF \CA + 2 \CA^2\Big)\rS^{(0),(1,2)(1,2)}_{12}(Y_{12})
\\[2mm]&
+ 2 \CA^2
  \Big(\rS^{(0),(1,3)(1,3)}_{12}(Y_{13}) + \rS^{(0),(2,3)(2,3)}_{12}(Y_{23})\Big)
\bigg\}\,.
\label{eq:I123j}
\esp
\eeq
Substituting the Laurent expansions of the kinematic functions, we
obtain
\beq
\bsp
\bI^{(0)}_{12}(p_1, p_2, p_3;\ep) &=
\left[\frac{\as}{2\pi}S_\ep \left(\frac{\mu^2}{Q^2}\right)^\ep\right]^2
\CF^2
\sum_{i=-4}^0
\ep^i
\\[2mm]&\times
\Big(
  \cI_{12,3j}^{(0,i)}
+ x\,\cI_{12,3j}^{(x,i)}
+ x^2\,\cI_{12,3j}^{(x^2,i)}
+ y\Nf\,\cI_{12,3j}^{(y,i)}
+ x y\Nf\,\cI_{12,3j}^{(x y,i)}
\Big)
\,.
\label{eq:serI123jet}
\esp
\eeq
We can compute the leading two terms in the $\ep$ expansion analytically:
\beq
\bsp
\bI^{(0)}_{12}(p_1, p_2, p_3;\ep) &=
\left[\frac{\as}{2\pi}S_\ep \left(\frac{\mu^2}{Q^2}\right)^\ep\right]^2
\CF^2\bigg\{
(6 + 2 x + x^2)\pole14
+\bigg[12 + {101 \over 6} x + {67 \over 12} x^2 
\\[2mm]&\qquad
- {13 \over 3} y\Nf - {3 \over 2} x y \Nf
-(4 - 4 x)\Sigma(y_0, D'_0-1)
\\[2mm]&\qquad
-(4 - 6x - x^2)\Sigma(y_0, D'_0)
-\bigg(8 + x - \frac52 x^2\bigg)\ln{y_{12}}
\\[2mm]&\qquad
-\bigg(4 x + \frac52 x^2\bigg)(\ln{y_{13}}+\ln{y_{23}})\bigg]\pole13 + \Oe{-2}\bigg\}\,.
\esp
\label{eq:I12_3j_analytic}
\eeq
In order to obtain \eqn{eq:I12_3j_analytic}, we used
$\ln Y_{ij} = \ln y_{ij} - \ln x_i - \ln x_j$.
The rest of the expansion coefficients are computed numerically. 
For purposes of demonstration, below we present numerical results 
in three specific phase space points. 
Since the result is insensitive to overall orientation, we will always 
choose the event to lie in the $x-y$ plane, with $p_1^\mu$ pointing in the 
positive $y$ direction.
In all cases we use the specific values of $\al_0 = y_0 = 1$, $d_0 = d'_0 = 3$.


\paragraph{Symmetric point.}
First, we consider the maximally symmetric configuration
\beq
\renewcommand{\arraystretch}{1.5}
\begin{tabular}{lcc @{\big(} l @{\;\;,\;\;} l @{\;\;,\;\;} l @{\;\;,\;\;} l @{\big)\,,}}
$p_1^\mu$ &=& $\sqrt{s}\,$ & $\frac13$ & 0 & $\frac13$ & 0\\
$p_2^\mu$ &=& $\sqrt{s}\,$ & $\frac13$ & $\frac{1}{2\sqrt{3}}$ & $-\frac16$ & 0\\
$p_3^\mu$ &=& $\sqrt{s}\,$ & $\frac13$ & $-\frac{1}{2\sqrt{3}}$ & $-\frac16$ & 0
\end{tabular}
\label{eq:symmPSpoint}
\eeq
which leads to the following values for the kinematic invariants ($i,j=1,2,3$ and $i\ne j$):
\beq
y_{ij} = \frac13\,,\qquad
x_i = \frac23\,,\qquad\mbox{and}\qquad
Y_{ij} = \frac34\,.
\eeq
The coefficients of the Laurent expansion of the insertion operator in the 
symmetric phase space point are shown in \tab{tab:3jet-symm}.
\begin{table}
\begin{center}
\renewcommand{\arraystretch}{1.5}
\begin{tabular}{|l|c|c|c|c|c|}
\hline\hline
$i$ & $-4$ & $-3$ & $-2$ & $-1$ & $0$ \\
\hline
$\cI_{12,3j}^{(0,i)}$ & 
	$6$ & 
	$34.12$ & 
	$82.98 \pm 0.25$ &
	$34.59 \pm 0.71$ &
	$-543.8 \pm 2.2$\\
\hline
$\cI_{12,3j}^{(x,i)}$ & 
	$2$ & 
	$9.721$ & 
	$1.209 \pm 0.52$ &
	$-142.2 \pm 1.5$ &
	$-696.6 \pm 4.8$\\
\hline
$\cI_{12,3j}^{(x^2,i)}$ & 
	$1$ & 
	$6.497$ & 
	$17.80 \pm 0.23$ &
	$15.87 \pm 0.79$ &
	$-47.93 \pm 2.9$\\
\hline
$\cI_{12,3j}^{(y,i)}$ & 
	$0$ & 
	$-\frac{13}{3}$ & 
	$-32.40 \pm 0.007$ &
	$-127.9 \pm 0.03$ &
	$-355.2 \pm 0.20$\\
\hline
$\cI_{12,3j}^{(xy,i)}$ & 
	$0$ & 
	$-\frac{3}{2}$ & 
	$-12.01 \pm 0.004$ &
	$-46.90 \pm 0.02$ &
	$-104.1 \pm 0.16$\\
\hline\hline
\end{tabular}
\caption{\label{tab:3jet-symm} Coefficients of the Laurent expansion of the $\cI_{12,3j}$
functions appearing in the insertion operator $\bI_{12}^{(0)}(p_1, p_2,p_3;\ep)$ for 
three-jet production in the symmetric phase space point. The numbers for $i=-4,-3$ 
are obtained by evaluating the appropriate analytic expressions. We used the 
parameters $\al_0 = y_0 = 1$, $d_0 = d'_0 = 3$.}
\end{center}
\end{table}

Finally, we show the value of the complete insertion operator in the symmetric 
phase space point for the case of QCD with $\Nf=5$ light flavours:
\beq
\bI_{12}^{(0)}(p_1, p_2,p_3;\ep) = 
	\frac{83}{3\ep^4}
	+\frac{132.3}{\ep^3}
	+\frac{114.5 \pm 4.6}{\ep^2}
	-\frac{1142. \pm 14.}{\ep}
	-\Big(6150. \pm 51.\Big) + \Oe{1}\,.
\eeq
In this equation the coefficients of $1/\ep^4$ and $1/\ep^3$ were obtained 
by evaluating the appropriate analytic expressions. As in the case of 
two-jet production, it is again instructive to compare the exact result with
one obtained by numerical computation. We find:
\bal
\bI_{12}^{(0)}(p_1, p_2,p_3;\ep)\big|_{\rm A} &= 
	\frac{27.67}{\ep^4}
	+\frac{132.3}{\ep^3} + \Oe{-2}\,,
\\[2mm] 
\bI_{12}^{(0)}(p_1, p_2,p_3;\ep)\big|_{\rm N} &= 
	\frac{27.658 \pm 0.027}{\ep^4}
	+\frac{132.4 \pm 1.9}{\ep^3} + \Oe{-2}\,.
\eal
As before, the results match within the uncertainty of the numerical calculation,
and the error estimate on the next-to-leading pole is again seen to be very 
conservative. We will reach similar conclusions in other examples as well.


\paragraph{Collinear point.}
Next, we choose a configuration where (in the rest frame of $Q^\mu$) we 
have a hierarchy of angles such that
\beq
\measuredangle(p_2,p_3) \ll \measuredangle(p_1,p_2), 
\measuredangle(p_1,p_3)\,,
\eeq
i.e.~where momenta $p_2^\mu$ and $p_2^\mu$ are close to being collinear. 
Specifically, we set
\beq
\renewcommand{\arraystretch}{1.5}
\begin{tabular}{lcc @{\big(} l @{\;\;,\;\;} l @{\;\;,\;\;} l @{\;\;,\;\;} l @{\big)\,,}}
$p_1^\mu$ &=& $\sqrt{s}\,$ & 0.49841 & 0. & 0.49841 & 0. \\
$p_2^\mu$ &=& $\sqrt{s}\,$ & 0.120923 & 0.0240634 & -0.118505 & 0. \\
$p_3^\mu$ &=& $\sqrt{s}\,$ & 0.380667 & -0.0240634 & -0.379905 & 0.
\end{tabular}
\label{eq:collPSpoint}
\eeq
which leads to the following values for the kinematic invariants ($i,j=1,2,3$ and $i\ne j$):
\bal
y_{12} &= 0.238667\,,&
y_{13} &= 0.758153\,,&
y_{23} &= 0.003180\,,
\nt\\[2mm] 
x_{1} &= 0.99682\,,&
x_{2} &= 0.241847\,,&
x_{3} &= 0.761333\,,
\\[2mm] 
Y_{12} &= 0.99\,,&
Y_{13} &= 0.999\,,&
Y_{23} &= 0.0172697\,.
\nt
\eal
The coefficients of the Laurent expansion of the insertion operator in the 
collinear phase space point are shown in \tab{tab:3jet-coll}.
\begin{table}
\begin{center}
\renewcommand{\arraystretch}{1.5}
\begin{tabular}{|l|c|c|c|c|c|}
\hline\hline
$i$ & $-4$ & $-3$ & $-2$ & $-1$ & $0$ \\
\hline
$\cI_{12,3j}^{(0,i)}$ & 
	$6$ & 
	$36.79$ & 
	$106.0 \pm 0.23$ &
	$120.6 \pm 0.68$ &
	$-431.0 \pm 2.0$\\
\hline
$\cI_{12,3j}^{(x,i)}$ & 
	$2$ & 
	$25.38$ & 
	$143.6 \pm 0.55$ &
	$537.3 \pm 1.6$ &
	$1505. \pm 5.3$\\
\hline
$\cI_{12,3j}^{(x^2,i)}$ & 
	$1$ & 
	$15.24$ & 
	$119.5 \pm 0.29$ &
	$660.5 \pm 1.1$ &
	$2903. \pm 4.9$\\
\hline
$\cI_{12,3j}^{(y,i)}$ & 
	$0$ & 
	$-\frac{13}{3}$ & 
	$-31.30 \pm 0.007$ &
	$-121.7 \pm 0.03$ &
	$-346.0 \pm 0.18$\\
\hline
$\cI_{12,3j}^{(xy,i)}$ & 
	$0$ & 
	$-\frac{3}{2}$ & 
	$-17.72 \pm 0.005$ &
	$-109.1 \pm 0.03$ &
	$-470.9 \pm 0.21$\\
\hline\hline
\end{tabular}
\caption{\label{tab:3jet-coll} Coefficients of the Laurent expansion of the $\cI_{12,3j}$
functions appearing in the insertion operator $\bI_{12}^{(0)}(p_1, p_2,p_3;\ep)$ for 
three-jet production in the collinear phase space point. The numbers for $i=-4,-3$ 
are obtained by evaluating the appropriate analytic expressions. We used the 
parameters $\al_0 = y_0 = 1$, $d_0 = d'_0 = 3$.}
\end{center}
\end{table}

Finally, we show the value of the complete insertion operator in the collinear
phase space point for the case of QCD with $\Nf=5$ light flavours:
\beq
\bI_{12}^{(0)}(p_1, p_2,p_3;\ep) = 
	\frac{83}{3\ep^4}
	+\frac{278.3}{\ep^3}
	+\frac{1601.3 \pm 5.2}{\ep^2}
	-\frac{7084. \pm 18.}{\ep}
	-\Big(26690. \pm 71.\Big) + \Oe{1}\,.
\eeq
We present next the comparison of the exact $1/\ep^4$ and $1/\ep^3$ 
pole coefficients given above with ones computed numerically. We find:
\bal
\bI_{12}^{(0)}(p_1, p_2,p_3;\ep)\big|_{\rm A} &= 
	\frac{27.67}{\ep^4}
	+\frac{278.3}{\ep^3} + \Oe{-2}\,,
\\[2mm] 
\bI_{12}^{(0)}(p_1, p_2,p_3;\ep)\big|_{\rm N} &= 
	\frac{27.658 \pm 0.027}{\ep^4}
	+\frac{278.4 \pm 2.1}{\ep^3} + \Oe{-2}\,.
\eal
We note that the two results match up to the uncertainty of the numerical 
computation, and as in previous examples, the error estimate on the 
next-to-leading pole is seen to be very conservative.


\paragraph{Soft point.}
Finally, we consider a configuration where (in the rest frame of $Q^\mu$) we 
have a hierarchy of energies such that
\beq
E_3 \ll E_1,E_2\,,
\eeq
i.e.~where momentum $p_3^\mu$ is close to being soft. 
Specifically, we set
\beq
\renewcommand{\arraystretch}{1.5}
\begin{tabular}{lcc @{\big(} l @{\;\;,\;\;} l @{\;\;,\;\;} l @{\;\;,\;\;} l @{\big)\,,}}
$p_1^\mu$ &=& $\sqrt{s}\,$ & 0.480625 & 0. & 0.480625 & 0.\\
$p_2^\mu$ &=& $\sqrt{s}\,$ & 0.487897 & -0.0308419 & -0.486921 & 0. \\
$p_3^\mu$ &=& $\sqrt{s}\,$ & 0.0314778 & 0.0308419 & 0.00629557 & 0.
\end{tabular}
\label{eq:softPSpoint}
\eeq
which leads to the following values for the kinematic invariants ($i,j=1,2,3$ and $i\ne j$):
\bal
y_{12} &= 0.937044\,,&
y_{13} &= 0.024207\,,&
y_{23} &= 0.038749\,,
\nt\\[2mm] 
x_{1} &= 0.961251\,,&
x_{2} &= 0.975794\,,&
x_{3} &= 0.062956\,,
\\[2mm] 
Y_{12} &= 0.999\,,&
Y_{13} &= 0.4\,,&
Y_{23} &= 0.630768\,.
\nt
\eal
The coefficients of the Laurent expansion of the insertion operator in the 
soft phase space point are shown in \tab{tab:3jet-soft}.
\begin{table}
\begin{center}
\renewcommand{\arraystretch}{1.5}
\begin{tabular}{|l|c|c|c|c|c|}
\hline\hline
$i$ & $-4$ & $-3$ & $-2$ & $-1$ & $0$ \\
\hline
$\cI_{12,3j}^{(0,i)}$ & 
	$6$ & 
	$25.85$ & 
	$34.59 \pm 0.23$ &
	$-84.25 \pm 0.66$ &
	$-566.8 \pm 1.9$\\
\hline
$\cI_{12,3j}^{(x,i)}$ & 
	$2$ & 
	$27.79$ & 
	$136.8 \pm 0.52$ &
	$330.6 \pm 1.4$ &
	$46.20 \pm 4.5$\\
\hline
$\cI_{12,3j}^{(x^2,i)}$ & 
	$1$ & 
	$21.02$ & 
	$195.4 \pm 0.26$ &
	$1174. \pm 0.96$ &
	$5354. \pm 4.1$\\
\hline
$\cI_{12,3j}^{(y,i)}$ & 
	$0$ & 
	$-\frac{13}{3}$ & 
	$-57.59 \pm 0.009$ &
	$-405.2 \pm 0.06$ &
	$-2119. \pm 0.34$\\
\hline
$\cI_{12,3j}^{(xy,i)}$ & 
	$0$ & 
	$-\frac{3}{2}$ & 
	$-24.07 \pm 0.006$ &
	$-194.7 \pm 0.05$ &
	$-1083. \pm 0.31$\\
\hline\hline
\end{tabular}
\caption{\label{tab:3jet-soft} Coefficients of the Laurent expansion of the $\cI_{12,3j}$
functions appearing in the insertion operator $\bI_{12}^{(0)}(p_1, p_2,p_3;\ep)$ for 
three-jet production in the soft phase space point. The numbers for $i=-4,-3$ are 
obtained by evaluating the appropriate analytic expressions. We used the parameters 
$\al_0 = y_0 = 1$, $d_0 = d'_0 = 3$.}
\end{center}
\end{table}

Finally, we show the value of the complete insertion operator in the soft
phase space point for the case of QCD with $\Nf=5$ light flavours:
\beq
\bI_{12}^{(0)}(p_1, p_2,p_3;\ep) = 
	\frac{83}{3\ep^4}
	+\frac{320.6}{\ep^3}
	+\frac{1995. \pm 4.9}{\ep^2}
	-\frac{8928. \pm 16.}{\ep}
	-\Big(32182. \pm 61.\Big) + \Oe{1}\,.
\eeq
We finish by comparing the exact coefficients of $1/\ep^4$ and $1/\ep^3$ 
that appear above with the values obtained via numerical computation. We find:
\bal
\bI_{12}^{(0)}(p_1, p_2,p_3;\ep)\big|_{\rm A} &= 
	\frac{27.667}{\ep^4}
	+\frac{320.56}{\ep^3} + \Oe{-2}\,,
\\[2mm] 
\bI_{12}^{(0)}(p_1, p_2,p_3;\ep)\big|_{\rm N} &= 
	\frac{27.658 \pm 0.027}{\ep^4}
	+\frac{320.6 \pm 1.9}{\ep^3} + \Oe{-2}\,.
\eal
Our conclusions are identical to those in the symmetric and collinear phase 
space point: the values match up to the numerical uncertainty and the error 
estimate on the next-to-leading pole is again shown to be very conservative.

We finish by briefly commenting on the size of numerical uncertainties. 
The uncertainties relevant for phenomenology are those associated with the 
complete $\bI_{12}^{(0)}$ insertion operator, in various phase space points. 
However, the requirements in terms of precision are different for the pole 
coefficients and the finite part.

On the one hand, the pole coefficients are only relevant for establishing the 
cancellation of all $\ep$-poles between the doubly-virtual cross section and 
various integrated subtraction terms. As stressed earlier, our subtraction scheme 
is fully local, hence this cancellation can be checked point by point in phase space 
for any specific process. From a practical point of view, it clearly suffices to demonstrate 
pole cancellation in a relatively small number of phase space points, thus the pole 
coefficients of $\bI_{12}^{(0)}$ have to be computed as precisely as feasible 
in a small set of points only. Because of this, the runtime of numerical integration 
is not an issue, and increased precision may be obtained simply by adjusting 
the parameters of the numerical integration to include more sampling points for 
each integral. 

On the other hand, the precision requirement on the finite part of the insertion 
operator is essentially set by the relative uncertainty associated with the numerical 
phase space integration of the doubly-virtual contribution. This is not expected to 
be below the per mille level, hence from a practical point of view it is pointless to 
evaluate the finite term of the insertion operator with a precision much greater than 
this. In all cases discussed above, the relative uncertainty of the finite part of the 
insertion operator is at the per mille level already.



\section{Conclusions}
\label{sec:conclusions}

In this work, we have performed the integration of the iterated 
singly-unresolved approximate cross section of the NNLO subtraction 
scheme of \refrs{Somogyi:2006cz,Somogyi:2006db,Somogyi:2006da}.
The final result can be written as the product (in colour space) of the Born 
cross section times a newly defined insertion operator, $\bI^{(0)}_{12}$.
The insertion operator depends on the colours, flavours and momenta of 
the final-state partons, and is an elaborate sum of many different terms, each 
corresponding to the integrated form of a specific iterated singly-unresolved 
subtraction term of \refr{Somogyi:2006da}.

We have also explicitly evaluated all integrated subtraction terms 
which are necessary to assemble the insertion operator for processes involving 
at most three hard partons in the final state. 
The knowledge of these integrals (i.e.~their Laurent expansions in $\ep$ 
to $\Oe{}$ accuracy) is necessary in order to make the subtraction scheme 
an effective tool, and we have computed them once and for all.

We have achieved this task by deriving Mellin--Barnes integral representations 
for all integrals under consideration. In principle, it is possible to evaluate all MB 
integrals via the residuum theorem, and in a subsequent step to obtain fully 
analytic expressions by performing the summation of nested sums over series of 
residua. However in practice, we have encountered several cases of higher order 
expansion coefficients, where the summation cannot be performed analytically 
with present methods. Therefore, in this paper, we have concentrated on the 
direct numerical evaluation of the MB integrals in the complex plane. 
All MB representations for both the numerical and, if available, the analytic 
expressions have been checked by an independent evaluation of the integrals using 
sector decomposition as in \refr{Somogyi:2008fc}.
We have found that all integrals contributing to the insertion operator 
are smooth functions of their variables (in the colloquial sense). For practical 
applications, this means that all integrals (in particular the finite in $\ep$ 
contributions) can be given either in terms of interpolating tables or simple 
fitting functions, which can be computed once and for all. We leave this step 
for later work. Finally we want to stress again that the tables we have shown 
here are for demonstration purposes only, and obtaining high resolution 
interpolating tables needed for the computation of an actual cross section is 
straightforward. Increasing the accuracy of each entry is feasible as well, the
best way of doing this is under investigation.

The integrals discussed in this paper appear when integrating the
subtraction terms that regularise the doubly-real NNLO correction to the
jet cross section, see \refr{Somogyi:2006db}. The final step in finishing
the definition of the subtraction scheme is the computation of the
integrated counterterms corresponding to the the doubly-unresolved
approximate cross section (those labeled by $\rm{A}_2$ in \refr{Somogyi:2006da}). 
In that case, the analytic structure of the integrals is essentially the 
same as those studied in this paper, though a few are admittedly somewhat 
more cumbersome. Nevertheless, we are confident that the techniques of the 
present paper will also be applicable to the computation of these remaining
contributions.



\acknowledgments

We thank S.-O.~Moch for useful discussions and his comments on the 
manuscript and B.~Tausk for useful discussions.
This work is supported in part by the
Deutsche Forschungsgemeinschaft in SFB/TR 9, the Swiss National
Science Foundation Joint Research Project SCOPES IZ73Z0\_1/28079,
the Hungarian Scientific Research Fund grant OTKA K-60432 and by the
T\'AMOP 4.2.1./B-09/1/KONV-2010-0007 project.





\providecommand{\href}[2]{#2}\begingroup\raggedright\endgroup




\appendix


\section{Summation over unresolved flavours}
\label{sec:flavoursum}

In this appendix, we discuss how to perform the summation over unobserved 
flavours in \eqn{eq:I2dsigRRA12Jm-1}. It turns out that we only need to consider 
three different cases explicitly, corresponding to sums involving two, three and four 
partons. All specific results are then easily obtained by appropriate substitutions.

%
%

\subsection{Generic flavour sums}
\label{sec:generic-flavour-sums}

Consider an $m$-parton configuration with $m_f$ quarks of flavour $f$,
$m_{\fb}$ antiquarks of flavour $f$ and $m_g$ gluons. From this 
configuration we can obtain an $(m+2)$-parton configuration in the 
following ways.
\begin{enumerate}
\item Increasing the number of gluons by two,
\beq
m_g \to m_g+2\,.
\label{eq:add-gg}
\eeq
\item Increasing the number of quarks and antiquarks of flavour $f$ by one each,
\beq
m_f \to m_f+1,\quad m_{\fb} \to m_{\fb}+1\,.
\label{eq:add-qq}
\eeq
\item Increasing the number of quarks and antiquarks by two each. The flavour of 
the two quarks may or may not be identical. We will refer to these two cases respectively
as the `equal flavour' (e.f.) and `unequal flavour' (u.f.) configurations,
\beq
\bsp
\begin{array}{llll}
m_g \to m_g-2,& \quad m_{f} \to m_{f}+2,& \quad m_{\fb} \to m_{\fb}+2\,,
&\qquad\mbox{e.f.}
\\[2mm]
m_g \to m_g-2,& \quad m_{f/f'} \to m_{f/f'}+1,& \quad m_{\fb/\fb'} \to m_{\fb/\fb'}+1\,,
&\qquad\mbox{u.f.}
\end{array}
\esp
\label{eq:add-qqqq}
\eeq
where $f$ and $f'$ are understood to be different quark flavours.
This case is relevant only for the doubly-collinear-type configuration, i.e.\ the 
sum involving four partons. 
 \end{enumerate}
The ratios of Bose symmetry factors for identical final state particles in the 
various cases are
\bal
\frac{\Sym{m}}{\Sym{m+2}^{(1)}} &
	= \frac{1}{(m_g+1)(m_g+2)}\,,
\label{eq:StoS-gg}
\\[2mm]
\frac{\Sym{m}}{\Sym{m+2}^{(2)}} &
	= \frac{1}{(m_f+1)(m_\fb+1)}\,,
\label{eq:StoS-qq}
\intertext{and finally}
\frac{\Sym{m}}{\Sym{m+2}^{(3)}} &= \left\{\begin{array}{ll}
	\displaystyle{\frac{m_g(m_g-1)}{(m_f+2)(m_f+1)(m_\fb+2)(m_\fb+1)}}
	&\qquad\mbox{e.f.}
	\\[6mm]
	\displaystyle{\frac{m_g(m_g-1)}{(m_f+1)(m_\fb+1)(m_{f'}+1)(m_{\fb'}+1)}}
	&\qquad\mbox{u.f.}
	\end{array}\right.
\label{eq:StoS-qqqq}
\eal
%


\paragraph{Two-parton flavour sums.}
Consider a generic integrated counterterm $[X^{(0)}_{kt}]_{\ldots}^{(\ldots)}$, which 
depends on two indices, $k$ and $t$. This integrated counterterm may or may not 
depend on the corresponding parton flavours and it may or may not carry an upper 
index, as explained in detail below \eqn{eq:intXikjl}. 
Examples would be e.g.\ $[\IcC{kt}{}\IcS{kt}{(0)}]_{f_k f_t}^{(j,l)}$ and 
$[\IcS{t}{}\IcS{rt}{(0)}]^{(j,l)}$. In the latter case, we see a situation where no flavour 
index is displayed, since both $r$ and $t$ are constrained to be gluons.
In what follows, we will discuss the most general case, when both flavour indices are 
explicit. 

Such terms necessarily appear in \eqn{eq:I2dsigRRA12Jm-1} under a double sum:
\beq
\sum_{\br{m+2}} \frac{1}{\Sym{m+2}}
\sum_{t} \sum_{k\ne t} [X^{(0)}_{kt}]_{f_k f_t}^{(\ldots)}\,.
\eeq
In this configuration, we go from $m$ to $(m+2)$ partons as in 
\eqns{eq:add-gg}{eq:add-qq}.
Then, since we are considering iterated singly-unresolved terms, both indices 
must correspond to unresolved partons, which means that they are either both 
gluons or a quark-antiquark pair. Now, decomposing the summation over $t$ and $k$ 
into sums in which the flavour (including specific quark flavour) of each index is fixed, 
we find
\beq
\bsp
&
\sum_{\br{m+2}} \frac{1}{\Sym{m+2}}
\sum_{t} \sum_{k\ne t} 
[X^{(0)}_{kt}]_{f_k f_t}^{(\ldots)} = 
\\[2mm]&=
	\sum_{\br{m}} \frac{1}{\Sym{m}}
		\frac{\Sym{m}}{\Sym{m+2}^{(1)}}
		\sum_{t} \sum_{k\ne t} 
		[X^{(0)}_{kt}]_{f_k f_t}^{(\ldots)}
		\delf{f_k}{g} \delf{f_t}{g} 
\\[2mm] &+
	\sum_{\br{m}} \frac{1}{\Sym{m}}
		\frac{\Sym{m}}{\Sym{m+2}^{(2)}}
		\sum_{f'} \sum_{t} \sum_{k\ne t} 
		[X^{(0)}_{kt}]_{f_k f_t}^{(\ldots)}
		\Big[\delf{f_k}{q_{f'}} \delf{f_t}{\qb_{f'}}
			+\delf{f_k}{\qb_{f'}} \delf{f_t}{q_{f'}}\Big]\,,
\esp
\label{eq:Xkt-FS1}
\eeq
where $\sum_{f'}$ stands for the explicit summation over specific quark flavours.
Performing the summation over $t$ and $k$ simply amounts to counting the number 
of ways in which we can assign the proper flavours to $k$ and $t$ in the appropriate 
$(m+2)$-parton configuration:
\beq
\sum_{t} \sum_{k\ne t} \ldots = 
	\#(f_t)_{m+2}\; \#(f_k;k\ne t)_{m+2} \ldots\,,
\label{eq:Xkt-count}
\eeq
where $\#(f_t)_{m+2}$ denotes the number of partons of flavour $f_t$ in the 
$(m+2)$-parton configuration, while $\#(f_k;k\ne t)_{m+2}$ is the number of 
partons, {\em different} form $t$, of flavour $f_k$ in the $(m+2)$-parton configuration. 
Note that $t$ and $k$ are assumed to be distinguishable, which is the generic case. 
Clearly we have
\beq
\bsp
&
\#(g)_{m+2}\; \#(g;k\ne t)_{m+2} = 
	(m_g+2)(m_g+1)\,,
\\[2mm]
&
\#(q_f)_{m+2}\; \#(\qb_f;k\ne t)_{m+2} = 
	(m_f+1)(m_\fb+1)\,.
\esp
\eeq
The case of $f_t=\qb_f$ and $f_k=q_f$ is obtained by exploiting symmetry of this factor under 
permutations of indices. Then, using \eqns{eq:StoS-gg}{eq:StoS-qq}, we find
\beq
\sum_{\br{m+2}} \frac{1}{\Sym{m+2}}
\sum_t \sum_{k\ne t}
[X^{(0)}_{kt}]_{f_k f_t}^{(\ldots)} =
	\sum_{\br{m}} \frac{1}{\Sym{m}}
	\bigg\{[X^{(0)}_{kt}]_{gg}^{(\ldots)}
		+\Nf \Big([X^{(0)}_{kt}]_{q\qb}^{(\ldots)}
			+[X^{(0)}_{kt}]_{\qb q}^{(\ldots)}\Big)
	\bigg\}\,.
\label{eq:Xkt-FS-fin}
\eeq
In writing \eqn{eq:Xkt-FS-fin}, we have used that $[X^{(0)}_{kt}]_{q\qb}^{(\ldots)}$ 
and $[X^{(0)}_{kt}]_{\qb q}^{(\ldots)}$ do not depend on the specific quark flavour (as 
implied by the notation), and hence the summation $\sum_{f'}$ in \eqn{eq:Xkt-FS1} 
may be performed, yielding the factor of $\Nf$. 

Defining the flavour summed counterterm as
\beq
\sum_{\br{m+2}} \frac{1}{\Sym{m+2}}
\sum_t \sum_{k\ne t}
[X^{(0)}_{kt}]_{f_k f_t}^{(\ldots)} \equiv
	\sum_{\br{m}} \frac{1}{\Sym{m}}
	\lbb X^{(0)}_{kt} \rbb^{(\ldots)}\,,
\label{eq:Xkt-FSICT1}
\eeq
we obtain : 
\beq
\bsp
\lbb X^{(0)}_{kt} \rbb^{(\ldots)} &= 
	[X^{(0)}_{kt}]_{gg}^{(\ldots)}
	+\Nf \Big([X^{(0)}_{kt}]_{q\qb}^{(\ldots)}
		+[X^{(0)}_{kt}]_{\qb q}^{(\ldots)}\Big)
\\[2mm]&=
	[X^{(0)}_{kt}]_{gg}^{(\ldots)}
	+2\Nf [X^{(0)}_{kt}]_{q\qb}^{(\ldots)}\,,
\esp
\label{eq:Xkt-FSICT2}
\eeq
where the second line follows, since
$[X^{(0)}_{kt}]_{q\qb}^{(\ldots)} = [X^{(0)}_{kt}]_{\qb q}^{(\ldots)}$ in all cases we 
need to consider.


\paragraph{Three-parton flavour sums.}
Next, consider a generic integrated counterterm $[X^{(0)}_{ktr}]_{\ldots}^{(\ldots)}$, 
depending on three indices, $k$, $t$ and $r$. Examples are e.g.\ 
$[\IcC{kt}{}\IcC{ktr}{(0)}]_{f_k f_t f_r}$ and $[\IcS{t}{}\IcSCS{ir;t}{(0)}]_{f_i f_r}^{(j,l)}$. As 
before, we will discuss the most general case, when all flavour indices are explicit.

These terms always appear in \eqn{eq:I2dsigRRA12Jm-1} under a triple sum:
\beq
\sum_{\br{m+2}} \frac{1}{\Sym{m+2}}
\sum_{t} \sum_{k\ne t} \sum_{r\ne k,t} [X^{(0)}_{ktr}]_{f_k f_t f_r}^{(\ldots)}\,.
\eeq
In this configuration, we again go from $m$ to $(m+2)$ partons as in 
\eqns{eq:add-gg}{eq:add-qq}.
Then, we decompose the summation over $t$, $k$ and $r$ into sums in which the 
flavour (including specific quark flavour) of each index is fixed. We obtain
\beq
\bsp
&
\sum_{\br{m+2}} \frac{1}{\Sym{m+2}}
\sum_{t} \sum_{k\ne t} \sum_{r\ne k,t} [X^{(0)}_{ktr}]_{f_k f_t f_r}^{(\ldots)} = 
\\[2mm] &=
	\sum_{\br{m}} \frac{1}{\Sym{m}}
		\frac{\Sym{m}}{\Sym{m+2}^{(1)}}
		\sum_{t} \sum_{k\ne t} \sum_{r\ne k,t}
		[X^{(0)}_{ktr}]_{f_k f_t f_r}^{(\ldots)}
\\[2mm]&\quad\times
		\Big[\Big(\delf{f_k}{q_f} \delf{f_t}{g} \delf{f_r}{g}
			+\delf{f_k}{g} \delf{f_t}{q_f} \delf{f_r}{g}
			+\delf{f_k}{g} \delf{f_t}{g} \delf{f_r}{q_f}
			+(q_f \leftrightarrow \qb_f)\Big)
			+\delf{f_k}{g} \delf{f_t}{g} \delf{f_r}{g} \Big]
\\[2mm] &+
	\sum_{\br{m}} \frac{1}{\Sym{m}}
		\frac{\Sym{m}}{\Sym{m+2}^{(2)}}
\\[2mm]&\quad\times\bigg\{
		\sum_{f'} \sum_{t} \sum_{k\ne t} \sum_{r\ne k,t}
		[X^{(0)}_{ktr}]_{f_k f_t f_r}^{(\ldots)}
\\[2mm]&\quad\quad\quad\times
		\Big[\delf{f_k}{q_{f'}} \delf{f_t}{\qb_{f'}} \delf{f_r}{g}
			+\delf{f_k}{q_{f'}} \delf{f_t}{g} \delf{f_r}{\qb_{f'}}
			+\delf{f_k}{g} \delf{f_t}{q_{f'}} \delf{f_r}{\qb_{f'}}
\\[2mm]&\quad\quad\quad\quad
			+\delf{f_k}{\qb_{f'}} \delf{f_t}{q_{f'}} \delf{f_r}{g}
			+\delf{f_k}{\qb_{f'}} \delf{f_t}{g} \delf{f_r}{q_{f'}}
			+\delf{f_k}{g} \delf{f_t}{\qb_{f'}} \delf{f_r}{q_{f'}}\Big]
\\[2mm] &\quad\quad+
	\sum_{f'\ne f} \sum_{t} \sum_{k\ne t} \sum_{r\ne k,t}
		[X^{(0)}_{ktr}]_{f_k f_t f_r}^{(\ldots)}
\\[2mm]&\quad\quad\quad\times
		\Big[\delf{f_k}{q_{f'}} \delf{f_t}{\qb_{f'}} \delf{f_r}{q_f}
			+\delf{f_k}{q_{f'}} \delf{f_t}{q_f} \delf{f_r}{\qb_{f'}}
			+\delf{f_k}{q_f} \delf{f_t}{q_{f'}} \delf{f_r}{\qb_{f'}}
\\[2mm]&\quad\quad\quad	\quad		
			+\delf{f_k}{\qb_{f'}} \delf{f_t}{q_{f'}} \delf{f_r}{q_f}
			+\delf{f_k}{\qb_{f'}} \delf{f_t}{q_f} \delf{f_r}{q_{f'}}
			+\delf{f_k}{q_f} \delf{f_t}{\qb_{f'}} \delf{f_r}{q_{f'}}						
			+(q_f \leftrightarrow \qb_f)\Big]
\\[2mm] &\quad\quad+
	\sum_{t} \sum_{k\ne t} \sum_{r\ne k,t}
		[X^{(0)}_{ktr}]_{f_k f_t f_r}^{(\ldots)}
\\[2mm]&\quad\quad\quad\times
		\Big[\delf{f_k}{q_{f}} \delf{f_t}{q_{f}} \delf{f_r}{\qb_f}
			+\delf{f_k}{q_{f}} \delf{f_t}{\qb_f} \delf{f_r}{q_{f}}
			+\delf{f_k}{\qb_f} \delf{f_t}{q_{f}} \delf{f_r}{\qb_{f}}
			+(q_f \leftrightarrow \qb_f)\Big]\bigg\}
\esp
\label{eq:Xktr-FS1} 
\eeq
Next, we use the flavour summation rules to rewrite the summation over the 
unobserved indices $k$, $t$ and $r$ in the $(m+2)$-parton configurations into 
a sum over a single index $\wti{ktr}$ in the $m$-parton configuration. We have
\beq
\sum_{t} \sum_{k\ne t} \sum_{r\ne k,t} \ldots =
\frac{\#(f_t)_{m+2}\; \#(f_k;k\ne t)_{m+2}\; \#(f_r;r\ne k,t)_{m+2}}
{\#(f_{ktr})_m} \sum_{\wti{ktr}} \ldots\,,
\label{eq:Xktr-count}
\eeq
where the notation is the same as in \eqn{eq:Xkt-count} and in particular
$\#(f_{ktr})_m$ is the number of partons with flavour $f_{ktr}$ in the 
$m$-parton configuration. Again, $t$, $k$ and $r$ are assumed to be distinguishable, 
which is the generic case. Then we have
\beq
\bsp
&
\frac{\#(q_f)_{m+2}\; \#(g;k\ne t)_{m+2}\; \#(g;r\ne k,t)_{m+2}}{\#(q_f)_m} 
	= (m_g+2)(m_g+1)\,.
\\[2mm]	
&
\frac{\#(g)_{m+2}\; \#(g;k\ne t)_{m+2}\; \#(g;r\ne k,t)_{m+2}}{\#(g)_m} 
	= (m_g+2)(m_g+1)\,.
\\[2mm]	
&
\frac{\#(q_{f'})_{m+2}\; \#(\qb_{f'};k\ne t)_{m+2}\; \#(g;r\ne k,t)_{m+2}}{\#(g)_m} 
	= (m_{f'}+1)(m_{\fb'}+1)\,.
\\[2mm]	
&
\frac{\#(q_{f'})_{m+2}\; \#(\qb_{f'};k\ne t)_{m+2}\; \#(q_f;r\ne k,t)_{m+2}}{\#(q_f)_m} 
	= (m_{f'}+1)(m_{\fb'}+1)\,.
\\[2mm]	
&
\frac{\#(q_{f})_{m+2}\; \#(q_{f};k\ne t)_{m+2}\; \#(\qb_f;r\ne k,t)_{m+2}}{\#(q_f)_m} 
	= (m_{f}+1)(m_{\fb}+1)\,.
\esp
\eeq
The rest of the cases are obtained by exploiting symmetry of this factor under 
permutations of indices. Then, using \eqns{eq:StoS-gg}{eq:StoS-qq}, we find
\beq
\bsp
&
\sum_{\br{m+2}} \frac{1}{\Sym{m+2}}
\sum_{t} \sum_{k\ne t} \sum_{r\ne k,t} [X^{(0)}_{ktr}]_{f_k f_t f_r}^{(\ldots)} = 
\\[2mm] &=
	\sum_{\br{m}} \frac{1}{\Sym{m}} \sum_{\wti{ktr}} \bigg\{
	\bigg[ [X^{(0)}_{ktr}]_{qgg}^{(\ldots)}
		+[X^{(0)}_{ktr}]_{gqg}^{(\ldots)}
		+[X^{(0)}_{ktr}]_{ggq}^{(\ldots)}
		+(\Nf-1)\Big([X^{(0)}_{ktr}]_{q' \qb' q}^{(\ldots)}
				+[X^{(0)}_{ktr}]_{q' q\qb'}^{(\ldots)}
\\[2mm] &\qquad\qquad\qquad\qquad
				+[X^{(0)}_{ktr}]_{q q' \qb'}^{(\ldots)}
				+[X^{(0)}_{ktr}]_{\qb' q' q}^{(\ldots)}
				+[X^{(0)}_{ktr}]_{\qb' q q'}^{(\ldots)}
				+[X^{(0)}_{ktr}]_{q \qb' q'}^{(\ldots)}\Big)
				+[X^{(0)}_{ktr}]_{q q \qb}^{(\ldots)}
\\[2mm] &\qquad\qquad\qquad\qquad
				+[X^{(0)}_{ktr}]_{q \qb q}^{(\ldots)}
				+[X^{(0)}_{ktr}]_{\qb q q}^{(\ldots)}\bigg]\delf{f_{ktr}}{q_f}
		+ (q_f \leftrightarrow \qb_f)
\\[2mm] &\qquad\qquad\qquad+
	\bigg[ [X^{(0)}_{ktr}]_{ggg}^{(\ldots)}	
		+\Nf \Big([X^{(0)}_{ktr}]_{q\qb g}^{(\ldots)}
				+[X^{(0)}_{ktr}]_{qg\qb}^{(\ldots)}
				+[X^{(0)}_{ktr}]_{gq\qb}^{(\ldots)}
				+[X^{(0)}_{ktr}]_{\qb qg}^{(\ldots)}
\\[2mm] &\qquad\qquad\qquad\qquad
				+[X^{(0)}_{ktr}]_{\qb gq}^{(\ldots)}
				+[X^{(0)}_{ktr}]_{g\qb q}^{(\ldots)}\Big)\bigg]\delf{f_{ktr}}{g}\bigg\}
\esp
\label{eq:Xktr-FS-fin}
\eeq
In obtaining \eqn{eq:Xktr-FS-fin}, we have used that whenever any of $t$, $k$ or $r$ 
are (anti)quarks, $[X^{(0)}_{ktr}]_{f_k f_t f_r}^{(\ldots)}$ does not depend on the specific 
quark flavour(s) (as implied by the notation), except that we have allowed for the 
possibility that the `equal flavour' and `unequal flavour' counterterms are different, e.g.\
$[X^{(0)}_{ktr}]_{q\qb q}^{(\ldots)} \ne [X^{(0)}_{ktr}]_{q'\qb' q}^{(\ldots)}$, which implies that 
the summations $\sum_{f'}$ in \eqn{eq:Xktr-FS1} may be performed, yielding the factors 
of $(\Nf-1)$ and $\Nf$. 

Let us define the flavour summed counterterms as follows:
\beq
\sum_{\br{m+2}} \frac{1}{\Sym{m+2}}
\sum_{t} \sum_{k\ne t} \sum_{r\ne k,t} [X^{(0)}_{ktr}]_{f_k f_t f_r}^{(\ldots)} = 
	\sum_{\br{m}} \frac{1}{\Sym{m}} \sum_{\wti{ktr}}
	\lbb X^{(0)}_{ktr}\rbb_{f_{ktr}}^{(\ldots)}.
\label{eq:Xktr-FSICT1}
\eeq
Then we find
\beq
\bsp
\lbb X^{(0)}_{ktr}\rbb_{q}^{(\ldots)} &=
[X^{(0)}_{ktr}]_{qgg}^{(\ldots)}
		+[X^{(0)}_{ktr}]_{gqg}^{(\ldots)}
		+[X^{(0)}_{ktr}]_{ggq}^{(\ldots)}
		+(\Nf-1)\Big([X^{(0)}_{ktr}]_{q' \qb' q}^{(\ldots)}
				+[X^{(0)}_{ktr}]_{q' q\qb'}^{(\ldots)}
\\[2mm] &
				+[X^{(0)}_{ktr}]_{q q' \qb'}^{(\ldots)}
				+[X^{(0)}_{ktr}]_{\qb' q' q}^{(\ldots)}
				+[X^{(0)}_{ktr}]_{\qb' q q'}^{(\ldots)}
				+[X^{(0)}_{ktr}]_{q \qb' q'}^{(\ldots)}\Big)
				+[X^{(0)}_{ktr}]_{q q \qb}^{(\ldots)}
				+[X^{(0)}_{ktr}]_{q \qb q}^{(\ldots)}
\\[2mm] &
				+[X^{(0)}_{ktr}]_{\qb q q}^{(\ldots)}\,,
\\[2mm]
\lbb X^{(0)}_{ktr}\rbb_{g}^{(\ldots)} &=
[X^{(0)}_{ktr}]_{ggg}^{(\ldots)}	
		+\Nf \Big([X^{(0)}_{ktr}]_{q\qb g}^{(\ldots)}
				+[X^{(0)}_{ktr}]_{qg\qb}^{(\ldots)}
				+[X^{(0)}_{ktr}]_{gq\qb}^{(\ldots)}
				+[X^{(0)}_{ktr}]_{\qb qg}^{(\ldots)}
				+[X^{(0)}_{ktr}]_{\qb gq}^{(\ldots)}
\\[2mm] &
				+[X^{(0)}_{ktr}]_{g\qb q}^{(\ldots)}\Big)\,.
\esp
\label{eq:Xktr-FSICT2}
\eeq
%


\paragraph{Four-parton flavour sums.}
Consider finally a generic integrated counterterm $[X^{(0)}_{ktir}]_{\ldots}^{(\ldots)}$, 
which depends on four indices, $k$, $t$, $i$ and $r$. Two examples are  
$[\IcC{kt}{}\IcC{ir;kt}{(0)}]_{f_k f_t; f_i f_r}$ and $[\IcC{kt}{}\IcS{t}{}\IcSCS{ir;t}{}\IcS{rt}{(0)}]$. 
We will discuss the most general case, when all flavour indices are explicit.

These terms always appear in \eqn{eq:I2dsigRRA12Jm-1} under a four-fold sum:
\beq
\sum_{\br{m+2}} \frac{1}{\Sym{m+2}}
\sum_{t} \sum_{k\ne t} \sum_{r\ne k,t} \sum_{i\ne k,t,r} [X^{(0)}_{ktir}]_{f_k f_t f_i f_r}^{(\ldots)}\,.
\eeq
In this configuration, we go from $m$ to $(m+2)$ partons as in 
\eqnss{eq:add-gg}{eq:add-qqqq}, 
i.e.\ case 3 must also be considered.
Decomposing the summation over $t$, $k$, $i$ and $r$ into sums in which the 
flavour (including specific quark flavour) of each index is fixed, we find
\beq
\bsp
&
\sum_{\br{m+2}} \frac{1}{\Sym{m+2}}
\sum_{t} \sum_{k\ne t} \sum_{r\ne k,t} \sum_{i\ne k,t,r} [X^{(0)}_{ktir}]_{f_k f_t f_i f_r}^{(\ldots)} = 
\\[2mm] &=
	\sum_{\br{m}} \frac{1}{\Sym{m}}
		\frac{\Sym{m}}{\Sym{m+2}^{(1)}}
		\sum_{t} \sum_{k\ne t} \sum_{r\ne k,t} \sum_{i\ne k,t,r}
		[X^{(0)}_{ktir}]_{f_k f_t f_i f_r}^{(\ldots)}
\\[2mm] &\quad\times
		\Big[\Big(\delf{f_k}{q_f} \delf{f_t}{g} \delf{f_i}{q_{f'}} \delf{f_r}{g}
			+\delf{f_k}{q_f} \delf{f_t}{g} \delf{f_i}{g} \delf{f_r}{q_{f'}}
			+\delf{f_k}{g} \delf{f_t}{q_f} \delf{f_i}{q_{f'}} \delf{f_r}{g}
\\[2mm]&\quad\quad\quad
			+\delf{f_k}{g} \delf{f_t}{q_f} \delf{f_i}{g} \delf{f_r}{q_{f'}}
			+ (q_f \leftrightarrow \qb_f)
			+ (q_{f'} \leftrightarrow \qb_{f'})
			+ (q_f \leftrightarrow \qb_f\, , \,q_{f'} \leftrightarrow \qb_{f'})\Big)
\\[2mm]&\quad\quad
			+\Big(\delf{f_k}{q_f} \delf{f_t}{g} \delf{f_i}{g} \delf{f_r}{g}
			+\delf{f_k}{g} \delf{f_t}{q_f} \delf{f_i}{g} \delf{f_r}{g}
			+\delf{f_k}{g} \delf{f_t}{g} \delf{f_i}{q_f} \delf{f_r}{g}
\\[2mm]&\quad\quad\quad
			+\delf{f_k}{g} \delf{f_t}{g} \delf{f_i}{g} \delf{f_r}{q_f}
			+ (q_f \leftrightarrow \qb_f)\Big)
			+\delf{f_k}{g} \delf{f_t}{g} \delf{f_i}{g} \delf{f_r}{g}\Big]
\\[2mm] &+
	\sum_{\br{m}} \frac{1}{\Sym{m}}
		\frac{\Sym{m}}{\Sym{m+2}^{(2)}}
		\sum_{f'} \sum_{t} \sum_{k\ne t} \sum_{r\ne k,t} \sum_{i\ne k,t,r}
		[X^{(0)}_{ktir}]_{f_k f_t f_i f_r}^{(\ldots)}
\\[2mm] &\quad\times
		\Big[\Big(\delf{f_k}{q_f} \delf{f_t}{g} \delf{f_i}{q_{f'}} \delf{f_r}{\qb_{f'}}
			+\delf{f_k}{q_f} \delf{f_t}{g} \delf{f_i}{\qb_{f'}} \delf{f_r}{q_{f'}}
			+\delf{f_k}{g} \delf{f_t}{q_f} \delf{f_i}{q_{f'}} \delf{f_r}{\qb_{f'}}
\\[2mm]&\quad\quad\quad
			+\delf{f_k}{g} \delf{f_t}{q_f} \delf{f_i}{\qb_{f'}} \delf{f_r}{q_{f'}}
			+\delf{f_k}{q_{f'}} \delf{f_t}{\qb_{f'}} \delf{f_i}{q_f} \delf{f_r}{g}
			+\delf{f_k}{\qb_{f'}} \delf{f_t}{q_{f'}} \delf{f_i}{q_f} \delf{f_r}{g}
\\[2mm]&\quad\quad\quad
			+\delf{f_k}{q_{f'}} \delf{f_t}{\qb_{f'}} \delf{f_i}{g} \delf{f_r}{q_f}
			+\delf{f_k}{\qb_{f'}} \delf{f_t}{q_{f'}} \delf{f_i}{g} \delf{f_r}{q_f}
			+(q_f \leftrightarrow \qb_f)\Big)
\\[2mm]&\quad\quad
			+\delf{f_k}{g} \delf{f_t}{g} \delf{f_i}{q_{f'}} \delf{f_r}{\qb_{f'}}
			+\delf{f_k}{g} \delf{f_t}{g} \delf{f_i}{\qb_{f'}} \delf{f_r}{q_{f'}}
			+\delf{f_k}{q_{f'}} \delf{f_t}{\qb_{f'}} \delf{f_i}{g} \delf{f_r}{g}
\\[2mm]&\quad\quad
			+\delf{f_k}{\qb_{f'}} \delf{f_t}{q_{f'}} \delf{f_i}{g} \delf{f_r}{g}\Big]
\\[2mm] &+
	\sum_{\br{m}} \frac{1}{\Sym{m}}
		\frac{\Sym{m}}{\Sym{m+2}^{(3)}}
		\sum_f \sum_{f'} \sum_{t} \sum_{k\ne t} \sum_{r\ne k,t} \sum_{i\ne k,t,r}
		[X^{(0)}_{ktir}]_{f_k f_t f_i f_r}^{(\ldots)}
\\[2mm] &\quad\times
		\Big[\delf{f_k}{q_f} \delf{f_t}{\qb_f} \delf{f_i}{q_{f'}} \delf{f_r}{\qb_{f'}}
			+\delf{f_k}{q_f} \delf{f_t}{\qb_f} \delf{f_i}{\qb_{f'}} \delf{f_r}{q_{f'}}
			+\delf{f_k}{\qb_f} \delf{f_t}{q_f} \delf{f_i}{q_{f'}} \delf{f_r}{\qb_{f'}}
\\[2mm]&\quad\quad
			+\delf{f_k}{\qb_f} \delf{f_t}{q_f} \delf{f_i}{\qb_{f'}} \delf{f_r}{q_{f'}}\Big]\,.
\esp
\label{eq:Xktir-FS1}
\eeq
Four-index subtraction terms only arise in conjunction with the double collinear 
limit and are always completely independent of the specific quark flavours. We have 
used these facts to write \eqn{eq:Xktir-FS1} in the above form. First, since the 
pairs of indices $k$, $t$ and $i$, $r$ will always correspond to true singly-unresolved 
collinear limits, we have discarded all terms where this is not the case. In effect, we 
have dropped all terms where both $k$ and $t$ or both $i$ and $r$ are (anti)quarks.
Second, complete independence of all counterterms on specific quark flavours implies 
that the `equal flavour' and `unequal flavour' ones are equal. 
E.g.\ $[X^{(0)}_{ktir}]_{qgq' \qb'} = [X^{(0)}_{ktir}]_{qgq \qb}$ and so on. We have used 
this fact in writing the equation, hence, in (\ref{eq:Xktir-FS1}), $f$ and $f'$ are not necessarily 
distinct flavours. Then using the flavour summation rules, we can rewrite the summation 
over the unobserved indices $k$, $t$, $i$ and $r$ in the $(m+2)$-parton configurations 
into sums over indices $\wti{kt}$ and $\wti{ir}$ in the $m$-parton configuration. 
We have
\beq
\bsp
\sum_{k} \sum_{t\ne k} \sum_{r\ne k,t} \sum_{i\ne k,t,r} \ldots &=
\frac{\#(f_t)_{m+2}\; \#(f_k;k\ne t)_{m+2}\; \#(f_r;r\ne k,t)_{m+2}\; \#(f_i;i\ne k,t,r)_{m+2}}
{\#(f_{kt})_m\; \#(f_{ir};\wti{ir}\ne\wti{kt})_m}
\\[2mm]&\times
\sum_{\wti{kt}} \sum_{\wti{ir}\ne\wti{kt}}\ldots\,,
\esp
\label{eq:Xktir-count}
\eeq
where the notation is the same as in \eqns{eq:Xkt-count}{eq:Xktr-count}. We assume that 
in general $k$, $t$, $i$ and $r$ are all distinguishable. Then we find
\beq
\bsp
&
\frac{\#(q_f)_{m+2}\; \#(g;k\ne t)_{m+2}\; \#(q_{f'};r\ne k,t)_{m+2}\; \#(g;i\ne k,t,r)_{m+2}}
{\#(q_f)_m\; \#(q_{f'};\wti{ir}\ne\wti{kt})_m} = 
	(m_g+2)(m_g+1)\,,
\\[2mm]
&
\frac{\#(q_f)_{m+2}\; \#(g;k\ne t)_{m+2}\; \#(g;r\ne k,t)_{m+2}\; \#(g;i\ne k,t,r)_{m+2}}
{\#(q_f)_m\; \#(g;\wti{ir}\ne\wti{kt})_m} = 
	(m_g+2)(m_g+1)\,,
\\[2mm]
&
\frac{\#(g)_{m+2}\; \#(g;k\ne t)_{m+2}\; \#(g;r\ne k,t)_{m+2}\; \#(g;i\ne k,t,r)_{m+2}}
{\#(g)_m\; \#(g;\wti{ir}\ne\wti{kt})_m} = 
	(m_g+2)(m_g+1)\,,
\\[2mm]
&
\frac{\#(q_f)_{m+2}\; \#(g;k\ne t)_{m+2}\; \#(q_{f'};r\ne k,t)_{m+2}\; \#(\qb_{f'};i\ne k,t,r)_{m+2}}
{\#(q_f)_m\; \#(g;\wti{ir}\ne\wti{kt})_m} = 
	(m_{f'}+1)(m_{\fb'}+1)\,,
\\[2mm]
&
\frac{\#(g)_{m+2}\; \#(g;k\ne t)_{m+2}\; \#(q_{f'};r\ne k,t)_{m+2}\; \#(\qb_{f'};i\ne k,t,r)_{m+2}}
{\#(g)_m\; \#(g;\wti{ir}\ne\wti{kt})_m} = 
	(m_{f'}+1)(m_{\fb'}+1)\,.
\esp
\label{eq:Xktir-counts}
\eeq
In case 3 we must remember that the counting is slightly different for the `equal flavour' 
and `unequal flavour' contributions even when the counterterms are the same. We have
\beq
\bsp
&
\frac{\#(q_f)_{m+2}\; \#(\qb_f;k\ne t)_{m+2}\; \#(q_{f};r\ne k,t)_{m+2}\; \#(\qb_{f};i\ne k,t,r)_{m+2}}
{\#(g)_m\; \#(g;\wti{ir}\ne\wti{kt})_m} =
\\[2mm]&\qquad=
	\frac{(m_f+2)(m_f+1)(m_\fb+2)(m_\fb+1)}{m_g(m_g-1)}\,,\qquad\qquad \mbox{e.f.}
\\[2mm]
&
\frac{\#(q_f)_{m+2}\; \#(\qb_f;k\ne t)_{m+2}\; \#(q_{f'};r\ne k,t)_{m+2}\; \#(\qb_{f'};i\ne k,t,r)_{m+2}}
{\#(g)_m\; \#(g;\wti{ir}\ne\wti{kt})_m} =
\\[2mm]&\qquad=
	\frac{(m_f+1)(m_\fb+1)(m_{f'}+1)(m_{\fb'}+1)}{m_g(m_g-1)}\,,\qquad\qquad \mbox{u.f.}
\esp
\eeq
By exploiting the symmetry of this factor under permutations of indices, we trivially 
obtain the rest of the cases as well. Finally, using \eqnss{eq:StoS-gg}{eq:StoS-qqqq}, 
we have
\beq
\bsp
&
\sum_{\br{m+2}} \frac{1}{\Sym{m+2}}
\sum_{t} \sum_{k\ne t} \sum_{r\ne k,t} \sum_{i\ne k,t,r} [X^{(0)}_{ktir}]_{f_k f_t f_i f_r}^{(\ldots)} = 
\\[2mm] &=
	\sum_{\br{m}} \frac{1}{\Sym{m}} \sum_{\wti{kt}} \sum_{\wti{ir}\ne\wti{kt}} \bigg\{
	\bigg[ [X^{(0)}_{ktir}]_{qgq'g}^{(\ldots)}
		+[X^{(0)}_{ktir}]_{qggq'}^{(\ldots)}
		+[X^{(0)}_{ktir}]_{gqq'g}^{(\ldots)}
		+[X^{(0)}_{ktir}]_{gqgq'}^{(\ldots)}\bigg] \delf{f_{kt}}{q} \delf{f_{ir}}{q}
\\[2mm] &\qquad\qquad\qquad\qquad\qquad
		+(q\leftrightarrow\qb)
		+(q'\leftrightarrow\qb')
		+(q\leftrightarrow\qb\, ,\, q'\leftrightarrow\qb')
\\[2mm] &\qquad\qquad\qquad\qquad\qquad
		+\bigg[ [X^{(0)}_{ktir}]_{qggg}^{(\ldots)}		
			+[X^{(0)}_{ktir}]_{gqgg}^{(\ldots)}		
			+\Nf\Big([X^{(0)}_{ktir}]_{qgq'\qb'}^{(\ldots)}	
				+[X^{(0)}_{ktir}]_{qg\qb' q'}^{(\ldots)}	
\\[2mm] &\qquad\qquad\qquad\qquad\qquad\qquad				
				+[X^{(0)}_{ktir}]_{gqq'\qb'}^{(\ldots)}	
				+[X^{(0)}_{ktir}]_{gq\qb' q'}^{(\ldots)}\Big)\bigg] \delf{f_{kt}}{q} \delf{f_{ir}}{g}
		+(q\leftrightarrow\qb)
\\[2mm] &\qquad\qquad\qquad\qquad\qquad
		+\bigg[ [X^{(0)}_{ktir}]_{ggqg}^{(\ldots)}		
			+[X^{(0)}_{ktir}]_{gggq}^{(\ldots)}		
			+\Nf\Big([X^{(0)}_{ktir}]_{q'\qb'qg}^{(\ldots)}	
				+[X^{(0)}_{ktir}]_{\qb' q'qg}^{(\ldots)}	
\\[2mm] &\qquad\qquad\qquad\qquad\qquad\qquad				
				+[X^{(0)}_{ktir}]_{q'\qb' gq}^{(\ldots)}	
				+[X^{(0)}_{ktir}]_{\qb' q' gq}^{(\ldots)}\Big)\bigg] \delf{f_{kt}}{g} \delf{f_{ir}}{q}
		+(q\leftrightarrow\qb)
\\[2mm] &\qquad\qquad\qquad\qquad\qquad
		+\bigg[ [X^{(0)}_{ktir}]_{gggg}^{(\ldots)}
		+\Nf\Big([X^{(0)}_{ktir}]_{ggq\qb}^{(\ldots)}	
				+[X^{(0)}_{ktir}]_{gg\qb q}^{(\ldots)}	
				+[X^{(0)}_{ktir}]_{q\qb gg}^{(\ldots)}	
\\[2mm] &\qquad\qquad\qquad\qquad\qquad\qquad				
				+[X^{(0)}_{ktir}]_{\qb q gg}^{(\ldots)}\Big)
		+\Nf^2\Big([X^{(0)}_{ktir}]_{q\qb q'\qb'}^{(\ldots)}
				+[X^{(0)}_{ktir}]_{q\qb \qb' q'}^{(\ldots)}
				+[X^{(0)}_{ktir}]_{\qb q q'\qb'}^{(\ldots)}
\\[2mm] &\qquad\qquad\qquad\qquad\qquad\qquad				
				+[X^{(0)}_{ktir}]_{\qb q \qb' q'}^{(\ldots)}\Big)\bigg] \delf{f_{kt}}{g} \delf{f_{ir}}{g}
	\bigg\}\,.
\esp
\label{eq:Xktir-FS-fin}
\eeq
We remind the reader that \eqn{eq:Xktir-FS-fin} was derived by using
that the counterterms are independent of specific quark flavours (as the notation 
implies), and further that the `equal flavour' and `unequal flavour' subtraction terms 
are equal. Then the sums $\sum_{f}$ and $\sum_{f'}$ in \eqn{eq:Xktir-FS1} may be 
performed, and we obtain the factors of $\Nf$ and $\Nf^2$ as shown. 

Finally, we define the flavour summed counterterms as 
\beq
\sum_{\br{m+2}} \frac{1}{\Sym{m+2}}
\sum_{t} \sum_{k\ne t} \sum_{r\ne k,t} \sum_{i\ne k,t,r} [X^{(0)}_{ktir}]_{f_k f_t f_i f_r}^{(\ldots)} = 
	\sum_{\br{m}} \frac{1}{\Sym{m}} \sum_{\wti{kt}} \sum_{\wti{ir}\ne\wti{kt}}
	\lbb X^{(0)}_{ktir} \rbb_{f_{kt} f_{ir}}^{(\ldots)}\,.
\label{eq:Xktir-FSICT1}
\eeq
We obtain
\beq
\bsp
\lbb X^{(0)}_{ktir} \rbb_{qq}^{(\ldots)} &=
	[X^{(0)}_{ktir}]_{qgq'g}^{(\ldots)}
	+[X^{(0)}_{ktir}]_{qggq'}^{(\ldots)}
	+[X^{(0)}_{ktir}]_{gqq'g}^{(\ldots)}
	+[X^{(0)}_{ktir}]_{gqgq'}^{(\ldots)}
\\[2mm]
\lbb X^{(0)}_{ktir} \rbb_{qg}^{(\ldots)} &=
	[X^{(0)}_{ktir}]_{qggg}^{(\ldots)}		
			+[X^{(0)}_{ktir}]_{gqgg}^{(\ldots)}
\\[2mm]&
			+\Nf\Big([X^{(0)}_{ktir}]_{qgq'\qb'}^{(\ldots)}	
				+[X^{(0)}_{ktir}]_{qg\qb' q'}^{(\ldots)}	
				+[X^{(0)}_{ktir}]_{gqq'\qb'}^{(\ldots)}	
				+[X^{(0)}_{ktir}]_{gq\qb' q'}^{(\ldots)}\Big)
\\[2mm]
\lbb X^{(0)}_{ktir} \rbb_{gq}^{(\ldots)} &=
	[X^{(0)}_{ktir}]_{ggqg}^{(\ldots)}		
			+[X^{(0)}_{ktir}]_{gggq}^{(\ldots)}		
\\[2mm]&	
			+\Nf\Big([X^{(0)}_{ktir}]_{q'\qb'qg}^{(\ldots)}	
				+[X^{(0)}_{ktir}]_{\qb' q'qg}^{(\ldots)}	
				+[X^{(0)}_{ktir}]_{q'\qb' gq}^{(\ldots)}	
				+[X^{(0)}_{ktir}]_{\qb' q' gq}^{(\ldots)}\Big)
\\[2mm]
\lbb X^{(0)}_{ktir} \rbb_{gg}^{(\ldots)} &=
	[X^{(0)}_{ktir}]_{gggg}^{(\ldots)}
		+\Nf\Big([X^{(0)}_{ktir}]_{ggq\qb}^{(\ldots)}	
				+[X^{(0)}_{ktir}]_{gg\qb q}^{(\ldots)}	
				+[X^{(0)}_{ktir}]_{q\qb gg}^{(\ldots)}	
				+[X^{(0)}_{ktir}]_{\qb q gg}^{(\ldots)}\Big)
\\[2mm]&
		+\Nf^2\Big([X^{(0)}_{ktir}]_{q\qb q'\qb'}^{(\ldots)}
				+[X^{(0)}_{ktir}]_{q\qb \qb' q'}^{(\ldots)}
				+[X^{(0)}_{ktir}]_{\qb q q'\qb'}^{(\ldots)}
				+[X^{(0)}_{ktir}]_{\qb q \qb' q'}^{(\ldots)}\Big)
\esp
\label{eq:Xktir-FSICT2}
\eeq
%

%
%

\subsection{Computing the flavour summed integrated counterterms}
\label{sec:specific-flavour-sum}

Using eqs.~(\ref{eq:Xkt-FSICT1}), (\ref{eq:Xkt-FSICT2}), (\ref{eq:Xktr-FSICT1}), 
(\ref{eq:Xktr-FSICT2}), (\ref{eq:Xktir-FSICT1}) and (\ref{eq:Xktir-FSICT2}), it is 
straightforward to compute all flavour summed integrated counterterms as 
presented in \sect{sec:flavsummedICTs}, after taking account of the following points.
\begin{itemize}
\item In \eqn{eq:I2dsigRRA12Jm-1}, most counterterms appear with some explicit 
overall factor, which must be included in the final result. E.g.\ for the collinear-triple 
collinear counterterm, this factor is $1/2$.
\item In certain cases, the ordering of some flavour indices may be meaningless, 
due to a symmetry of the integrated counterterms. E.g.\ in the collinear-triple collinear 
case, the integrated counterterm is symmetric in the first two indices, 
$[\IcC{kt}{}\IcC{ktr}{(0)}]_{qgq} = [\IcC{kt}{}\IcC{ktr}{(0)}]_{qgg}$ and so on. 
Hence, some terms that appear separately on the right hand sides of 
\eqns{eq:Xktr-FSICT2}{eq:Xktir-FSICT2} may be equal. See also the second line of 
\eqn{eq:Xkt-FSICT2}, where the appropriate symmetry is already taken into account.
\item In particular cases, some terms that appear on the right hand sides of 
eqs.~(\ref{eq:Xkt-FSICT2}), (\ref{eq:Xktr-FSICT2}) and (\ref{eq:Xktir-FSICT2}) may 
be zero for certain flavour assignments. E.g.\ in the collinear-triple collinear case, 
the first two indices must correspond to a true singly-collinear limit, hence 
$[\IcC{kt}{}\IcC{ktr}{(0)}]_{qq\qb} = 0$ and so on.
\item The `equal flavour' and `unequal flavour' counterterms in \eqn{eq:Xktr-FSICT2}
may actually be equal as e.g.\ in the collinear-triple collinear case, where
$ [\IcC{kt}{}\IcC{ktr}{(0)}]_{qq\qb} = [\IcC{kt}{}\IcC{ktr}{(0)}]_{qq' \qb'}$. Recall that 
this is already taken into account in \eqn{eq:Xktir-FSICT2}.
\end{itemize}
With these in mind, we easily find e.g.
\beq
\bsp
\Big(\IcC{kt}{}\IcC{ktr}{(0)}\Big)_q &= 
  [\IcC{kt}{}\IcC{ktr}{(0)}]_{qgg}
+ \frac12 [\IcC{kt}{}\IcC{ktr}{(0)}]_{ggq} 
+ \Nf [\IcC{kt}{}\IcC{ktr}{(0)}]_{q'\qb' q}
\label{eq:C-TC-again}
\,,
\\[2mm]
\Big(\IcC{kt}{}\IcC{ktr}{(0)}\Big)_g &= 
\frac12 [\IcC{kt}{}\IcC{ktr}{(0)}]_{ggg}
+ \Nf [\IcC{kt}{}\IcC{ktr}{(0)}]_{q\qb g}
+ 2 \Nf [\IcC{kt}{}\IcC{ktr}{(0)}]_{g q\qb}
\,,
\esp
\eeq
for the collinear-triple collinear flavour summed counterterms. The rest of the 
results in \sect{sec:flavsummedICTs} are obtained similarly.




\section{Modified doubly-real subtraction terms}
\label{sec:modsubterms}

We outline a simple modification to the NNLO subtraction scheme
presented in \refrs{Somogyi:2006da,Somogyi:2006db}.  Parts of these
modifications were presented previously: those relevant to the 
singly-unresolved approximate cross section $\dsiga{RR}{1}_{m+2}$ 
appearing in \eqn{eq:sigmaNNLOm+2}, and to the approximate cross 
sections in \eqn{eq:sigmaNNLOm+1}, were presented in \refr{Somogyi:2008fc}.  
In this appendix we describe the modification of the iterated 
singly-unresolved approximate cross section $\dsiga{RR}{12}_{m+2}$, 
which appears in \eqn{eq:sigmaNNLOm+2}.

Recall that the iterated singly-unresolved approximate cross section can 
be written symbolically as
\beq
\dsiga{RR}{12}_{m+2} = \PS{m}[\rd p_2]\bcA{12}\M{m+2}{(0)}\,,
\label{eq:dsigRRA12appx}
\eeq
where the iterated singly-unresolved approximation $\bcA{12}\M{m+2}{(0)}$ 
is a sum of a number of different collinear-, soft-, and soft-collinear-type 
terms (see \eqnss{eq:dsigRRA12Jm}{eq:CktStA2}). The precise definition of 
these terms involves the introduction of two momentum mappings
\beq
\mom{}_{n+1} \cmap{ir} \momb{(ir)}_{n}\,,
\qquad\mbox{and}\qquad
\mom{}_{n+1} \smap{r} \momb{(r)}_{n}\,,
\eeq
which are iterated in various combinations to produce appropriate mappings 
of $m+2\to m$ momenta. As discussed in \sect{sec:psfact}, all such mappings lead 
to an exact factorisation of the $m+2$ particle phase space, symbolically 
written as
\beq
\PS{m+2}(\mom{};Q) = \PS{m}(\momt{}_{m};Q)[\rd p_{1,m}][\rd p_{1,m+1}]\,.
\eeq
The exact form of the factorized phase spaces $[\rd p_{1,n}]$ ($n=m,m+1$) is 
given in \eqns{eq:dpir_n}{eq:dpr_n}, but their only feature which is relevant 
presently is that they carry a dependence on the number of partons, $n$, 
of the form
\bal
[\rd p^{(ir)}_{1,n}] &\propto (1-\al_{ir})^{2(n-1)(1-\ep)-1}\,,
\\[2mm]
[\rd p^{(r)}_{1,n}] &\propto (1-y_{rQ})^{(n-1)(1-\ep)-1}\,.
\eal
The subtraction terms, as originally defined in \refr{Somogyi:2006da} do 
not depend on the number of hard partons, thus the $m$-dependence of 
the factorized phase space measures is carried over to the integrated 
counterterms, where furthermore this dependence enters in a rather 
cumbersome way (see e.g.\ eqs.~(A.9) and (A.10) of \refr{Somogyi:2006cz}).

Thus, as in \refr{Somogyi:2008fc}, we reshuffle the $m$-dependence of the 
integrated counterterms into the subtraction terms themselves, where it 
appears in a very straightforward and harmless way, through factors of 
$(1-\al)$ and/or $(1-y)$ raised to $m$-dependent powers.
For easier reference, we gather the results in \tab{tab:modIT}, where together 
with the subtraction terms, we give the momentum mappings used to define 
the term(s) and the function which multiplies the original counterterm to produce 
the modified one. The $f$ functions appearing in \tab{tab:modIT} are defined as
\beq
f(z_0,z,p) = \Theta(z_0 - z)(1-z)^{-p}\,.
\label{eq:f_def}
\eeq
\begin{table}[t]
\renewcommand\arraystretch{1.5} 
\begin{center}
\footnotesize
\begin{tabular}{|c|c|c|}
\hline\hline
\multicolumn{3}{|c|}{Iterated collinear counterterms} \\
\hline\hline
Subtraction term & Momentum mapping & Function \\
\hline
\multirow{2}{4.5cm}{\centering
$\cC{kt}{}\cC{ktr}{(0,0)}$}
& 
\multirow{2}{5.3cm}{\centering
$\mom{} \cmap{kt} \momh{(kt)} \cmap{\wha{kt}\ha{r}} 
\momt{(\wha{kt}\ha{r},kt)}$}
&
$f(\al_0,\al_{kt},d(m,\ep))$ \\
& & $\times f(\al_0,\al_{\wha{kt}\ha{r}},d(m,\ep))$\\
\hline
\multirow{2}{4.5cm}{\centering
$\cC{kt}{}\cC{ir;kt}{(0,0)}$}
& 
\multirow{2}{5.3cm}{\centering
$\mom{} \cmap{kt} \momh{(kt)} \cmap{\ha{i}\ha{r}} 
\momt{(\ha{i}\ha{r},kt)}$}
&
$f(\al_0,\al_{kt},d(m,\ep))$ \\
& & $\times f(\al_0,\al_{\ha{i}\ha{r}},d(m,\ep))$\\
\hline
$\cC{kt}{}\cSCS{kt;r}{(0,0)}$, $\cC{kt}{}\cC{ir;kt}{}\cSCS{kt;r}{(0,0)}$,
&
\multirow{2}{5.3cm}{\centering
$\mom{} \cmap{kt} \momh{(kt)} \smap{\ha{r}} 
\momt{(\ha{r},kt)}$}
&
\multirow{2}{4cm}{\centering
$f(\al_0,\al_{kt},d(m,\ep))$ \\[2mm]
$\times f(y_0,y_{\ha{r}Q},d'(m,\ep))$}
\\
$\cC{kt}{}\cC{ktr}{}\cSCS{kt;r}{(0,0)}$ & & \\
\hline
\multirow{2}{4.5cm}{\centering
$\cC{kt}{}\cS{kt}{(0,0)}$, $\cC{kt}{}\cC{rkt}{}\cS{kt}{(0,0)}$}
&
\multirow{2}{5.3cm}{\centering
$\mom{} \cmap{kt} \momh{(kt)} \smap{\wha{kt}} \momt{(\wha{kt},kt)}$}
&
$f(\al_0,\al_{kt},d(m,\ep))$ \\
& & $\times f(y_0,y_{\wha{kt}Q},d'(m,\ep))$\\
\hline\hline
\multicolumn{3}{|c|}{Iterated soft counterterms} \\
\hline\hline
Subtraction term & Momentum mapping & Function \\
\hline
$\cS{t}{}\cC{irt}{(0,0)}$, $\cS{t}{}\cSCS{ir;t}{(0,0)}$,
&
\multirow{2}{5.3cm}{\centering
$\mom{} \smap{t} \momh{(t)} \cmap{\ha{i}\ha{r}} 
\momt{(\ha{i}\ha{r},t)}$}
&
\multirow{2}{4cm}{\centering
$f(y_0,y_{tQ},d'(m,\ep))$ \\[2mm]
$\times f(\al_0,\al_{\ha{i}\ha{r}},d(m,\ep))$}
\\
$\cS{t}{}\cC{irt}{}\cSCS{ir;t}{(0,0)}$ & & \\
\hline
$\cS{t}{}\cC{irt}{}\cS{rt}{(0,0)}$, $\cS{t}{}\cSCS{ir;t}{}\cS{rt}{(0,0)}$,
&
\multirow{2}{5.3cm}{\centering
$\mom{} \smap{t} \momh{(t)} \smap{\ha{r}} 
\momt{(\ha{r},t)}$}
&
\multirow{2}{4cm}{\centering
$f(y_0,y_{tQ},d'(m,\ep))$ \\[2mm]
$\times f(y_0,y_{\ha{r}Q},d'(m,\ep))$}
\\
$\cS{t}{}\cC{irt}{}\cSCS{ir;t}{}\cS{rt}{(0,0)}$,
$\cS{t}{}\cS{rt}{(0,0)}$ & & \\
\hline\hline
\multicolumn{3}{|c|}{Iterated soft-collinear counterterms} \\
\hline\hline
Subtraction term & Momentum mapping & Function \\
\hline
$\cC{it}{}\cS{t}{}\cC{irt}{(0,0)}$, $\cC{kt}{}\cS{t}{}\cSCS{ir;t}{(0,0)}$,
&
\multirow{2}{4.9cm}{\centering
$\mom{} \smap{t} \momh{(t)} \cmap{\ha{i}\ha{r}} 
\momt{(\ha{i}\ha{r},t)}$}
&
\multirow{2}{4cm}{\centering
$f(y_0,y_{kQ},d'(m,\ep))$ \\[2mm]
$\times f(\al_0,\al_{\ha{i}\ha{r}},d(m,\ep))$}
\\
$\cS{t}{}\cC{irt}{}\cSCS{ir;t}{(0,0)}$ & & \\
\hline
$\cC{rt}{}\cS{t}{}\cS{rt}{(0,0)}$ , 
$\cC{kt}{}\cS{t}{}\cC{krt}{}\cS{rt}{(0,0)}$,
&
\multirow{3}{4.9cm}{\centering
$\mom{} \smap{t} \momh{(t)} \smap{\ha{r}} 
\momt{(\ha{r},t)}$}
&
\multirow{3}{4cm}{\centering
$f(y_0,y_{tQ},d'(m,\ep))$ \\[2mm]
$\times f(y_0,y_{\ha{r}Q},d'(m,\ep))$}
\\
$\cC{rt}{}cS{t}{}\cC{krt}{}\cS{rt}{(0,0)}$,
$\cC{kt}{}\cS{t}{}\cS{rt}{(0,0)}$ & & \\
$\cC{kt}{}\cS{t}{}\cSCS{ir;t}{}\cS{rt}{(0,0)}$ & & \\
\hline\hline
\end{tabular}
\caption{\label{tab:modIT}
The modified iterated singly-unresolved subtraction terms 
are obtained from the original counterterms (first column) by multiplication 
with an appropriate function (last column). Also shown are the momentum
mappings used to define the subtraction terms (middle column). 
The $f(z_0,z,p)$ function is defined in \eqn{eq:f_def} while $d(m,\ep)$ 
and $d'(m,\ep)$ are defined in \eqn{eq:ddp}.
}
\end{center}
\end{table}

The pattern of modifications is hopefully clear: if the factorized phase space 
appropriate to a given subtraction term carries $m$-dependence through 
factors of $(1-\al)$ and/or $(1-y)$ respectively, it is multiplied by a factor/factors 
of $f(\al_0,\al,d(m,\ep))$ and/or $f(y_0,y,d'(m,\ep))$.
We emphasise that the form of the exponents $d(m,\ep)$ and $d'(m,\ep)$, 
is actually fixed by the prescription in  \refr{Somogyi:2008fc} (see eqs.~(3.2), (3.12) 
and (3.13) in particular) and the requirement that the modified subtraction terms 
should still correctly regularise all kinematic singularities. In fact, we must have
\beq
d(m,\ep) =  2m(1-\ep) - 2d_0\,,
\quad\mbox{and}\quad
d'(m,\ep) =  m(1-\ep) - d'_0\,,
\label{eq:ddp}
\eeq
where $d_0$ and $d'_0$ are the same constants which appear in 
eqs.~(3.2), (3.12) and (3.13) of \refr{Somogyi:2008fc}, i.e.
\beq
d_0 = D_0 + d_1\ep\,,
\quad\mbox{and}\quad
d'_0 = D'_0 + d'_1\ep\,,
\label{eq:d0s}
\eeq
where $D_0 ,D'_0 \ge 2$ are integers, while $d_1, d'_1$ are real. 
Also, the parameters $\alpha_0$ and $y_0$ must have the same 
values for all subtraction terms, including the singly-unresolved ones of 
\refr{Somogyi:2008fc}.

Finally, we note that the modifications introduced above do not spoil
any of the cancellations which take place among the original subtraction
terms, hence the modified counterterms are still a correct regulator of all 
kinematic singularities. This is not particularly hard to check explicitly, and
is actually a manifestation of the fact that the various momentum mappings 
obey several conditions in soft and/or collinear limits. As these were discussed 
in \refr{Somogyi:2006da}, we do not go into further details here.




\section{Basic collinear, soft and soft-collinear functions}
\label{sec:IJK}

Certain basic functions appear repeatedly during computations in this paper.
They all arise as integrals of various simple factors over factorized collinear 
or soft phase space measures.  Below we define and give explicit integral 
representations of these functions.
Some, notably $\cI$, $\cJ$ and $\cK$, have been considered previously
in \refr{Somogyi:2008fc}, albeit in somewhat more general forms. For 
completeness, we present these here as well, although only in the special 
cases used in this paper.

%
%

\subsection{Collinear functions}
\label{sec:I-type-fcns}

When computing the integral of the azimuthally averaged Altarelli--Parisi 
splitting functions over the factorized collinear phase space, the following 
function arises:
\beq
\bsp
\cI_1(y_{\wha{ir}Q},\ep,\al_0,d_0,k) &\equiv
\cI(y_{\wha{ir}Q},\ep,\al_0,d_0,0,k,0,1) =
\\[2mm] &= 
\frac{16\pi^2}{S_\ep}Q^{2\ep}
\int_1 [\rd p^{(ir)}_{1,m+1}] \frac{1}{s_{ir}} \tzz{r}{i}^k 
f(\al_0,\al_{ir},d(m,\ep))
\,,
\label{eq:cI-def}
\esp
\eeq
which, as indicated, is simply the $\cI$ function of \cite{Somogyi:2008fc}, 
for the special values of parameters $\kappa=\delta=0$ and $g_I=1$ (for their 
meaning see \refr{Somogyi:2008fc}).
The integral is over $[\rd p^{(ir)}_{1,m+1}]$, i.e.\ the factorized
measure obtained when going from $m+2$ to $m+1$ partons via the
collinear mapping of \refr{Somogyi:2006da}, which can explicitly be
written as
\beq
\bsp
[\rd p_{1,m+1}^{(ir)}(p_k,\ha{p}_{ir};Q)] &=
\frac{(Q^2)^{1-\ep}}{16\pi^2} S_\ep\,
(1-\al_{ir})^{2m(1-\ep)-1}\,y_{\wha{ir}Q}\,\rd\al_{ir}\,\rd v_{ir}
\\[2mm] &\times
 \al_{ir}^{-\ep}\,(\al_{ir}+(1-\al_{ir})y_{\wha{ir}Q})^{-\ep}\,
v_{ir}^{-\ep}\,(1-v_{ir})^{-\ep}
\\[2mm] &\times
\Theta(\al_{ir})\,\Theta(1-\al_{ir})\, \Theta(v_{ir})\,\Theta(1-v_{ir})
\,.
\label{eq:dpir_exp}
\esp
\eeq
The general collinear function was first computed analytically in
\refr{Aglietti:2008fe}.  For convenience, we present the integral
representation for the specific case, used in this paper:
\beq
\bsp
\cI_1(x;\ep,\al_0,d_0,k) &=
x \int_{0}^{\al_{0}} \rd{\al}\,
(1-\al)^{2d_0-1} \al^{-1-\ep} [\al+(1-\al) x]^{-1-\ep}
\\[2mm] &\qquad\times
\int_{0}^{1} \rd{v}\, v^{-\ep} (1-v)^{-\ep}
\left(\frac{\al+(1-\al) x v}{2\al+(1-\al) x}\right)^{\KK}
\,,
\label{eq:cI}
\esp
\eeq
where the factors
\beq
\al\,(\al+(1-\al)x)
\qquad \mbox{and} \qquad
{\al+(1-\al) x v \over 2\al+(1-\al) x}
\label{eq:z(al,v)}
\eeq
correspond to the collinear pole $y_{ir}$ and momentum fraction $\tzz{r}{i}$,
respectively (with $x \equiv x_{\wha{ir}}$, $\al \equiv \al_{ir}$ and
$v \equiv 1-v_{ir}$).

Among the iterated subtraction terms considered in this paper, we find two other 
basic integrals over the collinear phase space measure (\ref{eq:dpir_exp}). 
One of these is the Lorentz tensor
\beq
\cI^{\mu\nu}(\ha{p}_{ir}, Q;\ep,\al_0,d_0) = 
\frac{16\pi^2}{S_\ep}Q^{2\ep}
\int[\rd p^{(ir)}_{1,m+1}]
\frac{1}{s_{ir}}
\tzz{i}{r} \tzz{r}{i} \frac{4\kT{i,r}^\mu \kT{i,r}^\nu}{\kT{i,r}^2}
f(\al_0,\al_{ir},d(m,\ep))
\label{eq:cImunu-def}
\eeq
with kinematic dependence only on $\ha{p}_{ir}^\mu$ and $Q^\mu$.
The transverse momentum is defined to be orthogonal to both of these,
$\ha{p}_{ir}\ldot\kT{i,r} = Q\ldot\kT{i,r} = 0$ and contraction with
$g^{\mu\nu}$ replaces the fraction in the last factor with four. The
most general Lorentz structure that obeys these conditions is
\bal
\bsp
\cI^{\mu\nu} &= \frac{2}{1-\ep}
\left[g^{\mu\nu} 
	- {\ha{p}_{ir}^\mu Q^\nu + Q^\mu \ha{p}_{ir}^\nu \over \ha{p}_{ir}\ldot Q}
	+ {Q^2 \over (\ha{p}_{ir}\ldot Q)^2} \ha{p}_{ir}^\mu \ha{p}_{ir}^\nu
	\right]
\\[2mm]
&\times
\frac{16\pi^2}{S_\ep}Q^{2\ep}
\int [\rd p_{1,m+1}^{(ir)}(p_t,\ha{p}_{ir};Q)] 
{1\over s_{ir}}\tzz{r}{i}\tzz{i}{r}
f(\al_0,\al_{ir},d(m,\ep))
\,,
\label{eq:cImunu}
\esp
\eal
where the integral in the second line is clearly just
$[\cI_1(x_{\wha{ir}},\ep,\al_0,d_0,1)-\cI_1(x_{\wha{ir}},\ep,\al_0,d_0,2)]$.

The other one is
\beq
\cI_2(x_{\wha{ir}};\ep,\al_0,d_0) = 
\frac{16\pi^2}{S_\ep}Q^{2\ep}
\int[\rd p^{(ir)}_{1,m+1}]
\frac{1}{s_{ir}}
\frac{y_{ir}}{y_{iQ}^2}
f(\al_0,\al_{ir},d(m,\ep))
\,.
\label{eq:cI2-def}
\eeq
The definition of $\tzz{i}{r}$ \cite{Somogyi:2006da} implies
$y_{iQ} = \tzz{i}{r} (y_{iQ} + y_{rQ})$, and from the definition
of $\ha{p}_{ir}$ we have
$y_{iQ} + y_{rQ} = 2\al_{ir}+(1-\al_{ir})x_{\wha{ir}}$.
The momentum fraction can be expressed with the integration variables
as in \eqn{eq:z(al,v)}, therefore,
\beq
y_{iQ} = \al_{ir} + (1-\al_{ir})x_{\wha{ir}} v_{ir}\,,
\label{eq:yiQtoyirQ}
\eeq
hence
\beq
\bsp
\cI_2(x;\ep,\al_0,d_0) &=
x \int_{0}^{\al_{0}} \rd{\al}\,
(1-\al)^{2d_0-1} \al^{-\ep} [\al+(1-\al) x]^{-\ep}
\\[2mm] &\times
\int_{0}^{1} \rd{v}\, v^{-\ep} (1-v)^{-\ep}
[\al+(1-\al) x v]^{-2}
\,.
\label{eq:cI2}
\esp
\eeq
%

%
%

\subsection{Soft functions}
\label{sec:J-type-fcns}

When  integrating the eikonal factor $s_{ik}/(s_{ir}s_{kr})$ over the 
factorized soft phase space, we encounter the following function:
\beq
\bsp
\cJ_1(\Yh{i}{k};\ep,y_0,d'_0) &\equiv
\cJ(\Yh{i}{k};\ep,y_0,d'_0,0) =
\\[2mm]&=
-\frac{16\pi^2}{S_\ep}Q^{2\ep}
\int_1 [\rd p^{(r)}_{1,m+1}] \frac{s_{ik}}{s_{ir}s_{kr}}
f(y_0,y_{rQ},d'(m,\ep)) \,,
\label{eq:cJ-def}
\esp
\eeq
which, as shown, is simply the $\cJ$ function of \refr{Somogyi:2008fc}, for 
$\kappa=0$.
The integral is over $[\rd p^{(r)}_{1,m+1}]$, i.e.\ the factorized
measure obtained when going from $m+2$ to $m+1$ partons via the soft
mapping of \refr{Somogyi:2006da}. This measure can conveniently be
written using energy and angle variables in the centre of mass frame,
where 
\beq
\bsp
Q^\mu &= \sqrt{s}(1,\ldots)
\\[2mm]
p_r^\mu &= E_r(1,\mbox{`angles'},\sin\vth\sin\vphi\sin\eta,
               \sin\vth\sin\vphi\cos\eta,\sin\vth\cos\vphi,\cos\vth)\,,
\label{eq:softframe}
\esp
\eeq
where (here and below) the dots stand for vanishing components, while 
the notation 'angles' indicates the dependence of $p_r^\mu$ on
the $d-4$ angular variables that can be trivially integrated in all
relevant cases. In terms of the scaled energy-like variable $y_{rQ}$
and the angular variables $\vth$, $\vphi$ and $\eta$, the two-particle
phase space $\PS{2}(p_r,K;Q)$ is
\beq
\bsp
\PS{2}(p_r,K;Q) &=
     \frac{(Q^2)^{-\ep}}{16\pi^2} S_\ep (-\ep)\,4^\ep\,\rd y\,
     y^{1-2\ep} \delta(y - y_{\hr Q})
\\[2mm] &\times
     \rd(\cos\vth)\,\rd(\cos\vphi)\,\rd(\cos\eta)
     (\sin\vth)^{-2\ep} (\sin\vphi)^{-1-2\ep} (\sin\eta)^{-2-2\ep}\,.
\label{eq:dpr_exp}
\esp
\eeq
Often the integrand does not depend on all angles and we can integrate
out $\eta$,
\beq
\int_{-1}^1 \rd(\cos\eta) (\sin\eta)^{-2-2\ep} =
-\frac{2^{-2\ep}}{\ep}\,\frac{\Gamma^2(1-\ep)}{\Gamma(1-2\ep)}
\,,
\label{eq:Inteta}
\eeq
and $\vphi$,
\beq
{\Gamma^{2}(1-\ep) \over \pi \Gamma(1-2\ep)}
\int_{-1}^1 \rd(\cos\vphi) (\sin\vphi)^{-1-2\ep} =
2^{2\ep}
\,.
\label{eq:Intphi}
\eeq

The $\cJ_1$ function of \eqn{eq:cJ-def} was first computed analytically 
in \refr{Aglietti:2008fe}. We recall that it is conveniently evaluated in the 
frame \eqn{eq:softframe}, with the orientation fixed by
\beq
\ha{p}_i^\mu = E_{\ha{i}}(1,\ldots,1)\,,\qquad
\ha{p}_k^\mu = E_{\ha{k}}(1,\ldots,\sin\chi,\cos\chi)\,.
\label{eq:St-frame1}
\eeq
The precise definitions of $\ha{p}_i^\mu$ and $\ha{p}_k^\mu$ via the soft 
mapping \cite{Somogyi:2006da} imply
\beq
s_{ik} = (1-y_{rQ})s_{\ha{i}\ha{k}}\,,\qquad
s_{ir} = s_{\ha{i}r}\,,\qquad
s_{kr} = s_{\ha{k}r}\,,
\eeq
and expressing all two-particle invariants with integration variables, we find 
\cite{Aglietti:2008fe}
\beq
\cJ_1(Y;\ep,y_0,d'_0) =
-4Y {\Gamma^{2}(1-\ep) \over 2\pi\Gamma(1-2\ep)}
\Omega_{11}(\cos\chi(Y),1,1)
\int_{0}^{y_0} \rd{y}\, y^{-1-2\ep} (1-y)^{d'_0}
\,.
\label{eq:cJ}
\eeq
Above $Y\equiv \Yh{i}{k}$, while 
\beq
\cos\chi(Y) = 1-2Y\,,
\eeq
and the function $\Omega_{jl}$ denotes the angular integral
\beq
\bsp
\Omega_{jl}(\cos\chi,\beta_1,\beta_2) &\equiv \int_{-1}^{1}
\rd(\cos\vth)\,\rd(\cos\vphi)\,(\sin\vth)^{-2\ep} (\sin\vphi)^{-1-2\ep}
\\[2mm] &\times
(1-\beta_1\cos\vth)^{-j}
[1-\beta_2(\sin\chi\sin\vth\cos\vphi-\cos\chi\cos\vth)]^{-l}\,.
\esp
\label{eq:Omegajl-def}
\eeq
Presently we need the special case $\beta_1=\beta_2=1$, which we call 
the `massless' angular integral. The result of this angular integration 
is well-known \cite{vanNeerven:1985xr} and is proportional to a hypergeometric function.
Finally, using some hypergeometric identities and a one-dimensional 
integral representation of the hypergeometric function, we derive the 
following integral representation for $\cJ_1$, to be used in this paper:
\beq
\cJ_1(Y;\ep,y_0,d'_0) =
-Y^{-\ep} {2\Gamma^{2}(1-\ep) \over \Gamma(1-2\ep)}
\int_{0}^{y_0} \rd{y}\, y^{-1-2\ep} (1-y)^{d'_0}
\int_{0}^{1} \rd{t}\, t^{-1-\ep} \Big[1-(1-Y)t\Big]^{\ep}
\,.
\label{eq:cJ-final}
\eeq

In some cases the eikonal factor involves three momenta as in
\beq
\frac12 \calS_{(ik)l}(r) = \frac{s_{il} + s_{kl}}{(s_{ir} + s_{kr})s_{lr}}\,.
\eeq
Then the soft integral
\beq
\bsp
&
\cJm(Y_{(\ha{i}\ha{k})\ha{l},Q},\be_{(\ha{i}\ha{k})};\ep,y_0,d_0') = 
\\[2mm]&\qquad=
-\frac{16\pi^2}{S_\ep}Q^{2\ep} \int [\rd p^{(r)}_{1,m+1}(p_{r};Q)] 
\frac{s_{il} + s_{kl}}{(s_{ir} + s_{kr}) s_{lr}} f(y_0,y_{rQ},d'(m,\ep))
\label{eq:cJm-def}
\esp
\eeq
cannot be expressed with the soft function $\cJ$ any longer. In
\eqn{eq:cJm-def} $\be_{(\ha{i}\ha{k})}$ is the velocity of the momentum
$\ha{p}_i^\mu+\ha{p}_k^\mu$ in the centre of mass frame.
We evaluate $\cJm$ in the frame (\ref{eq:softframe}), with orientation
specified by
\beq
\ha{p}_i^\mu + \ha{p}_k^\mu = E_{(\ha{i}\ha{k})} (1,\ldots,\be_{(\ha{i}\ha{k})})\,,\qquad
\ha{p}_l^\mu = E_{\ha{l}} (1,\ldots,\sin\chi,\cos\chi)\,.
\label{eq:StCSirt-frame1}
\eeq
Using the precise definition of $\ha{p}_{j}^\mu$ ($j = i$, $k$, $l$), we find
\bal
s_{il} + s_{kl} &= (1-y_{rQ}) (s_{\ha{i}\ha{l}} + s_{\ha{k}\ha{l}})\,,
&
s_{ir} + s_{kr} &= s_{\ha{i}r} + s_{\ha{k}r}\,,
&
s_{lr} &= s_{\ha{l}r}\,.
\label{eq:StCSirt-inv}
\eal
Then we express all relevant two-particle invariants with integration
variables and obtain 
\beq
\cJm(Y,\be;\ep,y_0,d_0') = 
-4Y \frac{\Gamma^{2}(1-\ep)}{2\pi\Gamma(1-2\ep)}
\Omega_{11}(\cos\chi(Y,\be),\be,1)
\int_0^{y_0} \rd{y}\,  y^{-1-2\ep} (1- y)^{d'_0}
\,,
\label{eq:cJm}
\eeq
where we used the soft phase space \eqn{eq:dpr_exp} with $\eta$
integrated out (see \eqn{eq:Inteta}). In \eqn{eq:cJm} 
$Y\equiv Y_{(\ha{i}\ha{k})\ha{l},Q}$, $\be\equiv \be_{(\ha{i}\ha{k})}$, while
\beq
\cos\chi(Y,\be) = \frac{1 - 2Y}{\be}
\label{eq:coschi}
\eeq
and the function $\Omega_{jl}$ is defined in \eqn{eq:Omegajl-def} above.
Now we need the special case $\beta_1\equiv\beta<1$ and $\beta_2=1$, 
which we call the `one-mass' angular integral. 
The evaluation of this integral will be discussed elsewhere, and here 
we simply indicate that the result is proportional to an Appell function of the 
first kind. Finally, using  a one-dimensional integral representation of the Appell 
function, we obtain an integral representation for $\cJm$, similar 
to \eqn{eq:cJ}:
\bal
\bsp
\cJm(Y,\be;\ep,y_0,d_0') &=
-(2Y)^{-2\ep}{2\Gamma^{2}(1-\ep) \over \Gamma(1-2\ep)}
\int_{0}^{y_0} \rd{y}\, y^{-1-2\ep} (1-y)^{d'_0}
\\[2mm] &\times
\int_{0}^{1} \rd{t}\,t^{-1-2\ep}\,
\Big[1+\be-(1+\be-2Y)\,t\Big]^{\ep}\,\Big[1-\be-(1-\be-2Y)\,t\Big]^{\ep}\,.
\label{eq:cJm-final}
\esp
\eal
Setting $\be=1$, we see that $\cJm(Y,1;\ep,y_0,d_0') = \cJ_1(Y;\ep,y_0,d_0')$, as expected.

%
%

\subsection{Soft-collinear functions}
\label{sec:K-type-fcns}

The following function arises when integrating the collinear limit of the
eikonal factor, $2\tzz{i}{r}/(s_{ir}\tzz{r}{i})$, over the factorized
soft phase space (recall that $\tzz{i}{r}/\tzz{r}{i} = s_{iQ}/s_{rQ}$):
\beq
\cK_1(\ep,y_0,d'_0) \equiv
\cK(\ep,y_0,d'_0,0) =
\frac{16\pi^2}{S_\ep}Q^{2\ep}
\int_1 [\rd p^{(r)}_{1,m+1}]
\frac{2}{s_{ir}}\frac{\tzz{i}{r}}{\tzz{r}{i}}
f(y_0,y_{tQ},d'(m,\ep))\,.
\label{eq:cK-def}
\eeq
As indicated, this is just the $\cK$ function that was defined and computed in 
\refr{Somogyi:2008fc}, for $\kappa=0$.
The integral is over $[\rd p^{(r)}_{1,m+1}]$, i.e.\ the factorized
measure obtained when going from $m+2$ to $m+1$ partons via the
soft mapping. An integral representation for $\cK$ is easily derived 
in the frame of \eqns{eq:softframe}{eq:St-frame1} using
\beq
s_{ir} = s_{\ha{i}r}\,,\qquad
s_{iQ} = (1-y_{rQ})s_{\ha{i}Q} + s_{\ha{i}r}\,,
\eeq
which follow from the precise definition of $\ha{p}_i^\mu$ via the soft 
mapping. We find
\beq
\cK_1(\ep,y_0,d_0') = 
2^{2\ep} \int_0^{y_0} \rd{y}\, y^{-2\ep} (1-y)^{d'_0-1}
\int_{-1}^1 \rd(\cos\vth)\, (\sin\vth)^{-2\ep}
\left[1+ \frac{2(1-y)}{y(1-\cos\vth)}\right]
\,,
\label{eq:cK}
\eeq
which can also be written as
\beq
\cK_1(\ep,y_0,d'_0) = 2
\int_{0}^{y_0} \rd{y}\, y^{-1-2\ep} (1-y)^{d'_0-1}
\int_{0}^{1}\rd z \, z^{-1-\ep} (1 - z)^{-\ep} (1 - y + y z)
\,,
\label{eq:cKz}
\eeq
where we made the substitution $\cos\vth \to 1-2z$. Notice that the soft-collinear 
function $\cK$ is independent of the kinematics. 

In some cases, the collinear limit of the eikonal factor involves three
momenta as in 
\beq
\frac{2}{s_{(ir)t}} \frac{1-\tzz{t}{ir}}{\tzz{t}{ir}}
= \frac{2}{s_{it} + s_{rt}} \frac{y_{iQ} + y_{rQ}}{y_{tQ}}
\,.
\eeq
Then the soft-collinear integral
\beq
\cKm(\be_{(\ha{i}\ha{r})};\ep,y_0,d_0') = 
\frac{16\pi^2}{S_\ep}Q^{2\ep} \int [\rd p^{(t)}_{1,m+1}(p_{t};Q)] 
\frac{2}{s_{(ir)t}} \frac{1-\tzz{t}{ir}}{\tzz{t}{ir}}
f(y_0,y_{tQ},d'(m,\ep))
\label{eq:cKm-def}
\eeq
cannot be expressed with the soft-collinear function $\cK$ any longer. In
\eqn{eq:cKm-def}, $\be_{(\ha{i}\ha{r})}$ is the velocity of the momentum 
$\ha{p}_{i}^\mu + \ha{p}_{r}^\mu$ in the centre of mass frame.
Using the definition of the mapped momenta \cite{Somogyi:2006da}, we have
\beq
s_{it} + s_{rt} = s_{\ha{i}t} + s_{\ha{r}t}\,,\qquad
s_{iQ} + s_{rQ} = (1-y_{tQ}) (s_{\ha{i}Q} + s_{\ha{r}Q})
+ s_{\ha{i}t} + s_{\ha{r}t}
\,.
\eeq
Then we evaluate $\cKm$ in the frame given in 
\eqns{eq:softframe}{eq:StCSirt-frame1} (with the 
trivial replacement $k\to r$). Expressing all two-particle invariants with 
integration variables, we find the following integral representation for 
the `one-mass' soft-collinear integral:
\beq
\cKm(\be;\ep,y_0,d_0') = 
2^{2\ep} \int_0^{y_0} \rd{y}\, y^{-2\ep} (1-y)^{d'_0-1}
\int_{-1}^1 \rd(\cos\vth)\, (\sin\vth)^{-2\ep}
\left[1+\frac{2(1-y)}{y(1-\be\cos\vth)}\right]\,,
\label{eq:cKm}
\eeq
where $\be \equiv \be_{(\ha{i}\ha{r})}$.
We make the substitution $\cos\vth \to 1-2z$ to obtain the 
final form of the integral representation used in this paper:
\beq
\bsp
\cKm(\be;\ep,y_0,d_0') &= 
2 \int_{0}^{y_0} \rd{y}\, y^{-1-2\ep} (1-y)^{d'_0-1}
     \int_{0}^{1} \rd z\, [z (1 - z)]^{-\ep}
     {2 - y (1+\be-2\be z) \over 1-\be+2\be z}
\label{eq:cKm-final}
\,.
\esp
\eeq
For $\be = 1$, we recover the soft-collinear integral, 
$\cKm(1;\ep,y_0,d_0') = \cK_1(\ep,y_0,d_0')$.




\section{Integrating the collinear-type counterterms}
\label{sec:IntsCktA2}

In this appendix, we discuss the integration of the collinear-type 
counterterms of \sect{sec:IntsCktCTs}. In particular, we define and give an 
explicit integral representation of all $\cI^{(i)}_{\cC{}{}}$
functions ($i=1,\ldots,9$).

%
%

\subsection{Treatment of azimuthal correlations}
\label{sec:Intspindep}

Collinear subtraction terms contain azimuthal correlations if the 
factorisation formula in the corresponding collinear limit involves  
gluon splitting kernels. For the two-parton splitting $\wti{ij} \to i + j$, these 
azimuthal correlations involve a transverse momentum $\kT{i,j}^\mu$ that 
we always chose to be orthogonal to the parent momentum $\wti{p}_{ij}^\mu$. 
This condition is sufficient to prove that the integral of the spin-dependent 
and spin-averaged splitting kernels over the factorized phase space of the 
unresolved parton are the same \cite{Catani:1996vz}.
Therefore, one can always substitute the spin-dependent splitting kernels 
$\hP_{f_i f_j}^{(0)}(\tzz{j}{i};\kT{i,j};\ep)$ with their spin-averaged counterparts,
$\big\la\hP_{f_i f_j}^{(0)}(\tzz{j}{i};\kT{i,j};\ep)\big\ra \equiv 
P_{f_i f_j}^{(0)}(\tzz{j}{i};\ep)$, as done in \sect{sec:IntsCTs}. 

In the case of strongly-ordered three-parton splittings, one cannot
directly use the same argument. The splitting kernels in the integral of
the collinear-triple collinear subtraction, $\cC{kt}{}\cC{ktr}{(0,0)}$
depend on the transverse momentum $\kT{k,t}^\mu$ in two ways. One is
when the Lorentz index of the transverse momentum coincides with that
of the parent gluon as in the explicit
$\kT{k,t}^{\mu}\kT{k,t}^{\nu}/\kT{k,t}^2$
terms in the gluon splitting kernels:
\bal
&
\la\mu |\hP_{g_r q_k \qb_t}^{{\rm s.o.\,(0)}}
(\tzz{k}{t},\tzz{t}{k},\kT{k,t},\tzz{\wha{kt}}{\ha{r}}
,\tzz{\ha{r}}{\wha{kt}},\kT{\wha{kt},\ha{r}};\ep)|\nu\ra =
\nt \\[2mm] &\qquad=     2\CA\TR\Bigg[
-g^{\mu\nu}\Bigg(\frac{\tzz{\ha{r}}{\wha{kt}}}{\tzz{\wha{kt}}{\ha{r}}}
+ \frac{\tzz{\wha{kt}}{\ha{r}}}{\tzz{\ha{r}}{\wha{kt}}}
+\tzz{k}{t}\tzz{t}{k}
\frac{s_{\ha{r}\kT{k,t}}^2}{\kT{k,t}^2 s_{\wha{kt}\ha{r}}}\Bigg)
+ 4\tzz{k}{t}\tzz{t}{k}\frac{\tzz{\wha{kt}}{\ha{r}}}{\tzz{\ha{r}}{\wha{kt}}}
\frac{\kT{k,t}^{\mu}\kT{k,t}^{\nu}}{\kT{k,t}^2}\Bigg]
\nt \\[2mm] &\qquad
-\,4\CA(1-\ep)\tzz{\ha{r}}{\wha{kt}}\tzz{\wha{kt}}{\ha{r}}
P_{q_k \qb_t}^{(0)}(\tzz{k}{t},\tzz{t}{k},\kT{k,t};\ep)
\frac{\kT{\ha{r},\wha{kt}}^{\mu}\kT{\ha{r},\wha{kt}}^{\nu}}
{\kT{\ha{r},\wha{kt}}^2}
\label{eq:PgrqkqbtSO}
\intertext{and}
&
\la\mu |\hP_{g_k g_t g_r}^{{\rm s.o.\,(0)}}
(\tzz{k}{t},\tzz{t}{k},\kT{k,t},\tzz{\wha{kt}}{\ha{r}}
,\tzz{\ha{r}}{\wha{kt}},\kT{\wha{kt},\ha{r}};\ep)|\nu\ra =
4\CA^2\Bigg[
- g^{\mu\nu}\Bigg(\frac{\tzz{\ha{r}}{\wha{kt}}}{\tzz{\wha{kt}}{\ha{r}}}
+ \frac{\tzz{\wha{kt}}{\ha{r}}}{\tzz{\ha{r}}{\wha{kt}}}\Bigg)
\nt \\[2mm] &\qquad\times\,
\Bigg(\frac{\tzz{k}{t}}{\tzz{t}{k}} +
\frac{\tzz{t}{k}}{\tzz{k}{t}}\Bigg)
+ g^{\mu\nu} \tzz{k}{t}\tzz{t}{k}\frac{1-\ep}{2}
  \frac{s_{\ha{r}\kT{k,t}}^2}{\kT{k,t}^2 s_{\wha{kt}\ha{r}}}
- 2(1-\ep)\tzz{k}{t}\tzz{t}{k}
\frac{\tzz{\wha{kt}}{\ha{r}}}{\tzz{\ha{r}}{\wha{kt}}}
\frac{\kT{k,t}^{\mu}\kT{k,t}^{\nu}}{\kT{k,t}^2}\Bigg]
\nt \\[2mm] &\qquad
-\,4\CA(1-\ep)\tzz{\ha{r}}{\wha{kt}}\tzz{\wha{kt}}{\ha{r}}
P_{g_k g_t}^{(0)}(\tzz{k}{t},\tzz{t}{k},\kT{k,t};\ep)
\frac{\kT{\ha{r},\wha{kt}}^{\mu}\kT{\ha{r},\wha{kt}}^{\nu}}
{\kT{\ha{r},\wha{kt}}^2}\,.
\label{eq:PgkgtgrSO}
\eal
This transverse momentum is not orthogonal to $\wti{p}_{ktr}^\mu$, as
defined originally in \refr{Somogyi:2006da}.  Nevertheless, when integrating 
these subtraction terms, we can still substitute the spin-dependent splitting 
kernels with the spin-averaged ones, as we now show.

Recall that the strongly-ordered three-parton splitting kernel appears in the 
collinear-triple collinear subtraction term in the form
\beq
\cC{kt}{}\cC{ktr}{(0,0)} \propto
\bra{m}{(0)}{(\momt{(\wha{kt}\ha{r},kt)}{})}
\hP_{f_k f_t f_r}^{{\rm s.o.\,(0)}}
\ket{m}{(0)}{(\momt{(\wha{kt}\ha{r},kt)}{})}\,,
\label{eq:CktCktr-structure}
\eeq
and the bra-ket expression above has the following precise meaning:
\beq
\bsp
&
\bra{m}{(0)}{} \hP_{f_k f_t f_r}^{{\rm s.o.\,(0)}} \ket{m}{(0)}{} =
\\[2mm]&\qquad
	\bra{m}{(0)}{(\momt{}{})}\mu\ra d_{\mu\mu'}(\wti{p}_{ktr},n_1)
	\la\mu'|\hP_{f_k f_t f_r}^{{\rm s.o.\,(0)}}|\nu'\ra
	d_{\nu'\nu}(\wti{p}_{ktr},n_2) \la\nu\ket{m}{(0)}{(\momt{}{})}\,,
\label{eq:braket}
\esp
\eeq
where $d_{\mu\mu'}(\wti{p}_{ktr},n_1)$ and $d_{\nu\nu'}(\wti{p}_{ktr},n_2)$ 
are gluon polarisation tensors  with (time-like) gauge vectors, $n_1$ and $n_2$. 
Hence, the object to integrate is not simply
$\la\mu|\hP_{f_k f_t f_r}^{{\rm s.o.\,(0)}}|\nu\ra$, but rather the contraction
\beq
d_{\mu\mu'}(\wti{p}_{ktr},n_1)
	\la\mu'|\hP_{f_k f_t f_r}^{{\rm s.o.\,(0)}}|\nu'\ra
d_{\nu'\nu}(\wti{p}_{ktr},n_2)\,,
\eeq
that is clearly orthogonal to $\wti{p}_{ktr}^\mu$ (because of the presence of
the polarisation tensors). Thus, by the usual arguments, the
azimuthal correlations present in $\cC{kt}{}\cC{ktr}{(0,0)}$ vanish after
integration over the phase space of the unresolved parton.
However, we must still be careful to compute the average over the 
polarisations correctly. When $k_{\perp}\ldot \wti{p}\ne 0$, we have
\beq
d_{\mu\mu'}(\wti{p},n_1)\frac{k_\perp^{\mu'} k_\perp^{\nu'}}{k_\perp^2}
d_{\nu'\nu}(\wti{p},n_2) = 
	\frac{1}{k_\perp^2}
	\left(\kT{\mu}-\frac{k_\perp\ldot \wti{p}}{\wti{p}\ldot n_1} n_{1\mu}\right)
	\left(\kT{\nu}-\frac{k_\perp\ldot \wti{p}}{\wti{p}\ldot n_2} n_{2\nu}\right)
	+\ldots\,,
\label{eq:dkkd}
\eeq
where the dots stand for terms proportional to $\wti{p}^\mu$ or $\wti{p}^\nu$ 
which vanish after contraction with the matrix element, by gauge invariance. 
Thus we find ($\wti{n}$ is a further time-like gauge vector)
\beq
\bsp
&
\left\la 
	d_{\mu\mu'}(\wti{p},n_1) \frac{k_\perp^{\mu'}k_\perp^{\nu'}}{k_\perp^2}
	d_{\nu'\nu}(\wti{p},n_2)
\right\ra \equiv
	\frac{1}{2(1-\ep)} d_{\mu\nu}(\wti{p},\wti{n}) 
	\frac{d_{\mu\mu'}(\wti{p},n_1)k_\perp^{\mu'}k_\perp^{\nu'}d_{\nu'\nu}(\wti{p},n_2)}
	{k_\perp^2}
\\[2mm] &\qquad=
	\frac{-1}{2(1-\ep)}\frac{1}{k_\perp^2}
	\left[k_\perp^2 
	- \frac{k_\perp\ldot \wti{p}}{\wti{p}\ldot n_1} k_\perp\ldot n_1
	- \frac{k_\perp\ldot \wti{p}}{\wti{p}\ldot n_2} k_\perp\ldot n_2
	+ \frac{(k_\perp\ldot \wti{p})^2}{(\wti{p}\ldot n_1)(\wti{p}\ldot n_2)} n_1\ldot n_2\right]\,.
\label{eq:spinaveraging}
\esp
\eeq
\Eqn{eq:spinaveraging} shows that the advantage of having $k_\perp\ldot
\wti{p}= 0$ actually lies in the trivial azimuthal averaging. However, this
can also be arranged if $k_\perp\ldot \wti{p}\ne 0$. For example, choosing gauge
vectors $n_1^\mu$ and $n_2^\mu$ such that $n_1^\mu+n_2^\mu\propto k_\perp^\mu$ 
(and of course $n_1^2=n_2^2=0$), we find that the above average just reduces 
to $-1/[2(1-\ep)]$, which is the usual result. 
Since $k_\perp^2<0$, such nonzero $n_1^\mu$ and $n_2^\mu$ always exist. 
Indeed, in any Lorentz frame we have the parametrisation 
$k_\perp^\mu = K(1,\beta\vec{v})$, 
where $\vec{v}^2=1$ and $\be^2\ne 1$. Then setting e.g.
\beq
n_1^\mu = \frac12 K(1+\beta)(1,\vec{v})\,,\qquad\mbox{and}\qquad
n_2^\mu = \frac12 K(1-\beta)(1,-\vec{v})\,,
\label{eq:gauge-vectors}
\eeq
we clearly have $n_1^\mu$ and $n_2^\mu$ nonzero, $n_1^2=n_2^2=0$, and 
$n_1^\mu+n_1^\mu=k_\perp^\mu$, as required. We make essentially such 
a choice in our numerical code.

A shorter, though less transparent proof is to observe that if we 
change this single troublesome term to
\beq
{\kT{k,t}^\mu \kT{k,t}^\nu \over \kT{k,t}^2} \to
	{\kTt{k,t}^\mu \kTt{k,t}^\nu \over \kTt{k,t}^2}\,,
\label{eq:kTktsub}
\eeq
where
\beq
\kTt{k,t}^\mu = \kT{k,t}^\mu 
	- {\kT{k,t}\ldot \wti{p}_{ktr} \over \wti{p}_{ktr}\ldot Q} Q^\mu\,,
\eeq
then obviously $\kTt{k,t}\ldot \wti{p}_{ktr}=0$, and all the usual arguments 
apply. What is not immediately obvious, is that this modification does not ruin 
any of the delicate cancellations between the various counterterms in any IR 
limit, and hence is allowed.
This can be established as follows: in the centre of mass frame of $Q$, 
the replacement of \eqn{eq:kTktsub} simply amounts to 
\beq
\kT{k,t} = -K(1,\be\vec{v}) \to \kTt{k,t} = -K(1+\Delta,\be\vec{v})\,,
\eeq
with some $\Delta$. Then, continuing to choose the gauge vectors as in 
\eqn{eq:gauge-vectors}, we have
\beq
\bsp
n_1^\mu &= \frac12 K(1+\beta)(1,\vec{v}) \to \ti{n}_1^\mu = \frac12 K(1+\Delta+\beta)(1,\vec{v})\,,
\\[2mm]
n_2^\mu &= \frac12 K(1-\beta)(1,-\vec{v})  \to \ti{n}_2^\mu = \frac12 K(1+\Delta-\beta)(1,\vec{v})\,,
\esp
\eeq
i.e.\ only the normalisation of the gauge vectors is changed by the replacement 
in \eqn{eq:kTktsub}. However, this implies that \eqn{eq:dkkd} is actually {\em unchanged}. 
Indeed, recalling that $k_\perp^\mu = n_1^\mu + n_2^\mu$, we find
\beq
\bsp
d_{\mu\mu'}(\wti{p},n_1)\frac{k_\perp^{\mu'} k_\perp^{\nu'}}{k_\perp^2}
d_{\nu'\nu}(\wti{p},n_2) &= 
	d_{\mu\mu'}(\wti{p},n_1)\frac{(n_1^{\mu'}+n_2^{\mu'}) (n_1^{\nu'}+n_2^{\nu'})}{(n_1+n_2)^2}
	d_{\nu'\nu}(\wti{p},n_2)
\\[2mm]&=
	d_{\mu\mu'}(\wti{p},n_1)\frac{n_2^{\mu'} n_1^{\nu'}}{2(n_1\ldot n_2)}
	d_{\nu'\nu}(\wti{p},n_2)\,,
\esp
\eeq
and the last expression is clearly seen to be invariant under (independent) 
rescalings of $n_1$ and $n_2$. Hence, the replacement in \eqn{eq:kTktsub} is 
completely harmless. (However, notice that the proof requires the specific 
choice of gauge vectors as in \eqn{eq:gauge-vectors}.)
 
The other occurrence of $\kT{k,t}$ in the strongly-ordered kernels is
in the ratio $\displaystyle \frac{s_{\ha{r}\kT{k,t}}^2}{\kT{k,t}^2}$, also present in the
quark splitting kernels.
Examining the explicit forms of the strongly-ordered splitting kernels,
we find that this ratio always appears in the form
\beq
\frac{\ha{p}_r^\mu \ha{p}_r^\nu}{s_{\wha{kt}\ha{r}}}
\cI^{\mu\nu}(\ha{p}_{kt}, Q)
\,,
\eeq
where the integral $\cI^{\mu\nu}$ is defined in \eqn{eq:cImunu-def} and
computed in \eqn{eq:cImunu}. Contracting the latter with
$\ha{p}_r^\mu \ha{p}_r^\nu/s_{\wha{kt}\ha{r}}$, we obtain
\beq
\frac{2}{1-\ep}\left( \frac{Q^2 s_{\wha{kt}\ha{r}}}{s_{\wha{kt}Q}^2}
-\frac{s_{\ha{r}Q}}{s_{\wha{kt}Q}} \right)
\frac{16\pi^2}{S_\ep}Q^{2\ep}
\int [\rd p^{(kt)}_{1,m+1}] \frac{1}{s_{kt}}\,\tzz{t}{k}\,\tzz{k}{t}
f(\al_0,\al_{ir},d(m,\ep))
\,.
\eeq
Observing that
\beq
\frac{s_{\ha{r}Q}}{s_{\wha{kt}Q}}=
\frac{\tzz{\ha{r}}{\wha{kt}}}{\tzz{\wha{kt}}{\ha{r}}}\,,
\eeq
we find that when integrating the strongly-ordered splitting kernels over
the factorized phase space, the integrals of 
\beq
\frac{1-\ep}{2}
\frac{s_{\ha{r}\kT{k,t}}^2}{\kT{k,t}^2 s_{\wha{kt}\ha{r}}}
\qquad {\rm and}\qquad
\frac{y_{\wha{kt}\ha{r}}}{y_{\wha{kt}Q}^2}
- \frac{\tzz{\ha{r}}{\wha{kt}}}{\tzz{\wha{kt}}{\ha{r}}}
\label{eq:replace}
\eeq
are equal, so we can substitute the former with the latter, which we
implement in the next subsection, where we give the spin-averaged
splitting kernels explicitly.
%

%
%

\subsection{Explicit forms of the spin-averaged splitting kernels}
\label{sec:APfcns}

For the sake of completeness we present the explicit expressions for the
spin-averaged splitting kernels used in the integrals of the
subtraction terms.

The azimuthally averaged Altarelli--Parisi splitting kernels are well known:
\bal
P^{(0)}_{gg}(z) &=
2\CA \left[\frac{1}{z}+\frac{1}{1-z}-2+z-z^2\right]\,,
\label{eq:P0gg}
\\[2mm]
P^{(0)}_{q\qb}(z;\ep) &=
\TR \left[1-\frac{2}{1-\ep}\left(z-z^2\right)\right]\,,
\label{eq:P0qq}
\\[2mm]
P^{(0)}_{qg}(z;\ep) &=
\CF \left[\frac{2}{z}-2 + (1-\ep)z\right]\,.
\label{eq:P0qg}
\eal
In our convention the ordering of the labels on the splitting-kernels
is usually meaningless, but in \eqn{eq:P0qg} $z$ refers to the momentum
fraction of the second label. In other words $P^{(0)}_{gq}(z;\ep) =
P^{(0)}_{qg}(1-z;\ep)$. The other two cases are symmetric with respect
to $z \leftrightarrow 1-z$.

In the strongly-ordered kernels $\hP_{f_k f_t f_r}^{{\rm s.o.\,(0)}}$
the ordering matters, too. As a result, the same triple-parton
splitting function may have different strongly-ordered limits, which
can be distinguished by the momentum labels in the kernel, once the
ordering of the limits is fixed by the momentum mapping,
$\wti{ktr} \to \wha{kt} + \ha{r} \to (k + t) + r$ in our convention. We
always choose $z = \tzz{t}{k}$ and $\ha{z} = \tzz{\ha{r}}{\wha{kt}}$ as
independent variables.  For quark splitting we have
\bal
P_{q g g}^{{\rm s.o.\,(0)}}(z,\ha{z};\ep) &=
P_{q g}^{(0)}(z;\ep) P_{q g}^{(0)}(\ha{z};\ep)
\,,
\label{eq:PqkgtgrSO_avg}
\\[2mm]
P_{g g q}^{{\rm s.o.\,(0)}}(z,\ha{z},R;\ep) &=
P_{g g}^{(0)}(z) P_{q g}^{(0)}(1-\ha{z};\ep)
- \CA\,\CF z(1-z) b^{(0)}_{gg} b^{(0)}_{qg} R
\,,
\label{eq:qrgkgtSO_avg}
\\[2mm]
P_{q' \qb' q}^{{\rm s.o.\,(0)}}(z,\ha{z},R;\ep) &=
P_{q \qb}^{(0)}(z;\ep) P_{q g}^{(0)}(1-\ha{z};\ep)
- \CA\,\CF z(1-z) b^{(0)}_{q\qb} b^{(0)}_{qg} R
\,,
\label{eq:PqrqbkqtSO_avg}
\eal
while for gluon splitting we find
\bal
P_{g g g}^{{\rm s.o.\,(0)}}(z,\ha{z},R) &=
P_{g g}^{(0)}(z) P_{g g}^{(0)}(\ha{z}) - \CA^2 z(1-z) (b^{(0)}_{gg})^2 R\,,
\label{eq:PgkgtgrSO_avg}
\\[2mm]
P_{q \qb g}^{{\rm s.o.\,(0)}}(z,\ha{z},R;\ep) &=
P_{q \qb}^{(0)}(z;\ep) P_{g g}^{(0)}(\ha{z})
- \CA^2 z(1-z) b^{(0)}_{q\qb} b^{(0)}_{gg} R\,,
\label{eq:PgrqkqbtSO_avg}
\\[2mm]
P_{g q \qb}^{{\rm s.o.\,(0)}}(z,\ha{z};\ep) &=
P_{q g}^{(0)}(1-z;\ep) P_{q \qb}^{(0)}(\ha{z};\ep)\,,
\label{eq:PgkqtqbrSO_avg}
\eal
where the constants $b^{(0)}_{f_1 f_2}$ are given in \eqn{eq:bkt}.
\Eqnss{eq:PqkgtgrSO_avg}{eq:PgkqtqbrSO_avg} can also be written in a
unified form,
\beq
P_{f_k f_t f_r}^{{\rm s.o.\,(0)}}(z,\ha{z},R;\ep) =
P_{f_k f_t}^{(0)}(z;\ep) P_{f_{kt} f_r}^{(0)}(\ha{z};\ep)
- \delta_{f_{kt} g}
C_{f_{kt}} C_{f_{ktr}} z(1-z) b^{(0)}_{f_k f_t} b^{(0)}_{f_{kt} f_r} R\,.
\label{eq:Psounified}
\eeq
We see that the second term is present only
if the three-parton splitting involves a two-parton sub-splitting
with parent gluon.

%
%

\subsection{Collinear-triple collinear counterterm}
\label{sec:IntsCktCktr}

The collinear-triple collinear counterterm involves two successive
collinear mappings, which leads to exact phase space factorisation
in the iterated form
\beq
\PS{m+2}{}(\mom{};Q) = \PS{m}{}(\momt{(\wha{kt}\ha{r},kt)}{};Q)
[\rd p_{1;m}^{(\wha{kt}\ha{r})}(\ha{p}_{r},\wti{p}_{\wha{kt}\ha{r}};Q)]
[\rd p_{1;m+1}^{(kt)}(p_k,\ha{p}_{kt};Q)]\,.
\label{eq:psfactCktCktr}
\eeq
The one-parton factorized phase spaces
$[\rd p_{1;m+1}^{(kt)}(p_k,\ha{p}_{kt};Q)]$ and
$[\rd p_{1;m}^{(\wha{kt}\ha{r})}(\ha{p}_{r},\wti{p}_{\wha{kt}\ha{r}};Q)]$
are given explicitly by \eqn{eq:dpir_exp} after appropriate changes in 
labelling, including the replacement $m\to m-1$ in the second case.
The Altarelli-Parisi splitting functions and the factor $z(1-z)$ in
\eqn{eq:Psounified} can be expressed as linear combinations of powers
of momentum fractions. Consequently, the integral over the factorized
phase space $[\rd p_{1,m+1}^{(kt)}]$ in the integrated collinear-triple
collinear counterterm is written in terms of collinear functions
$\cI_1(x_{\wha{kt}},\ep,\al_0,d_0;k)$ of \eqn{eq:cI-def}.
In order to compute the subsequent integrals over
$[\rd p^{(\wha{kt}\ha{r})}_{1,m}]$,
\bal
\bsp
\Big\{\cI^{(1)}_{\cC{}{}},\,
  \cI^{(2)}_{\cC{}{}},\,
  \cI^{(3)}_{\cC{}{}}\Big\} =
\frac{16\pi^2}{S_\ep}Q^{2\ep}
\int &[\rd p^{(\wha{kt}\ha{r})}_{1,m}]
\frac{1}{s_{\wha{kt}\ha{r}}}
\Bigg\{(1-\tzz{\ha{r}}{\wha{kt}})^l,\,\tzz{\ha{r}}{\wha{kt}}^l,\,
\frac{y_{\wha{kt}\ha{r}}}{y_{\wha{kt}Q}^2}\Bigg\}
\\[2mm] &\times
f(\al_0,\al_{\wha{kt}\ha{r}},d(m,\ep))\,
\cI_1(x_{\wha{kt}},\ep,\al_0,d_0;k)
\,,
\label{eq:IC2C3types}
\esp
\eal
the variable $x_{\wha{kt}} = y_{\wha{kt}Q}$ needs to be
expressed in terms of $\wti{p}_{ktr}^\mu$ instead of $\ha{p}_{kt}^\mu$ 
(see \eqn{eq:yiQtoyirQ} with proper changes in labelling),
\beq
x_{\wha{kt}} = y_{\wha{kt}Q} = \al_{\wha{kt}\ha{r}}
+ (1-\al_{\wha{kt}\ha{r}})x_{\wti{ktr}} v_{\wha{kt}\ha{r}}\,.
\eeq
Then using the abbreviations $\al=\al_{\wha{kt}\ha{r}}$,
$v=v_{\wha{kt}\ha{r}}$ (the integration variable corresponding to
$1-\tzz{\ha{r}}{\wha{kt}}$) and $x \equiv x_{\wti{ktr}}$, 
the integral representations (\ref{eq:cI}) and (\ref{eq:cI2}), we find that
the integrated collinear-triple collinear counterterm can be
expressed using the following three types of integrals:
\bal
\cI^{(1)}_{\cC{}{}}(x;\ep,\al_0,d_0;k,l) &=
	x \int_{0}^{\al_{0}} \rd{\al}
	(1-\al)^{2d_0-3+2\ep} \al^{-1-\ep}
	(\al+(1-\al) x)^{-1-\ep}
\nt\\[2mm] &\qquad\times
	\int_{0}^{1} \rd{v}\,v^{-\ep} (1-v)^{-\ep}
	\left({\al+(1-\al) x v \over 2\al+(1-\al) x}\right)^{\LL}
\label{eq:IC1}
\\[2mm] &\qquad\times
	\cI_1(\al+(1-\al) x v,\ep,\al_0,d_0;k)
\,,\qquad k,\,l=-1,0,1,2
\,,
\nt
\\[2mm]
\cI^{(2)}_{\cC{}{}}(x;\ep,\al_0,d_0;k,l) &=
	x \int_{0}^{\al_{0}} \rd{\al}
	(1-\al)^{2d_0-3+2\ep} \al^{-1-\ep}
	(\al+(1-\al) x)^{-1-\ep}
\nt\\[2mm] &\qquad\times
	\int_{0}^{1} \rd{v}\,v^{-\ep} (1-v)^{-\ep}
	\left({\al+(1-\al) x (1-v) \over 2\al+(1-\al) x}\right)^{\LL}
\label{eq:IC2}
\\[2mm] &\qquad\times
	\cI_1(\al+(1-\al) x v,\ep,\al_0,d_0;k)
\,,\qquad k,\,l=-1,0,1,2
\,,
\nt
\intertext{and}
\cI^{(3)}_{\cC{}{}}(x;\ep,\al_0,d_0;k) &=
	x \int_{0}^{\al_{0}} \rd{\al}
	(1-\al)^{2d_0-3+2\ep} \al^{-\ep} (\al+(1-\al) x)^{-\ep}
\nt\\[2mm] &\qquad\times
	\int_{0}^{1} \rd{v}\,v^{-\ep} (1-v)^{-\ep}
	(\al+(1-\al) x v)^{-2}
\label{eq:IC3}
\\[2mm] &\qquad\times
	\cI_1(\al+(1-\al) x v,\ep,\al_0,d_0;k)
\,,\qquad k=1,2
\,.
\nt
\eal
We can use the relations
\bal
\cI^{(2)}_{\cC{}{}}(x;\ep,\al_{0},d_0;k,0) &=
\cI^{(1)}_{\cC{}{}}(x;\ep,\al_{0},d_0;k,0)\,,
\\[2mm]
\cI^{(2)}_{\cC{}{}}(x;\ep,\al_{0},d_0;k,1) &=
\cI^{(1)}_{\cC{}{}}(x;\ep,\al_{0},d_0;k,0) -
\cI^{(1)}_{\cC{}{}}(x;\ep,\al_{0},d_0;k,1)\,,
\intertext{and}
\cI^{(2)}_{\cC{}{}}(x;\ep,\al_{0},d_0;k,2) &=
\cI^{(1)}_{\cC{}{}}(x;\ep,\al_{0},d_0;k,0) -
2 \cI^{(1)}_{\cC{}{}}(x;\ep,\al_{0},d_0;k,1) 
\nt\\[2mm]&+
\cI^{(1)}_{\cC{}{}}(x;\ep,\al_{0},d_0;k,2)\,,
\eal
to reduce the explicit computation of the $\cI^{(2)}_{\cC{}{}}$ integral
to the case $l=-1$.

In terms of the functions $\cI^{(i)}_{\cC{}{}}$ ($i = 1$, 2 and 3) we
find the result given in \eqn{eq:IcCktIcCktrresult}.

%
%

\subsection{Collinear-double collinear counterterm}
\label{sec:IntsCktCirkt}

The collinear-double collinear counterterm also involves two successive
collinear mappings, which leads to exact phase space factorisation
in an iterated form similar to that in \eqn{eq:psfactCktCktr}.
The integral over the factorized phase space measure
$[\rd p_{1,m+1}^{(kt)}]$ leads to the same collinear integrals as in
\eqn{eq:cI-def}. Then the necessary integrals over
$[\rd p^{(\ha{i}\ha{r})}_{1,m}]$ are
\bal
\bsp
\cI^{(4)}_{\cC{}{}} =
\frac{16\pi^2}{S_\ep}Q^{2\ep}
\int &[\rd p^{(\ha{i}\ha{r})}_{1,m}]
\frac{1}{s_{\ha{i}\ha{r}}}
\tzz{\ha{r}}{\ha{i}}^l,\, f(\al_0,\al_{\ha{i}\ha{r}},d(m,\ep))
\\[2mm] &\times
\cI_1(x_{\wha{kt}},\ep,\al_0,d_0;k)
\,.
\label{eq:IC2C22ndint}
\esp
\eal
Again, $x_{\wha{kt}}$ needs to be expressed in
terms of $\wti{p}_{kt}^\mu$ instead of $\ha{p}_{kt}^\mu$,
\beq
x_{\wha{kt}} = \frac{2\ha{p}_{kt}\ldot Q}{Q^2} 
	= \frac{2(1-\al_{\ha{i}\ha{r}})\wti{p}_{kt}\ldot Q}{Q^2}
	= (1-\al_{\ha{i}\ha{r}})x_{\wti{kt}}\,.
\eeq
Setting $\al\equiv\al_{\ha{i}\ha{r}}$, $v\equiv v_{\ha{i}\ha{r}}$,
$x\equiv x_{\wti{kt}}$, $y = x_{\wti{ir}}$, and using the
$v\leftrightarrow 1-v$ symmetry of the integration measure, we find
that the integrated collinear-double collinear counterterm
can be expressed as a linear combination of the integrals
\beq
\bsp
\cI^{(4)}_{\cC{}{}}(x,y;\ep,\al_{0},d_0;k,l) &=
	y \int_{0}^{\al_{0}} \rd{\al}
	(1-\al)^{2d_0-3+2\ep} \al^{-1-\ep} [\al+(1-\al)y]^{-1-\ep}
\\[2mm] &\qquad\times 
	\int_{0}^{1} \rd{v}\, v^{-\ep} (1-v)^{-\ep}
	\left({\al+(1-\al)y v \over 2\al+(1-\al)y}\right)^{\LL}
\\[2mm] &\qquad\times 
	\cI_1((1-\al) x,\ep,\al_0,d_0;k)
\,,\qquad k,\,l=-1,0,1,2
\,.
\label{eq:IC4}
\esp
\eeq

In terms of the functions
$\cI^{(4)}_{\cC{}{}}(x,y;\ep,\al_{0},d_0;k,l)$ we find
the result given in \eqn{eq:CktCirktresult}.

%
%

\subsection{Collinear-soft-collinear-type counterterms}
\label{sec:IntsCktCSktr}

The collinear-soft-collinear-type counterterms in eqns.\
(\ref{eq:IcCktIcSCSktr}), (\ref{eq:IcCktIcCirktIcSCSktr}) and 
(\ref{eq:IcCktIcCktrIcSCSktr}), involve a collinear mapping followed by
a soft mapping of the phase space, which leads to an exact
factorisation of the original $m+2$-particle phase space in the form
\beq
\PS{m+2}(\mom{}) = \PS{m}(\momt{(\ha{r},kt)})
[\rd p_{1,m}^{(\ha{r})}(\ha{p}_r;Q)]
[\rd p_{1,m+1}^{(kt)}(p_t,\ha{p}_{kt};Q)]\,,
\label{eq:psfactCktCSktr}
\eeq
where the factorized phase space measures $[\rd p_{1,m+1}^{(kt)}]$ and
$[\rd p_{1,m}^{(\ha{r})}]$ are given in \eqns{eq:dpir_exp}{eq:dpr_exp} after
appropriate changes in labelling, including the replacement $m\to m-1$ in the 
second case. The integral over $[\rd p_{1,m+1}^{(kt)}]$ gives the same collinear 
function as in \eqn{eq:cI-def}.  In order to compute the subsequent integrals over 
the measure $[\rd p_{1,m}^{(\ha{r})}]$,
\beq
\bsp
\Big\{\cI^{(5)}_{\cC{}{}},\,
  \cI^{(6)}_{\cC{}{}},\,
  \cI^{(7)}_{\cC{}{}}\Big\} =
-\frac{16\pi^2}{S_\ep}Q^{2\ep} \int &[\rd p^{(\ha{r})}_{1,m}]
\Bigg\{\frac{1}{2} \calS_{\ha{j}\ha{l}}(\ha{r}),\,
\frac{2}{s_{\ha{i}\ha{r}}}
\frac{\tzz{\ha{i}}{\ha{r}}}{\tzz{\ha{r}}{\ha{i}}},\,
\frac{2}{s_{\wha{kt}\ha{r}}}
\frac{\tzz{\wha{kt}}{\ha{r}}}{\tzz{\ha{r}}{\wha{kt}}}
\Bigg\}
\\[2mm] &\times
f(y_0,y_{\ha{r}Q},d'(m,\ep))\,\cI_1(x_{\wha{kt}},\ep,\al_0,d_0;k)
\,,
\label{eq:C2CS32ndint}
\esp
\eeq
we have to express the invariants of the dependent momenta (with hat)
with those of the independent ones (with tilde):
\beq
\bsp
s_{\ha{j}\ha{l}} &= (1-y_{\hr Q})s_{\wti{j}\wti{l}}\,,\qquad
s_{\ha{k}\ha{r}} = s_{\wti{k}\ha{r}}\,,
\quad \mbox{for} \quad \ha{k} = \ha{i},\:  \ha{j},\:  \wha{kt},\:  \ha{l},\: 
\\[2mm]
s_{\ha{k}Q} &= (1-y_{\hr Q})s_{\wti{k}Q} + s_{\wti{k}\ha{r}}\,,
\quad \mbox{for}\quad \ha{k} = \ha{i},\:  \wha{kt},
\esp
\eeq
which also implies
\beq
x_{\wha{kt}} = (1-y_{\hr Q})x_{\wti{kt}} + y_{\wti{kt}\ha{r}}\,.
\eeq
Furthermore,
\beq
\frac{\tzz{\ha{i}}{\ha{r}}}{\tzz{\ha{r}}{\ha{i}}} = 
\frac{y_{\ha{i}Q}}{y_{\ha{r}Q}} = 
\frac{(1-y_{\hr Q})y_{\wti{i}Q} + y_{\wti{i}\ha{r}}}{y_{\ha{r}Q}}
\,,
\label{eq:zratio}
\eeq
with a similar expression for $\tzz{\wha{kt}}{\ha{r}}/\tzz{\ha{r}}{\wha{kt}}$.

To write explicit integral representations of $\cI^{(i)}_{\cC{}{}}$ ($i=5,6$ and $7$), 
we choose the specific Lorentz frame of \eqn{eq:softframe}
with a different orientation for each function.


\paragraph{Integrated collinear-soft-collinear counterterm.} 
Here and in the following, we will use the partial fraction identity below 
to disentangle the singularities associated with the factors of $1/s_{\ha{j}\ha{r}}$ 
and $1/s_{\ha{l}\ha{r}}$, appearing in the eikonal factor (first term in the 
braces in \eqn{eq:C2CS32ndint}):
\beq
\bsp
{s_{\ha{j}\ha{l}} \over s_{\ha{j}\ha{r}} s_{\ha{l}\ha{r}}} &=
\frac{1-y_{\ha{r}Q}}{Q^2} 
{y_{\wti{j}\wti{l}} \over y_{\wti{j}\ha{r}} y_{\wti{l}\ha{r}}} =
\frac{1-y_{\ha{r}Q}}{Q^2}\,4\Yt{j}{l}
\,{y_{\wti{j}Q}y_{\wti{l}Q} \over 4\,y_{\wti{j}\ha{r}} y_{\wti{l}\ha{r}}}=
\\[2mm] &
= \frac{1-y_{\ha{r}Q}}{Q^2}\,4\Yt{j}{l}
	\left({y_{\wti{j}Q} \over 2y_{\wti{j}\ha{r}}} 
		+ {y_{\wti{l}Q} \over 2y_{\wti{l}\ha{r}}} \right)
	\left({2y_{\wti{j}\ha{r}} \over y_{\wti{j}Q}} 
		+ {2y_{\wti{l}\ha{r}} \over y_{\wti{l}Q}} \right)^{-1}
\,.
\label{eq:partialfrac}
\esp
\eeq
This is useful when computing the integral via iterated sector decomposition, 
while the original form is better suited to derive the Mellin--Barnes representation. 
Since `undoing' the partial fractioning is trivial, below we will show the more 
elaborate form of the integrals, which are directly suited to treatment with sector 
decomposition.

The convenient frame for integrating the first term in
\eqn{eq:partialfrac} is
\bal
\bsp
&\;\;
\ti{p}_j^\mu = E_{\wti{j}}(1,\dots,1)\,, \qquad
\ti{p}_l^\mu = E_{\wti{l}}(1,\ldots,\sin\chi_{\wti{l}},\cos\chi_{\wti{l}})\,, \\[2mm]
&\;\;
\ti{p}_{kt}^\mu = E_{\wti{kt}}(1,\ldots,\sin\phi_{\wti{kt}}\sin\chi_{\wti{kt}},
\cos\phi_{\wti{kt}}\sin\chi_{\wti{kt}},\cos\chi_{\wti{kt}})
\,,
\label{eq:frameCktCSktr}
\esp
\eal
while for the second term we choose a frame where $j$ and $l$ are
interchanged as compared to \eqn{eq:frameCktCSktr}.
In terms of the scaled energy-like variable $y_{\ha{r}Q}$ and the
angular variables $\vth$, $\vphi$ and $\eta$, the two-particle phase
space $\PS{2}(\ha{p}_r,K;Q)$ is given by \eqn{eq:dpr_exp}.
The two-particle invariants $y_{\wti{j}\ha{r}}$,  $y_{\wti{l}\ha{r}}$,
$y_{\wti{kt}\ha{r}}$ have to be expressed in terms of the integration
variables, i.e.
\bal
\frac{2 y_{\wti{j}\ha{r}}}{y_{\wti{j}Q}} &= y_{\hr Q} (1 - \cos\vth)\,,
\label{eq:IC5vars1}
\\[2mm]
\frac{2 y_{\wti{l}\ha{r}}}{y_{\wti{l}Q}} &= y_{\hr Q} 
(1 - \sin\chi_{\wti{l}} \sin\vth \cos\vphi - \cos\chi_{\wti{l}} \cos\vth)\,,
\\[2mm]
\bsp
y_{\wti{kt}\ha{r}} &= 
	{1\over 2} y_{\wti{kt}Q} y_{\hr Q} 
	(1 - \sin\phi_{\wti{kt}} \sin\chi_{\wti{kt}} \sin\vth \sin\vphi \cos\eta
\\[2mm]& \qquad \qquad
	-\cos\phi_{\wti{kt}} \sin\chi_{\wti{kt}} \sin\vth \cos\vphi
	-\cos\chi_{\wti{kt}} \cos\vth)\,.
\esp
\eal
Furthermore, writing out the definition, \eqn{eq:xi_YijQdef}, of all $\Y{i}{k}$'s 
in the specific Lorentz frames, we easily find that the fixed angles can be 
expressed with invariants as 
\bal
&\cos\chi_{\wti{l}} = \cos\chi(\Y{j}{l})\,,
&\cos\chi_{\wti{kt}} = \cos\chi(Y_{\wti{j}\wti{kt},Q})\,,
\\[2mm]
&\cos\phi_{\wti{kt}} =
\cos\phi(\Y{j}{l},Y_{\wti{j}\wti{kt},Q},Y_{\wti{l}\wti{kt},Q})\,,
\label{eq:IC5vars2}
\eal
with
\beq
\bsp
&\cos\chi(Y) = 1 - 2 Y\,,
\qquad
\sin\chi(Y) = 2\sqrt{Y (1-Y)}\,,
\\[2mm]
&\cos\phi(Y_1,Y_2,Y_3) = 
{Y_1 + Y_2 - Y_3 - 2 Y_1 Y_2 \over 2\sqrt{Y_1 (1-Y_1) Y_2 (1-Y_2)}}\,.
\label{eq:coschi(Y)}
\esp
\eeq
Using (\ref{eq:dpr_exp}) and the expressions for the ratios of kinematic 
invariants in \eqnss{eq:IC5vars1}{eq:IC5vars2}, we find the following 
explicit expression for $\cI^{(5)}_{\cC{}{}}$:
\beq
\bsp
\cI^{(5)}_{\cC{}{}}(x,Y_1,Y_2,Y_3;&\ep,\al_0,d_0,y_0,d'_0;k) =
-\left({-2^{2\ep}\ep \over 2\pi}\right) 4Y_1
\int_{0}^{y_0} \rd{y}\, y^{-1-2\ep} (1-y)^{d'_0-1+\ep} 
\\[2mm]& \times
\int_{-1}^{1} \rd{(\cos\vth)}\, \rd{(\cos\vphi)}\, \rd{(\cos\eta)}
(\sin\vth)^{-2\ep}
\\[2mm]& \quad\times
(\sin\vphi)^{-1-2\ep} (\sin\eta)^{-2-2\ep}
(1-\cos\vth)^{-1}
\\[2mm]& \quad\times
\Big(2-\cos\vth-\sin\chi(Y_1) \sin\vth \cos\vphi-\cos\chi(Y_1)
\cos\vth\Big)^{-1}
\\[2mm]& \quad\times
\Big[
\cI_1\Big([(1-y) x + y_{\wti{kt}\ha{r}}(y,x,Y_1,Y_2,Y_3,\vth,\vphi)];\ep,\al_0,d_0;k\Big)
\\[2mm]& \qquad +
\cI_1\Big([(1-y) x + y_{\wti{kt}\ha{r}}(y,x,Y_1,Y_3,Y_2,\vth,\vphi)];\ep,\al_0,d_0;k\Big)
\Big]\,,
\label{eq:IC5}
\esp
\eeq
which we need for $k=-1$, 0, 1, 2. 
Above, $x \equiv x_{\wti{kt}}$, $Y_1 \equiv \Y{j}{l}$, $Y_2 \equiv \Y{j}{kt}$, and 
$Y_3 \equiv \Y{l}{kt}$. The second term in the squared brackets in \eqn{eq:IC5}
corresponds to the second term in the partial fractions of
\eqn{eq:partialfrac}. The integral representations of the two terms are formally 
identical, only the kinematic variables $Y_2$ and $Y_3$ are interchanged.

In terms of the functions $\cI^{(5)}_{\cC{}{}}$ we find the result given in 
\eqn{eq:IcCktIcSCSktrresult}.  
In writing \eqn{eq:IC5}, we have tacitly assumed that $\ti{p}_j^\mu$, $\ti{p}_l^\mu$ 
and $\ti{p}_{kt}^\mu$ are all distinct (massless) momenta, whose kinematics is 
furthermore unconstrained. For processes involving two or three hard final state 
partons, this is not the case, so the integral $\cI^{(5)}_{\cC{}{}}$ with full kinematic 
dependence, as written above, first appears in computing NNLO corrections to 
processes with at least four hard final state partons.

When there are only three hard partons in the final state, 
the kinematics of the event is constrained because
momentum conservation forces the final state momenta 
to be coplanar. Thus, in \eqn{eq:frameCktCSktr} we have 
$\sin\phi_{\wti{kt}}=0$, and the parametrisation of $\ti{p}_j$, 
$\ti{p}_l$ and $\ti{p}_{kt}$ simplifies accordingly:
\bal
\bsp
& \wti{p}_j^\mu = E_{\wti{j}} (1,\dots,1)\,, \qquad
\wti{p}_l^\mu = E_{\wti{l}} (1,\ldots,\sin\chi_{\wti{l}},\cos\chi_{\wti{l}})\,, \\[2mm]
& \wti{p}_{kt}^\mu = E_{\wti{kt}} (1,\ldots,-\sin\chi_{\wti{kt}},\cos\chi_{\wti{kt}})\,,
\label{eq:frameCktCSktr-3jet}
\esp
\eal
where we choose $\cos\phi_{\wti{kt}}=-1$, so that we may assume 
$\sin\chi_{\wti{l}}$ and $\sin\chi_{\wti{kt}}$ to be nonnegative.
As a result of the constrained kinematics, we can first of all perform 
the $\cos\eta$ integration in \eqn{eq:dpr_exp} using \eqn{eq:Inteta}. 
Second, out of the four kinematic invariants ($x_{\wti{kt}}$, $\Y{j}{l}$, 
$\Y{j}{kt}$, and $\Y{l}{kt}$) of the general case in \eqn{eq:IC5}, only two 
are independent. (Momentum conservation implies three constraints 
among the five variables $E_{\wti{j}}$, $E_{\wti{l}}$, $E_{\wti{kt}}$, 
$\cos\chi_{\wti{l}}$ and $\cos\chi_{\wti{kt}}$.) However, it is convenient
to leave the formal dependence on all four variables, with the
constraints
\beq
Y_{\wti{l}\wti{kt},Q} = Y_{\wti{j}\wti{l},Q} + Y_{\wti{j}\wti{kt},Q} -
2Y_{\wti{j}\wti{l},Q} Y_{\wti{j}\wti{kt},Q}
+ 2 \sqrt{Y_{\wti{j}\wti{l},Q} (1-Y_{\wti{j}\wti{l},Q}) Y_{\wti{j}\wti{kt},Q}
(1-Y_{\wti{j}\wti{kt},Q})}\,,
\eeq
and
\beq
x_{\wti{kt}} = 
{\sqrt{\Y{j}{l}(1-\Y{j}{l})} \over
	\Y{j}{l}\sqrt{Y_{\wti{j}\wti{kt},Q}(1-Y_{\wti{j}\wti{kt},Q})}
	+Y_{\wti{j}\wti{kt},Q}\sqrt{\Y{j}{l}(1-\Y{j}{l})}}
\,.
\eeq
The physical region for $\Y{j}{l}$ and $Y_{\wti{j}\wti{kt},Q}$ is
$0\le\Y{j}{l},Y_{\wti{j}\wti{kt},Q}\le1$ and $\Y{j}{l} + Y_{\wti{j}\wti{kt},Q} \ge 1$.
The two-particle invariant $y_{\wti{kt}\ha{r}}$ becomes independent of
$\phi_{\wti{kt}}$,
\bal
y_{\wti{kt}\ha{r}} &= 
{1\over 2} y_{\wti{kt}Q} y_{\hr Q} 
(1 + \sin\chi_{\wti{kt}} \sin\vth \cos\vphi - \cos\chi_{\wti{kt}} \cos\vth)\,,
\eal
therefore, the dependence on $\Y{j}{l}$, $Y_{\wti{j}\wti{kt}Q}$ and 
$Y_{\wti{l}\wti{kt}Q}$ enters only through the angles
\beq
\cos\chi_{\wti{l}} = \cos\chi(Y_{\wti{j}\wti{l},Q})\,,\qquad
\cos\chi_{\wti{kt}} = \cos\chi(Y_{\wti{j}\wti{kt},Q})
\eeq
for the first term in the partial fraction and the angles
\beq
\cos\chi_{\wti{j}} = \cos\chi(Y_{\wti{j}\wti{l},Q})\,,\qquad
\cos\chi_{\wti{kt}} = \cos\chi(Y_{\wti{l}\wti{kt},Q})
\eeq
in the second one (with \eqn{eq:coschi(Y)} for $\cos\chi(Y)$).
Integrating out $\cos\eta$ as in \eqn{eq:Inteta}, the integral
(\ref{eq:IC5}) simplifies to
\beq
\bsp
\cI^{(5)}_{\cC{}{},3j}(x,Y_1,Y_2,Y_3;&\ep,\al_0,d_0,y_0,d'_0;k) =
	-4Y_1 {\Gamma^{2}(1-\ep) \over 2\pi \Gamma(1-2\ep)}
	\int_{0}^{y_0} \rd{y}\, y^{-1-2\ep} (1-y)^{d'_0-1+\ep} 
\\[2mm]& \times
	\int_{-1}^{1} \rd{(\cos\vth)}\, \rd{(\cos\vphi)}\,
	(\sin\vth)^{-2\ep} (\sin\vphi)^{-1-2\ep}
(1-\cos\vth)^{-1}
\\[2mm]& \quad\times
\Big(2-\cos\vth-\sin\chi(Y_1) \sin\vth \cos\vphi-\cos\chi(Y_1)
\cos\vth\Big)^{-1}
\\[2mm]& \quad\times
\Big[
\cI_1\Big([(1-y) x + y_{\wti{kt}\ha{r}}(y,x,Y_2,\vth,\vphi)];\ep,\al_0,d_0;k\Big)
\\[2mm]& \qquad +
\cI_1\Big([(1-y) x + y_{\wti{kt}\ha{r}}(y,x,Y_3,\vth,\vphi)];\ep,\al_0,d_0;k\Big)
\Big]
\,,
\label{eq:IC53j}
\esp
\eeq
($k =-1$, 0, 1, 2) and it replaces the function $\cI^{(5)}_{\cC{}{}}$
in \eqn{eq:IcCktIcSCSktrresult}.

Further simplifications emerge if the three labels $j$, $l$ and $kt$ are not 
all distinct and e.g.\ $l=kt$, which is the only case relevant for processes with 
only two hard final state partons. (Note that $j\ne l$, so up to $j\leftrightarrow l$ 
interchange, this is the only option.) 
Then $Y_{\wti{l}\wti{kt}Q} \to Y_{\wti{l}\wti{l}Q}= 0$ 
and the integral (\ref{eq:IC5})  depends only on two kinematic variables,
$x_{\wti{kt}}$ and $\Y{j}{l} = Y_{\wti{j}\wti{kt},Q}$.  The
parametrisation of the two hard momenta $\wti{p}_j$ and $\wti{p}_{kt}$
in \eqn{eq:frameCktCSktr} simplifies to:
\beq
\wti{p}_j^\mu = E_{\wti{j}} (1,\dots,1)\,, \qquad
\wti{p}_{kt}^\mu = E_{\wti{kt}} (1,\ldots,\sin\chi_{\wti{kt}},\cos\chi_{\wti{kt}})\,,
\label{eq:frameCktCSktr-2jet}
\eeq
so
\bal
y_{\wti{kt}\ha{r}} &= 
{1\over 2} y_{\wti{kt}Q} y_{\hr Q} 
(1 - \sin\chi_{\wti{kt}} \sin\vth \cos\vphi - \cos\chi_{\wti{kt}} \cos\vth)\,,
\eal
and the dependence on $Y_{\wti{j}\wti{kt}Q}$ enters through the angle
\beq
\cos\chi_{\wti{kt}} = \cos\chi(Y_{\wti{j}\wti{kt},Q})
\eeq
for the first term in the partial fraction and through
\beq
\cos\chi_{\wti{j}} = \cos\chi(Y_{\wti{j}\wti{kt},Q})
\eeq
in the second one (with \eqn{eq:coschi(Y)} for $\cos\chi(Y)$).
Integrating out $\cos\eta$ as in \eqn{eq:Inteta}, the integral
(\ref{eq:IC5}) reduces to
\beq
\bsp
\cI^{(5)}_{\cC{}{},2j}(x,Y_1;\ep,\al_0,d_0,y_0,d'_0;k) &=
	-4Y_1 {\Gamma^{2}(1-\ep) \over 2\pi \Gamma(1-2\ep)}
	\int_{0}^{y_0} \rd{y}\, y^{-1-2\ep} (1-y)^{d'_0-1+\ep} 
\\[2mm]& \times
	\int_{-1}^{1} \rd{(\cos\vth)}\, \rd{(\cos\vphi)}\,
	(\sin\vth)^{-2\ep} (\sin\vphi)^{-1-2\ep}
(1-\cos\vth)^{-1}
\\[2mm]& \quad\times
\Big(2-\cos\vth-\sin\chi(Y_1) \sin\vth \cos\vphi-\cos\chi(Y_1)
\cos\vth\Big)^{-1}
\\[2mm]& \quad\times
\Big[
\cI_1\Big([(1-y) x + y_{\wti{kt}\ha{r}}(y,x,Y_1,\vth,\vphi)];\ep,\al_0,d_0;k\Big)
\\[2mm]& \qquad +
\cI_1\Big([(1-y) x + y_{\wti{kt}\ha{r}}(y,x,0,\vth,\vphi)];\ep,\al_0,d_0;k\Big)
\Big]
\,,
\label{eq:IC5R}
\esp
\eeq
($k =-1$, 0, 1, 2) and it replaces the function $\cI^{(5)}_{\cC{}{}}$
in \eqn{eq:IcCktIcSCSktrresult}.


\paragraph{Rest of the integrated collinear-soft-collinear-type terms.}
The convenient frame for integrating the second term in the braces in
\eqn{eq:C2CS32ndint} is
\beq
\ti{p}_i^\mu = E_{\wti{i}} (1,\dots,1)\,, \qquad
\ti{p}_{kt}^\mu = E_{\wti{kt}} (1,\ldots,\sin\chi_{\wti{kt}},\cos\chi_{\wti{kt}})
\,.
\label{eq:frameCktCirktCSktr}
\eeq
The two-particle invariants are expressed with the integration
variables as
\bal
\frac{2 y_{\wti{i}\ha{r}}}{y_{\wti{i}Q}} &=  y_{\ha{r}Q} (1 - \cos\vth)\,,
\\[2mm]
y_{\wti{kt}\ha{r}} &= {1\over 2} y_{\wti{kt}Q} y_{\ha{r}Q} 
(1 - \sin\chi_{\wti{kt}} \sin\vth \cos\vphi - \cos\chi_{\wti{kt}} \cos\vth)\,,
\eal
where
\beq
\cos\chi_{\wti{kt}} = \cos\chi(\Y{i}{kt})\,,
\eeq
with \eqn{eq:coschi(Y)} for $\cos\chi(Y)$.
Then, using \eqn{eq:zratio} the integral in
\eqn{eq:IcCktIcCirktIcSCSktr} can be expressed as a linear combination
of the integrals
\beq
\bsp
\cI^{(6)}_{\cC{}{}}(x,Y;&\ep,\al_0,d_0,y_0,d'_0;k) =
	{\Gamma^{2}(1-\ep) \over \pi \Gamma(1-2\ep)}
	\int_{0}^{y_0} \rd{y}\, y^{-2\ep} (1-y)^{d'_0-2+\ep} 
\\[2mm]& \qquad\times
	\int_{-1}^{1} \rd{(\cos\vth)}\, \rd{(\cos\vphi)}\, 
	(\sin\vth)^{-2\ep} (\sin\vphi)^{-1-2\ep} 
	\,{2-y(1 + \cos\vth) \over y (1 - \cos\vth)} 
\\[2mm]& \qquad\times
\cI_1\Big([(1-y) x + y_{\wti{kt}\ha{r}}(y,x,Y,\vth,\vphi)];
\ep,\al_0,d_0;k\Big)
\,,
\qquad k=-1,0,1,2\,,
\label{eq:IC6}
\esp
\eeq
as in \eqn{eq:IcCktIcCirktIcSCSktrresult}.

Finally, the integral of the third term in the braces in
\eqn{eq:C2CS32ndint} is obtained by setting $\ha{i} = \wha{kt}$ in the
previous case, which implies $\Y{i}{kt} \to \Y{i}{i} = 0$, so 
the integration over $\vphi$ can be evaluated using \eqn{eq:Intphi} and
the integral in \eqn{eq:IcCktIcCktrIcSCSktr} can be expressed as 
a linear combination of the integrals
\beq
\bsp
\cI^{(6)}_{\cC{}{}}(x,0;\ep,\al_0,d_0,y_0,d'_0;k) &=
	2^{2\ep} \int_{0}^{y_0} \rd{y}\, y^{-2\ep} (1-y)^{d'_0-2+\ep} 
\\[2mm]& \qquad\times
	\int_{-1}^{1} \rd{(\cos\vth)}\,
	(\sin\vth)^{-2\ep} \,{2-y(1 + \cos\vth) \over y (1 - \cos\vth)} 
\\[2mm]& \qquad\times
\cI_1\Big([(1-y) x + y_{\wti{kt}\ha{r}}(y,x,0,\vth,\vphi)];
\ep,\al_0,d_0;k\Big)
\,,
\label{eq:IC6R}
\esp
\eeq
($k=-1$, 0, 1, 2) as in \eqn{eq:IcCktIcCktrIcSCSktrresult}.

%
%

\subsection{Integrated collinear-double soft-type counterterms}
\label{sec:IntsCktSkt}

The collinear-double soft counterterm is defined by an iterated
application of a collinear and a soft momentum mapping, which results in
an exact factorisation of the original $(m+2)$-particle phase space
very similar to that in \eqn{eq:psfactCktCSktr}. The difference is
that in the present case the soft measure involves the momentum
$\ha{p}_{kt}^\mu$ instead of $\ha{p}_r^\mu$.  The integrand of the collinear
integral is the spin-dependent splitting kernel of gluon splitting,
with Lorentz structure
\beq
\frac{1}{\CA}
\la\mu|\ha{P}^{(0)}_{f_k f_t}(\tzz{k}{t},\tzz{t}{k},\kT{k,t};\ep)|\nu\ra\ =
-g^{\mu\nu} a^{(0)}_{f_k f_t}(\tzz{t}{k})
- b^{(0)}_{f_k f_t}\frac{1-\ep}{2}\tzz{k}{t}
  \tzz{t}{k} {4 \kT{k,t}^\mu \kT{k,t}^\nu \over \kT{k,t}^2}\,,
\label{eq:P0munu}
\eeq
where
\bal
a^{(0)}_{gg}(z) &= 2 \left({z\over 1-z} + {1-z \over z}\right)\,, &
a^{(0)}_{q\qb}(z) &= \frac{\TR}{\CA}\,,
\label{eq:bdef}
\eal
and $b^{(0)}_{f_k f_t}$ is defined in \eqn{eq:bkt}, hence, the
integral over the phase space measure $[\rd p_{1,m+1}^{(kt)}]$ involves
both collinear functions $\cI$ and $\cI^{\mu\nu}$.  Introducing the
abbreviation
\beq
\bsp
\cI_{f_k f_t}^{\mu\nu} =
\frac{16\pi^2}{S_\ep}Q^{2\ep}
\int &[\rd p_{1,m+1}^{(kt)}(p_t,\ha{p}_{kt};Q)] {1\over s_{kt}} 
f(\al_0,\al_{kt},d(m,\ep))
\\[2mm] &\times
\frac{1}{\CA}
\la\mu|\ha{P}^{(0)}_{f_k f_t}(\tzz{k}{t},\tzz{t}{k},\kT{k,t};\ep)|\nu\ra
\,,
\label{eq:Idef}
\esp
\eeq
and using the result in \eqn{eq:cImunu}, we find that
\beq
\bsp
\cI_{f_k f_t}^{\mu\nu} &=
-g^{\mu\nu} \frac{16\pi^2}{S_\ep}Q^{2\ep}
\int [\rd p_{1,m+1}^{(kt)}] {1\over s_{kt}}
f(\al_0,\al_{kt},d(m,\ep))
\\[2mm]&\qquad\qquad\qquad\qquad\times
\Big(a^{(0)}_{f_k f_t}(\tzz{t}{k}) +b^{(0)}_{f_k f_t}\tzz{k}{t}\tzz{t}{k}\Big)
\\[2mm]&
+b^{(0)}_{f_k f_t}
\left(
{\ha{p}_{kt}^\mu Q^\nu + Q^\mu \ha{p}_{kt}^\nu \over \ha{p}_{kt}\ldot Q}
- {Q^2 \over (\ha{p}_{kt}\ldot Q)^2} \ha{p}_{kt}^\mu \ha{p}_{kt}^\nu
\right)
\\[2mm]&\qquad\times
\frac{16\pi^2}{S_\ep}Q^{2\ep}
\int [\rd p_{1,m+1}^{(kt)}] {1\over s_{kt}}
\tzz{k}{t}\tzz{t}{k}
f(\al_0,\al_{kt},d(m,\ep))
\,,
\label{eq:Ifinal2}
\esp
\eeq
where in the first integral we recognise the spin-averaged splitting kernel,
\beq
a^{(0)}_{f_k f_t}(\tzz{t}{k}) + b^{(0)}_{f_k f_t}\tzz{t}{k}(1-\tzz{t}{k}) =
P^{(0)}_{f_k f_t}(\tzz{t}{k};\ep)\,.
\eeq
Therefore, $\cI_{f_k f_t}^{\mu\nu}$ can be expressed as a linear
combination of collinear functions $\cI$, with Lorentz structure
exhibited in \eqn{eq:Ifinal2}. After contraction with
$\calS^{\mu\nu}_{\ha{j}\ha{l}}(\wha{kt})$ (see \eqn{eq:IcCktIcS}), we
obtain
\beq
\bsp
\frac12 \calS^{\mu\nu}_{\ha{j}\ha{l}}(\wha{kt}) \cI_{f_k f_t\,\mu\nu} &=
- \frac12 \calS_{\ha{j}\ha{l}}(\wha{kt})
\,\frac{16\pi^2}{S_\ep}Q^{2\ep}
\int [\rd p^{(kt)}_{1,m+1}]
{1\over s_{kt}} 
P^{(0)}_{f_k f_t}(\tzz{t}{k};\ep) 
f(\al_0,\al_{kt},d(m,\ep))
\\[2mm]&
+ b^{(0)}_{f_k f_t}
\left({s_{\ha{j}Q} \over s_{\wha{kt}\ha{j}} s_{\wha{kt}Q}}
	+{s_{\ha{l}Q} \over s_{\wha{kt}\ha{l}} s_{\wha{kt}Q}}
	-{2 Q^2 \over s_{\wha{kt}Q}^2}
\right)
\\[2mm]& \qquad\times
\frac{16\pi^2}{S_\ep}Q^{2\ep}
\int [\rd p^{(kt)}_{1,m+1}]
{1\over s_{kt}} \tzz{t}{k}(1-\tzz{t}{k})
f(\al_0,\al_{kt},d(m,\ep))
\,.
\esp
\label{eq:ICktSkt-2}
\eeq
In the integral of the complete collinear-double soft subtraction, 
\eqn{eq:ICktSkt-2} is multiplied with terms that are symmetric with
respect to the interchange $j\leftrightarrow l$, and summed over both $j$
and $l$ (cf.~\eqns{eq:CktA2}{eq:I2dsigRRA12Jm-1}). Therefore, the
integrals of the terms
\beq
{s_{\ha{j}Q} \over s_{\wha{kt}\ha{j}} s_{\wha{kt}Q}}
\qquad\mbox{and}\qquad
{s_{\ha{l}Q} \over s_{\wha{kt}\ha{l}} s_{\wha{kt}Q}}
\eeq
in the parenthesis give identical contributions, and it is sufficient
to evaluate three types of integrals:
\beq
\bsp
\Big\{\cI^{(7)}_{\cC{}{}},\,
  \cI^{(8)}_{\cC{}{}},\,
  \cI^{(9)}_{\cC{}{}}\Big\} =
\frac{16\pi^2}{S_\ep}Q^{2\ep}
\int &[\rd p^{(\wha{kt})}_{1,m}]\frac{1}{s_{\wha{kt}\ha{r}}}
\left\{-{s_{\ha{j}\ha{l}} \over s_{\ha{j}\wha{kt}} s_{\ha{l}\wha{kt}}},\,
{2 s_{\ha{j}Q} \over s_{\wha{kt}\ha{j}} s_{\wha{kt}Q}},\,
-{2 Q^2 \over s_{\wha{kt}Q}^2}
\right\}
\\[2mm] &\times
f(y_0,y_{\wha{kt}Q},d'(m,\ep))
\,\cI(x_{\wha{kt}};\ep,\al_0,d_0;k)
\,.
\label{eq:IC7-IC9}
\esp
\eeq

To compute the integrals in \eqn{eq:IC7-IC9}, first we have to express
the invariants of the dependent momenta with those of independent ones:
\beq
\bsp
s_{\ha{j}\ha{l}} &= (1-y_{\wha{kt}Q}) s_{\wti{j}\wti{l}}
\,,\qquad
s_{\ha{j}\wha{kt}} = s_{\wti{j}\wha{kt}}
\,,\qquad
s_{\ha{l}\wha{kt}} = s_{\wti{l}\wha{kt}}
\,,
\\[2mm] 
s_{\ha{j}Q} &= (1-y_{\wha{kt}Q}) s_{\wti{j}Q} + s_{\wti{j}\wha{kt}}
\,,\qquad
s_{\ha{l}Q} = (1-y_{\wha{kt}Q}) s_{\wti{l}Q} + s_{\wti{l}\wha{kt}}
\,.
\esp
\eeq
Then we can proceed  as usual by writing the factorized phase space
$[\rd p^{(kt)}_{1,m+1}]$ explicitly. We choose a Lorentz frame defined by
\beq
\ti{p}_j^\mu = E_{\wti{j}} (1,\dots,1)\,, \qquad
\ti{p}_{l}^\mu = E_{\wti{l}} (1,\ldots,\sin\chi_{\wti{l}},\cos\chi_{\wti{l}})\,,
\eeq
where the two-particle invariants are expressed
in terms of the integration variables as
\beq
\frac{2 y_{\wti{j}\wha{kt}}}{y_{\wti{j}Q}} = y_{\wha{kt}Q} (1 - \cos\vth)\,,
\qquad
\frac{2 y_{\wti{l}\wha{kt}}}{y_{\wti{l}Q}} = y_{\wha{kt}Q}
(1 - \sin\chi_{\wti{l}} \sin\vth \cos\vphi - \cos\chi_{\wti{l}} \cos\vth)\,,
\eeq
with
\beq
\cos\chi_{\wti{l}} = \cos\chi(\Y{j}{l}) \,,
\eeq
and \eqn{eq:coschi(Y)} for $\cos\chi(Y)$.

For the eikonal factor (first term in the braces in \eqnss{eq:IC7}{eq:IC9})
we use the partial fraction identity (\ref{eq:partialfrac}) (with the
substitution $\ha{r} \to \wha{kt}$), and the $j\leftrightarrow l$
symmetry to integrate only one term. Thus we are left with the
following three integrals:
\bal
\cI^{(7)}_{\cC{}{}}(Y;\ep,\al_{0},d_0;k) &=
-4Y {\Gamma^{2}(1-\ep) \over 2\pi \Gamma(1-2\ep)}
\int_{0}^{y_0} \rd{y}\, y^{-1-2\ep} (1-y)^{d'_0-1+\ep}
\nt\\[2mm]&\times
\int_{-1}^{1} \rd{(\cos\vth)}\, \rd{(\cos\vphi)}\,
(\sin\vth)^{-2\ep} (\sin\vphi)^{-1-2\ep}
\frac{2}{1-\cos\vth}
\label{eq:IC7}
\\[2mm]&\qquad\times
\Big(2-\cos\vth 
 - \sin\chi(Y) \sin\vth \cos\vphi - \cos\chi(Y) \cos\vth)
\Big)^{-1}
\nt\\[2mm]&\qquad\times
\cI_1(y;\ep,\al_0,d_0;k) \,,\qquad k = -1,0,1,2
\nt
\intertext{(here $Y$ corresponds to $\Y{j}{l}$),}
\cI^{(8)}_{\cC{}{}}(\ep,\al_{0},d_0;k) &= 
2^{2\ep} \int_{0}^{y_0} \rd{y}\, y^{-1-2\ep} (1-y)^{d'_0-1+\ep}
\nt\\[2mm]&\times
\int_{-1}^{1} \rd{(\cos\vth)}\, (\sin\vth)^{-2\ep}
\frac{2-y(1+\cos\vth)}{1-\cos\vth}
\label{eq:IC8}
\\[2mm]&\qquad\times
\cI_1(y;\ep,\al_0,d_0;k) \,,\qquad k = -1,0,1,2,
\nt
\intertext{and}
\cI^{(9)}_{\cC{}{}}(\ep,\al_{0},d_0;k) &= 
-2 {\Gamma^{2}(1-\ep) \over \Gamma(2-2\ep)} 
\int_{0}^{y_0} \rd{y}\, y^{-1-2\ep} (1-y)^{d'_0-2+\ep}
\label{eq:IC9}
\\[2mm]&\qquad\qquad\qquad\qquad\times
\cI_1(y;\ep,\al_0,d_0;k) \,,\qquad k = 1,2.
\nt
\eal
(The cases $k = -1,0$ in \eqn{eq:IC8} are needed for \eqn{eq:CktCrktSkt},
see next paragraph.) The final result is presented in
\eqn{eq:IcCktIcSresult}.

For the integrated collinear-triple collinear-double soft subtraction
in \eqn{eq:CktCrktSkt} the phase space factorisation is
the same as for the integrated collinear-double soft subtraction, 
and we need to evaluate the integrals
\beq
\bsp
\frac{16\pi^2}{S_\ep}Q^{2\ep}
\int &[\rd p^{(\wha{kt})}_{1,m}]
\left\{\frac{2}{s_{\wha{kt}\ha{r}}}
\frac{\tzz{\ha{r}}{\wha{kt}}}{\tzz{\wha{kt}}{\ha{r}}},\,
-{2 Q^2 \over s_{\wha{kt}Q}^2}
\right\}
\\[2mm] &\times
f(y_0,y_{\wha{kt}Q},d'(m,\ep))
\,\cI_1(x_{\wha{kt}};\ep,\al_0,d_0;k)
\,.
\esp
\eeq
Recalling the definition of the momentum fractions, in the first term we
recognise twice the second one in \eqn{eq:IC7-IC9} (after replacing
$j$ with $r$), while the second term is equal to the last one in
\eqn{eq:IC7-IC9}.  Thus we do not have to compute any new integrals,
we can express this integrated counterterm as a linear
combination of the integrals defined in \eqns{eq:IC8}{eq:IC9}.
The final result is given in \eqn{eq:CktCrktSktresult}.




\section{Integrating the soft-type terms}
\label{sec:IntsStA2}

In this appendix, we discuss the integration of the soft-type 
counterterms of \sect{sec:IntsStCTs}. In particular, we define and give an 
explicit integral representation of all $\cI^{(i)}_{\cS{}{}}$
functions ($i=1,\ldots,12$).

%
%

\subsection{Soft-triple collinear-type counterterms}
\label{sec:StCirt-type}

There are three integrated counterterms that involve a soft mapping of
the momenta, followed by a collinear one, which leads to an exact
factorisation of the original $m+2$-particle phase space in the form
\beq
\PS{m+2}(\mom{}) =
\PS{m}(\momt{(\ha{i}\ha{r},t)}_m)
[\rd p^{(\ha{i}\ha{r})}_{1,m}(\ha{p}_{r},\wti{p}_{ir};Q)]
[\rd p^{(t)}_{1,m+1}(p_{t};Q)]\,.
\label{eq:psfact_CitStCirt-type}
\eeq
The factorized phase space measures $[\rd p^{(t)}_{1,m+1}(p_{t};Q)]$ and
$[\rd p^{(\ha{i}\ha{r})}_{1,m}(\ha{p}_{r},\wti{p}_{ir};Q)]$ are given in
\eqns{eq:dpr_exp}{eq:dpir_exp}, with appropriate changes in labelling and 
the replacement of $m\to m-1$ in the second case.


\paragraph{Integrated soft-triple collinear counterterm.} 
We begin by noting that the soft functions $P^{(\rm{S})}_{f_i f_r f_t}$ 
(originally defined in \refr{Somogyi:2006da}), which appear in \eqn{eq:IcStIcCirt}  
can be written in the following unified form:
\beq
\bsp
P^{(\rm{S})}_{f_i f_r f_t} &=
  (C_{f_{it}} + C_{f_{rt}} - C_{f_{ir}})
\frac{s_{ir}}{s_{it} s_{rt}} 
+ (C_{f_{ir}} + C_{f_{it}} - C_{f_{rt}})
\frac{1}{s_{it}} \frac{\tzz{i}{rt}}{\tzz{t}{ir}} 
\\[2mm]&
+ (C_{f_{ir}} + C_{f_{rt}} - C_{f_{it}})
\frac{1}{s_{rt}} \frac{\tzz{r}{it}}{\tzz{t}{ir}}
\,.
\label{eq:PSirt}
\esp
\eeq
This general form hides the fact that $f_t = g$, which also implies 
$C_{f_{it}} = C_{f_{i}}$ and $C_{f_{rt}} = C_{f_{r}}$ (used in
\eqn{eq:softcollresults}).  Furthermore, according to our definition of
the momentum fractions, we have
\beq
\frac{\tzz{i}{jk}}{\tzz{j}{ik}} = \frac{y_{iQ}}{y_{jQ}}
= \frac{\tzz{i}{j}}{\tzz{j}{i}}
\label{eq:zratio1}
\,,
\eeq
for any $i$, $j$ and $k$. 
Therefore, performing the integration over the soft phase space 
$[\rd p^{(t)}_{1,m+1}(p_{t};Q)]$ first, we find that the three terms 
in \eqn{eq:PSirt} all lead to known functions.
The eikonal term gives $(-1)$ times the soft function
$\cJ_1(Y,\ep,y_0,d'_0)$, while both soft collinear terms give $\frac12$
times the same soft-collinear function, $\cK_1(\ep,y_0,d'_0)$. 
For these functions we have the integral representations discussed in
\appx{sec:IJK}.  Hence, we can express the integrated
soft-triple collinear counterterms as linear combinations of two
types of basic integrals,
\bal
\cI^{(1)}_{\cS{}{}} &= \frac{16\pi^2}{S_\ep}Q^{2\ep}
	\int [\rd p^{(\ha{i}\ha{r})}_{1,m}(\ha{p}_{r},\wti{p}_{ir};Q)] 
	\frac{1}{s_{\ha{i}\ha{r}}} \tzz{\ha{r}}{\ha{i}}^\KK
	f(\al_0,\al,d(m,\ep))\,\frac{1}{2} \cK_1(\ep,y_0,d'_0)\,,
\label{eq:IS1-def}
\intertext{and}
\cI^{(2)}_{\cS{}{}} &= -\frac{16\pi^2}{S_\ep}Q^{2\ep}
	\int [\rd p^{(\ha{i}\ha{r})}_{1,m}(\ha{p}_{r},\wti{p}_{ir};Q)] 
	\frac{1}{s_{\ha{i}\ha{r}}} \tzz{\ha{r}}{\ha{i}}^\KK
	f(\al_0,\al,d(m,\ep))\,\cJ_1(\Yh{i}{r};\ep,y_0,d'_0)
\,.
\label{eq:IS2-def}
\eal

The soft-collinear function $\cK$ does not depend on any kinematic 
variable, therefore, it factorizes completely from the integral over
$[\rd p^{(\ha{i}\ha{r})}_{1,m}]$. The remaining integral can be
expressed with the collinear function $\cI(x;\ep,\al_0,d_0-1+\ep,0,k,0,1)$
(the parameter $d_0$ is shifted because this integral corresponds to $m$
partons in the final state), so
\beq
\cI^{(1)}_{\cS{}{}}(x;\ep,\al_0,d_0,y_0,d'_0;k) = 
\frac{1}{2} \cK_1(\ep,y_0,d'_0)\,\cI_1(x;\ep,\al_0,d_0-1+\ep;k)\,.
\label{eq:IS1}
\eeq

The soft integral depends on $\Yh{i}{r}$, that has to be expressed 
with the integration variables of $[\rd p^{(\ha{i}\ha{r})}_{1,m}]$:
\beq
\Yh{i}{r}(x_{\wti{ir}},\al,v) =
{\al[\al+(1-\al)x_{\wti{ir}}] \over
[\al+(1-\al)x_{\wti{ir}} v] [\al+(1-\al)x_{\wti{ir}} (1-v)]}
\,,
\label{eq:Y2avStCirt2}
\eeq
where we used the usual abbreviations $\al=\al_{\ha{i}\ha{r}}$ and
$v=v_{\ha{i}\ha{r}}$. Clearly, $\Yh{i}{r}(x_{\wti{ir}};\al,v)$, as well as the 
phase space measure in \eqn{eq:dpir_exp} is symmetric in $v \leftrightarrow 1-v$. 
Thus, it is sufficient to consider only the following coupled collinear and soft integral,
\beq
\bsp
\cI^{(2)}_{\cS{}{}}(x;\ep,\al_0,d_0,y_0,d'_0;k) &=
-x \int_{0}^{\al_{0}} \rd{\al}\,
(1-\al)^{2d_0-3+2\ep} \al^{-1-\ep} [\al+(1-\al) x]^{-1-\ep} 
\\[2mm] &\qquad\times
\int_{0}^{1} \rd{v}\, v^{-\ep} (1-v)^{-\ep}
\left({\al+(1-\al) x v \over 2\al+(1-\al) x}\right)^{\KK}	
\\[2mm] &\qquad\times
\cJ_1(\Yh{i}{r}(x,\al,v);\ep,y_0,d'_0)\,,\qquad k=-1,0,1,2\,.
\label{eq:IS2}
\esp
\eeq
In terms of the integrals $\cI^{(1)}_{\cS{}{}}$ and $\cI^{(2)}_{\cS{}{}}$
the integrated soft-triple collinear subtraction term can be written as
in \eqn{eq:IcStIcCirtresult}.


\paragraph{Integrated soft-soft-collinear counterterm.} 
To obtain the counterterm in \eqn{eq:IcStIcCSirt}, we again perform 
the integration  over the factorized soft phase space 
$[\rd p^{(t)}_{1,m+1}(p_{t};Q)]$ first. 
Now, we must distinguish two cases:
(i) $j$, $l$ and $ir$ are three distinct labels,
(ii) $j$ or $l$ coincide with $ir$. (We always have $j\ne l$.) 
In the first case, the integral over the factorized soft phase space 
simply leads to the $\cJ_1$ function of \eqn{eq:cJ-def}, and we find 
that the integrated soft-soft-collinear counterterm can be expressed 
as a linear combination of the integrals
\beq
\bsp
\cI^{(3)}_{\cS{}{}} =
\frac{16\pi^2}{S_\ep}Q^{2\ep}
\int &[\rd p^{(\ha{i}\ha{r})}_{1,m}]
\frac{1}{s_{\ha{i}\ha{r}}} 
\tzz{\ha{r}}{\ha{i}}^k
f(\al_0,\al_{\ha{i}\ha{r}},d(m,\ep))
\\[2mm] &\times
\cJ_1(\Yh{j}{l};\ep,y_0,d'_0,0)
\,.
\label{eq:IS3-def}
\esp
\eeq
Using the collinear mapping formula
$\ha{p}_{j,l}^\mu = (1-\al_{\ha{i}\ha{r}}) \wti{p}_{j,l}^\mu$, we find
\beq
\Yh{j}{l} = \frac{s_{\ha{j}\ha{l}}}{s_{\ha{j}Q} s_{\ha{l}Q}} 
= \frac{(1-\al_{\ha{i}\ha{r}})^2 s_{\wti{j}\wti{l}}}
{(1-\al_{\ha{i}\ha{r}})^2 s_{\wti{j}Q} s_{\wti{l}Q}}
= \Yt{j}{l}\,. 
\eeq
Consequently, the collinear and soft integrals decouple, and we obtain
\beq
\cI^{(3)}_{\cS{}{}}(x,Y;\ep,\al_0,d_0,y_0,d'_0;k) =
\cI_1(x;\ep,\al_0,d_0-1+\ep;k)\,\cJ_1(Y;\ep,y_0,d'_0)
\,.
\label{eq:IS3}
\eeq

In case (ii) if e.g.\ $j = (ir)$, the eikonal factor $\calS_{(ir)l}(t)$
evaluates as in \eqn{eq:softcolleikonal}, and the integral over the 
soft phase space leads to the one-mass soft function $\cJm$, defined 
in \eqn{eq:cJm-def} and computed in \eqn{eq:cJm}. Hence we find that 
the integrated soft-soft-collinear counterterm can be expressed as a 
linear combination of the integrals
\beq
\bsp
\cI^{(4)}_{\cS{}{}} =
\frac{16\pi^2}{S_\ep}Q^{2\ep}
\int &[\rd p^{(\ha{i}\ha{r})}_{1,m}]
\frac{1}{s_{\ha{i}\ha{r}}} 
\tzz{\ha{r}}{\ha{i}}^k
f(\al_0,\al_{\ha{i}\ha{r}},d(m,\ep))
\\[2mm] &\times
\cJm(Y_{(\ha{i}\ha{r})\ha{l},Q},\be_{(\ha{i}\ha{r})};\ep,y_0,d_0')
\,.
\label{eq:IS4-def}
\esp
\eeq
This time the collinear
integral does not decouple because the parameters
$Y_{(\ha{i}\ha{r})\ha{l},Q}$ and $\be_{(\ha{i}\ha{r})}$ depend
on the collinear integration variables. Using the definition of
$\wti{p}_{ir}^\mu$ and $\wti{p}_l^\mu$, we have $\ha{p}_{(ir)}^\mu =
(1-\al_{\ha{i}\ha{r}})\wti{p}_{ir}^\mu + \al_{\ha{i}\ha{r}} Q^\mu$ and
$\ha{p}_l^\mu = (1-\al_{\ha{i}\ha{r}})\wti{p}_{l}^\mu$. Consequently,
\bal
\bsp
Y_{(\ha{i}\ha{r})\ha{l},Q}(y_{\wti{ir}Q},\Yt{ir}{l},\al_{\ha{i}\ha{r}})
&= \frac12 \left(1-\be_{(\ha{i}\ha{r})}(y_{\wti{ir}Q},\al_{\ha{i}\ha{r}})
(1 - 2\Y{ir}{l})\right)
\,,\\[2mm]
\be_{(\ha{i}\ha{r})}(y_{\wti{ir}Q},\al_{\ha{i}\ha{r}}) &=
{(1-\al_{\ha{i}\ha{r}})y_{\wti{ir}Q} \over
2\al_{\ha{i}\ha{r}} + (1-\al_{\ha{i}\ha{r}}) y_{\wti{ir}Q}}\,,
\label{eq:Y-be}
\esp
\eal
and
\beq
\bsp
\cI^{(4)}_{\cS{}{}}(x,Y;\ep,\al_0,d_0,y_0,d'_0;k) &=
x  \int_{0}^{\al_{0}} \rd{\al}\,
(1-\al)^{2d_0-3+2\ep} \al^{-1-\ep} [\al+(1-\al) x]^{-1-\ep} 
\\[2mm] &\qquad\times	
\int_{0}^{1} \rd{v}\, v^{-\ep} (1-v)^{-\ep}
\left({\al+(1-\al) x v \over 2\al+(1-\al) x}\right)^{\KK}	
\\[2mm] &\qquad\times	
\cJm(Y_{(\ha{i}\ha{r})\ha{l},Q}(x,Y,\al),\be_{(\ha{i}\ha{r})}(x,\al);\ep,y_0,d_0')
\,,
\label{eq:IS4}
\esp
\eeq
($k=-1,0,1,2$).

In terms of the integrals $\cI^{(3)}_{\cS{}{}}$ and
$\cI^{(4)}_{\cS{}{}}$, the integrated soft-soft-collinear counterterm
can be expressed as in \eqn{eq:IcStIcCSirtresult}.


\paragraph{Integrated soft-triple collinear-soft-collinear counterterm.} 
When computing the integrals in \eqn{eq:IcStIcCirtIcCSirt},
we first perform the integral over the soft phase space. Using 
the same frame as in \eqn{eq:StCSirt-frame1}, we find that the 
integral leads to the one-mass soft-collinear function $\cKm$ 
defined in \eqn{eq:cKm-def} and computed in \eqn{eq:cKm-final}.
Thus the integrated soft-triple collinear-soft-collinear counterterm
can be expressed as linear combination of the integrals
\beq
\bsp
\cI^{(5)}_{\cS{}{}} =
\frac{16\pi^2}{S_\ep}Q^{2\ep}
\int &[\rd p^{(\ha{i}\ha{r})}_{1,m}]
\frac{1}{s_{\ha{i}\ha{r}}} 
\tzz{\ha{r}}{\ha{i}}^k
f(\al_0,\al_{\ha{i}\ha{r}},d(m,\ep))
\\[2mm] &\times
\cKm(\be_{(\ha{i}\ha{r})}(x,\al);\ep,y_0,d_0')
\,.
\label{eq:IS5-def}
\esp
\eeq
The soft-collinear integral
$\cKm(\be)$ does not decouple from the collinear one because
$\be=\be_{(\ha{i}\ha{r})}(y_{\wti{ir}Q},\al_{\ha{i}\ha{r}})$ depends on
the collinear integration variable $\al_{\ha{i}\ha{r}}$ as in
\eqn{eq:Y-be}. Thus, 
\beq
\bsp
\cI^{(5)}_{\cS{}{}}(x;\ep,\al_0,d_0,y_0,d'_0;k) &=
	x  \int_{0}^{\al_{0}} \rd{\al}\,
	(1-\al)^{2d_0-3+2\ep} \al^{-1-\ep} [\al+(1-\al) x]^{-1-\ep} 
\\[2mm] &\qquad\times	
	\int_{0}^{1} \rd{v}\,v^{-\ep} (1-v)^{-\ep}
	\left({\al+(1-\al) x v \over 2\al+(1-\al) x}\right)^{\KK}	
\\[2mm] &\qquad\times
\cKm(\be_{(\ha{i}\ha{r})}(x,\al);\ep,y_0,d_0')
\,,\qquad k=-1,0,1,2\,.
\label{eq:IS5}
\esp
\eeq

In terms of the integrals $\cI^{(5)}_{\cS{}{}}$,
we find the result as in \eqn{eq:IcStIcCirtIcCSirtresult}.

%
%

\subsection{The soft-double soft-type counterterms}
\label{sec:StCirtSrt-type}

The remaining soft-type integrated counterterms involve two successive
soft momentum mappings, which leads to an exact factorisation of the
original $m+2$-particle phase space in the form
\beq
\PS{m+2}(\mom{}) =
     \PS{m}(\momt{(\ha{r},t)}_m) [\rd
p^{(\ha{r})}_{1,m}(p_{\ha{r}};Q)]
     [\rd p^{(t)}_{1,m+1}(p_{t};Q)]\,.
\label{eq:psfact_CktStSrt-type}
\eeq
The factorized phase space measures $[\rd p^{(\ha{r})}_{1,m}]$
and $[\rd p^{(t)}_{1,m+1}]$ are given in \eqn{eq:dpr_exp}, with 
appropriate changes in labelling (including $m\to m-1$ in the first case).
We often need to express the two-particle scaled invariants of the
momenta after the first mapping (`hatted momenta') with the final momenta
obtained after the second mapping (`tilded momenta').
The relevant formulae, collected here for later reference, are:
\bal
\bsp
y_{\ha{i}\ha{l}} &= (1-y_{\hr Q}) y_{\wti{i}\wti{l}}\,,\qquad
y_{\ha{k}\ha{r}} = y_{\wti{k}\ha{r}}\,,\qquad
\\[2mm]
y_{\ha{k}Q} &= (1-y_{\hr Q}) y_{\wti{k}Q} + y_{\wti{k}\hr}\,,\qquad
k = i,\,j,\,l\,.
\label{eq:hattotilde}
\esp
\eal
%


\paragraph{Integrated soft-triple collinear-double soft counterterms.} 
In computing the integrals in \eqn{eq:IcStIcCirtIcSrt}, we first perform
the integration over the soft phase space factor $[\rd p^{(t)}_{1,m+1}]$.
This integration leads to either the $\cJ_1$ soft function, or the $\cK_1$ 
soft-collinear function of \sect{sec:IJK}.
Then the remaining integral over $[\rd p^{(\ha{r})}_{1,m}]$ involves
\beq
\bsp
\Big\{\cI^{(6)}_{\cS{}{}},\,\cI^{(7)}_{\cS{}{}} \Big\} =
\frac{16\pi^2}{S_\ep}Q^{2\ep}
\int &[\rd p^{(\ha{r})}_{1,m}] 
\frac{2}{s_{\ha{i}\ha{r}}}\frac{\tzz{\ha{i}}{\ha{r}}}{\tzz{\ha{r}}{\ha{i}}}
f(y_0,y_{\ha{r}Q},d'(m,\ep))
\\[2mm] & \times
\Big\{\cK_1(\ep,y_0,d'_0),\,
-\cJ_1(\Yh{i}{r};\ep,y_0,d'_0)
\Big\}
\,.
\label{eq:IS6IS7-def}
\esp
\eeq

When the result of the first integration is a $\cK_1$ function, it
decouples from the second integral, which gives a soft-collinear
function again.  Thus we find that all terms in the integrand of the
type
\beq
\frac{1}{s_{jl}} \frac{\tzz{j}{il}}{\tzz{l}{ij}}
\frac{2}{s_{\ha{i}\ha{j}}}
\frac{\tzz{\ha{i}}{\ha{j}}}{\tzz{\ha{j}}{\ha{i}}}
\eeq
integrate to the product of two soft-collinear integrals:
\beq
\cI^{(6)}_{\cS{}{}}(\ep,y_0,d'_0) =
\frac{1}{2} \cK_1(\ep,y_0,d'_0-1+\ep) \cK_1(\ep,y_0,d'_0)\,.
\label{eq:IS6}
\eeq

On the other hand, when the result of the first integration is a soft function 
$\cJ_1$, the two integrals do not decouple. In order to compute the second 
integral over the phase space factor $[\rd p^{(\ha{r})}_{1,m}]$, we choose the 
usual frame (\ref{eq:softframe}), with orientation specified by
\beq
\wti{p}_i^\mu = E_{\wti{i}} (1,\ldots,1)\,.
\eeq
Using \eqn{eq:hattotilde}, we compute
\beq
\Yh{i}{r} = \frac{y_{\ha{i}\ha{r}}}{y_{\ha{i}Q} y_{\ha{r}Q}}
= {y_{\ti{i}\ha{r}} \over [(1-y_{\ha{r}Q})y_{\ti{i}Q} + y_{\ti{i}\ha{r}}] y_{\ha{r}Q}}
= {1-\cos\vth \over 2-y_{\ha{r}Q} (1+\cos\vth)}\,.
\label{eq:Y2yctStCirtSrt2}
\eeq
Then we find that the term in the integrand of the type
\beq
\frac{s_{ir}}{s_{it}s_{rt}}
\frac{2}{s_{\ha{i}\ha{j}}}
\frac{\tzz{\ha{i}}{\ha{j}}}{\tzz{\ha{j}}{\ha{i}}}
\eeq
leads to the integral
\beq
\bsp
\cI^{(7)}_{\cS{}{}}(\ep,y_0,d'_0) &= -
2 \int_{0}^{y_0} \rd y\,  y^{-1-2\ep} (1- y)^{d'_0-2+\ep} 
\\[2mm] &\qquad\times
\int_{0}^{1}\rd z\, z^{-1-\ep} (1 - z)^{-\ep}
(1-y +y z)
\\[2mm] &\qquad\times
\cJ_1\left(\frac{z}{1 - y + y z};\ep,y_0,d'_0\right)\,.
\label{eq:IS7}
\esp
\eeq
To write \eqn{eq:IS7} in the above form, we made the usual substitution 
of $\cos\vth\to 1-2z$.

In terms of the integrals $\cI^{(6)}_{\cS{}{}}(\ep,y_0,d'_0)$ and
$\cI^{(7)}_{\cS{}{}}(\ep,y_0,d'_0)$ the integrated counterterm in
\eqn{eq:IcStIcCirtIcSrt} can be expressed as in 
\eqn{eq:IcStIcCirtIcSrtresult}.


\paragraph{Integrated soft-soft-collinear-double soft counterterm.} 
In the case of the integral in \eqn{eq:IcStIcSCSirtIcSrt}, we begin by 
integrating the eikonal function $\frac12 \calS_{jl}(t)$ over the soft 
phase space $[\rd p^{(t)}_{1,m+1}]$. However, whenever $j$
or $l$ is equal to $ir$ (recall that $j\ne l$), the eikonal factor evaluates as 
in \eqn{eq:softcolleikonal}. Therefore, we have to distinguish two cases: 
(i) $j$, $l$ and $ir$ are three distinct labels, thus the integration over the first 
soft phase space leads to a $\cJ_1$ soft function and we obtain the integral
\beq
\bsp
\cI^{(8)}_{\cS{}{}} =
\frac{16\pi^2}{S_\ep}Q^{2\ep}
\int &[\rd p^{(\ha{r})}_{1,m}]
\frac{2}{s_{\ha{i}\ha{r}}}
\frac{\tzz{\ha{i}}{\ha{r}}}{\tzz{\ha{r}}{\ha{i}}}
f(y_0,y_{\ha{r}Q},d'(m,\ep))
\\[2mm] &\times
\cJ_1\Big(\Yh{j}{l};\ep,y_0,d_0'\Big)
\esp
\eeq
and (ii) $j$ or $l$ coincide with $(ir)$, hence the integration over the first 
soft phase space leads to a $\cJm$ one-mass soft function and we find the integral 
(choosing $j=ir$ for concreteness)
\beq
\bsp
\cI^{(9)}_{\cS{}{}} =
\frac{16\pi^2}{S_\ep}Q^{2\ep}
\int &[\rd p^{(\ha{r})}_{1,m}]
\frac{2}{s_{\ha{i}\ha{r}}}
\frac{\tzz{\ha{i}}{\ha{r}}}{\tzz{\ha{r}}{\ha{i}}}
f(y_0,y_{\ha{r}Q},d'(m,\ep))
\\[2mm] &\times
\cJm\Big(Y_{(\ha{i}\ha{r})\ha{l}Q},\be_{(\ha{i}\ha{r})};\ep,y_0,d_0'\Big)
\,.
\esp
\eeq
To proceed, we must express the parameters of the soft functions with 
independent momenta.
In the first case, using \eqn{eq:hattotilde}, we can express $\Yh{j}{l}$ as
\beq
\Yh{j}{l} = 4 \Yt{j}{l}
{1-y_{\ha{r}Q} \over [2(1-y_{\ha{r}Q}) + 2 {y_{\wti{j}\ha{r}} / y_{\wti{j}Q}}]
[2(1-y_{\ha{r}Q}) + 2 {y_{\wti{l}\ha{r}} / y_{\wti{l}Q}}]}\,.
\label{eq:Y2yctStCSirtSrt1}
\eeq
In the second case the one-mass soft function also depends on the
velocity of the momentum $\ha{p}_i^\mu + \ha{p}_r^\mu$,
\beq
\be_{(\ha{i}\ha{r})} = \sqrt{1 - {4 y_{\ha{i}\ha{r}} \over y_{(\ha{i}\ha{r})Q}^2}}
\,.
\label{eq:beir}
\eeq
Again using \eqn{eq:hattotilde}, we find
\bal
\be_{(\ha{i}\ha{r})} &= 
{\sqrt{\left[2(1-y_{\ha{r}Q}) + {2 y_{\wti{i}\ha{r}} / y_{\wti{i}Q}}
+ {2 y_{\ha{r}Q} / y_{\wti{i}Q}}\right]^2
-{16 y_{\wti{i}\ha{r}} / y_{\wti{i}Q}^2}}
	\over
2(1-y_{\ha{r}Q}) + {2 y_{\wti{i}\ha{r}}/y_{\wti{i}Q}} + {2 y_{\ha{r}Q}/y_{\wti{i}Q}}}
\,,
\label{eq:behtot}
\\[2mm]
Y_{(\ha{i}\ha{r})\ha{l},Q} &=
{4(1-y_{\ha{r}Q}) \Yt{i}{l}
  +{4 y_{\wti{l}\ha{r}}/(y_{\wti{i}Q} y_{\wti{l}Q}})
		\over
\left[2(1-y_{\ha{r}Q}) + {2 y_{\wti{i}\ha{r}}/y_{\wti{i}Q}}
+ 2 y_{\ha{r}Q}/y_{\wti{i}Q}\right]
\left[2(1-y_{\ha{r}Q}) + {2 y_{\wti{l}\ha{r}}/y_{\wti{l}Q}}\right]}\,.
\label{eq:YhtoYt}
\eal

Turning to the integral over $[\rd p^{(\ha{r})}_{1,m}]$, we use two
different orientations of frames in the two cases.
In the first case, we fix the orientation such that
\beq
\wti{p}_{j}^\mu = E_{\wti{j}} (1,\ldots,1)\,,\qquad
\wti{p}_{l}^\mu = E_{\wti{l}} (1,\ldots,\sin\chi_{\wti{l}},\cos\chi_{\wti{l}})\,,
\eeq
that implies
\beq
\frac{2 y_{\wti{j}\ha{r}}}{y_{\wti{j}Q}} = y_{\ha{r}Q} (1 - \cos\vth)\,, 
\qquad 
\frac{2 y_{\wti{l}\ha{r}}}{y_{\wti{l}Q}} = y_{\ha{r}Q}
(1 - \sin\chi_{\wti{l}}\sin\vth \cos\vphi - \cos\chi_{\wti{l}}\cos\vth)
\,,
\eeq
where
\beq
\cos\chi_{\wti{l}} = \cos\chi(\Yt{j}{l})
\eeq
with \eqn{eq:coschi(Y)} for $\cos\chi(Y)$.
Thus we see that although the integral over $[\rd p^{(\ha{r})}_{1,m}]$ 
is of soft-collinear-type, we cannot trivially integrate over $\vphi$
because $\Yh{j}{l}$ depends on it. Thus
\beq
\bsp
\cI^{(8)}_{\cS{}{}}(Y,\ep,y_0,d'_0) &=
{1 \over \pi}{\Gamma^{2}(1-\ep) \over \Gamma(1-2\ep)} 
\int_{0}^{y_0} \rd y\, y^{-1-2\ep} (1-y)^{d'_0-2+\ep} 
\\[2mm] &\times
\int_{-1}^{1}\rd{(\cos\vth)}\, \rd{(\cos\vphi)}\, 
(\sin\vth)^{-2\ep} (\sin\vphi)^{-1-2\ep}
\\[2mm] &\qquad\times	
{2-y(1+\cos\vth) \over 1-\cos\vth}
\,\cJ_1\Big(\Yh{j}{l}(Y,y,\vth,\vphi);\ep,y_0,d_0'\Big)
\,,
\label{eq:IS8}
\esp
\eeq 
where the explicit dependence of the argument of the soft function
($\Yh{j}{l}$ in \eqn{eq:Y2yctStCSirtSrt1}) on the integration
variables is
\beq
\bsp
\Yh{j}{l}(Y,y,\vth,\vphi) &= 4 Y (1-y)
\Big(2 - y (1 + \cos\vth)\Big)^{-1}
\\[2mm] &\times
\Big(2 - y (1 + \sin\chi(Y)\sin\vth \cos\vphi + \cos\chi(Y)\cos\vth)\Big)^{-1}
\,.
\esp
\eeq

In the second case, the orientation of the frame is fixed by
\beq
\wti{p}_{i}^\mu = E_{\wti{i}} (1,\ldots,1)\,,\qquad
\wti{p}_{l}^\mu = E_{\wti{l}} (1,\ldots,\sin\chi_{\wti{l}},\cos\chi_{\wti{l}})\,,
\eeq
that implies
\beq
{2 y_{\wti{i}\ha{r}} \over y_{\wti{i}Q}} = y_{\ha{r}Q}(1 - \cos\vth)\,,
\qquad 
\frac{2 y_{\wti{l}\ha{r}}}{y_{\wti{l}Q}} = y_{\ha{r}Q}
(1 - \sin\chi_{\wti{l}}\sin\vth \cos\vphi - \cos\chi_{\wti{l}}\cos\vth)
\,,
\label{eq:yratios}
\eeq
where
\beq
\cos\chi_{\wti{l}} = \cos\chi(\Yt{j}{l})
\eeq
with \eqn{eq:coschi(Y)} for $\cos\chi(Y)$. Then 
\beq
\bsp
&
\cI^{(9)}_{\cS{}{}}(x, Y,\ep,y_0,d'_0) =
	{1 \over \pi} {\Gamma^{2}(1-\ep) \over \Gamma(1-2\ep)} 
	\int_{0}^{y_0} \rd y\, y^{-1-2\ep} (1-y)^{d'_0-2+\ep} 
\\[2mm] &\qquad\times
	\int_{-1}^{1}\rd{(\cos\vth)}\, \rd{(\cos\vphi)}
	\,(\sin\vth)^{-2\ep} (\sin\vphi)^{-1-2\ep}
	\,{2-y(1+\cos\vth) \over 1-\cos\vth}
\\[2mm] &\qquad\times	
\,\cJm\Big(Y_{(\ha{i}\ha{r})\ha{l},Q}(x,Y,y,\vth,\vphi),\be_{(\ha{i}\ha{r})}(y,\vth,\vphi);
\ep,y_0,d_0'\Big)
\,,
\label{eq:IS9}
\esp
\eeq 
where $\be_{(\ha{i}\ha{r})}$ and $Y_{(\ha{i}\ha{r})\ha{l},Q}$ are given in
\eqns{eq:behtot}{eq:YhtoYt}, with the ratios of invariants in \eqn{eq:yratios}
($x$ corresponds to $y_{\wti{i}Q}$, $Y$ to $\Yt{i}{l}$ and $y$ to
$y_{\ha{r}Q}$).  

Our final result for the integrated counterterm
defined in \eqn{eq:IcStIcSCSirtIcSrt}
is presented in \eqn{eq:IcStIcSCSirtIcSrtresult}.


\paragraph{Integrated soft-triple collinear-soft-collinear-double soft counterterm.} 
The integral over the soft phase space factor $[\rd p^{(t)}_{1,m+1}]$
in \eqn{eq:IcStIcCirtIcSCSirtIcSrt} is a one-mass soft-collinear
integral $\cKm$ of \eqn{eq:cKm-def}, which is computed in \eqn{eq:cKm-final},
where the velocity of the momentum $\ha{p}_i^\mu + \ha{p}_r^\mu$ is
found in \eqn{eq:beir}.  Then the integral over the measure
$[\rd p^{(\ha{r})}_{1,m}]$,
\beq
\bsp
\cI^{(10)}_{\cS{}{}} =
\frac{16\pi^2}{S_\ep}Q^{2\ep}
\int &[\rd p^{(\ha{r})}_{1,m}]
\frac{2}{s_{\ha{i}\ha{r}}}
\frac{\tzz{\ha{i}}{\ha{r}}}{\tzz{\ha{r}}{\ha{i}}}
f(y_0,y_{\ha{r}Q},d'(m,\ep))
\\[2mm] &\times
\cKm\Big(\be_{(\ha{i}\ha{r})};\ep,y_0,d'_0\Big)
\,,
\esp
\eeq
is again of soft-collinear-type, but the two integrals are coupled
through $\be_{(\ha{i}\ha{r})}$ that depends on the integration variables 
of the second integral as in \eqn{eq:behtot}.  The ratio of the invariants
$2 y_{\wti{i}\ha{r}}/y_{\wti{i}Q}$ is
\beq
\frac{2 y_{\wti{i}\ha{r}}}{y_{\wti{i}Q}} = y_{\ha{r}Q} (1 - \cos\vth)
\,,
\eeq
in a frame whose orientation is fixed by setting 
\beq
\wti{p}_{i}^\mu = E_{\wti{i}} (1,\ldots,1)\,.
\eeq
Thus, $[\IcS{t}{}\IcC{irt}{}\IcSCS{ir;t}{}\IcS{rt}{(0)}]$ is equal to
the integral
\beq
\bsp
\cI^{(10)}_{\cS{}{}}(x;\ep,y_0,d'_0) &=
2 \int_{0}^{y_0} \rd y\,  y^{-1-2\ep} (1- y)^{d'_0-2+\ep} 
\\[2mm] &\qquad\times
\int_0^1\rd z\, z^{-1-\ep} (1 - z)^{-\ep}
(1-y +y z)
\\[2mm] &\qquad\times
\cKm\Big(\be(x,y,z);\ep,y_0,d'_0\Big)\,,
\label{eq:IS10}
\esp
\eeq
with 
$x$ corresponding to $y_{\wti{i}Q}$, $y$ to $y_{\ha{r}Q}$ and  
\beq
\be(x,y,z) = \frac{\sqrt{(1-y+yz+y/x)^2-4y z/x}}{1-y+y z +y/x}
\,,
\eeq
where, as usual, we set $\cos\vth \to 1-2z$.

Our final result for the integrated counterterm
defined in \eqn{eq:IcStIcCirtIcSCSirtIcSrt}
is presented in \eqn{eq:IcStIcCirtIcSCSirtIcSrtresult}.


\paragraph{Integrated soft-double soft counterterms.} 
There are two types of integrated soft-double soft counterterms:
an `abelian' one in \eqn{eq:IcStIcSrtab} and a `non-abelian' one in
\eqn{eq:IcStIcSrtnab}.

Let us first consider the `abelian' case. Performing the integral over
$[\rd p^{(t)}_{1,m+1}]$ first, we obtain a soft function
$-\cJ_1(\Yh{j}{l},\ep,y_0,d'_0)$. In order to compute the integral over
$[\rd p^{(\ha{r})}_{1,m}]$,
\beq
\bsp
\cI^{(11)}_{\cS{}{};ik,jl} =
-\frac{16\pi^2}{S_\ep}Q^{2\ep}
\int &[\rd p^{(\ha{r})}_{1,m}]
\frac{1}{4} \calS_{\ha{i}\ha{k}}(\ha{r})
f(y_0,y_{\ha{r}Q},d'(m,\ep))
\\[2mm] &\times
\cJ\Big(\Yh{j}{l};\ep,y_0,d'_0\Big)
\,,
\label{eq:StSrt-raw}
\esp
\eeq
we must first express $\Yh{j}{l}$ in terms of $\wti{p}_j^\mu$ and $\wti{p}_l^\mu$
as in \eqn{eq:Y2yctStCSirtSrt1} ($j,l\ne r$), which is seen to depend on the 
integration variables in $[\rd p^{(\ha{r})}_{1,m}]$ through the appearance of 
$y_{\wti{j}\hr}$ and $y_{\wti{l}\hr}$ in the denominator. Thus, the
integration over $[\rd p^{(\ha{r})}_{1,m}]$, which is of soft-type, is nontrivial. 
Also, although $i\ne k$ and $j\ne l$, there is no restriction on whether or not 
$i,k$ is equal to $j,l$. Thus we must consider the following three cases:
(i) all of  $i$, $k$, $j$ and $l$ are distinct, (ii) only three of the four indices 
are distinct and, e.g.\ $l=k$, and (iii) only two indices are distinct and e.g.\ 
$j=i$ and $l=k$.

Case (i) requires at least four hard partons in the final state. Hence the 
corresponding integrated counterterm, $[\IcS{t}{}\IcS{rt}{(0)}]^{(i,k)(j,l)}$ 
with all labels distinct, does not enter a computation of two- or three-jet 
quantities, and we will not consider it in this paper.

In case (ii), we have $\Yh{j}{l} \to \Yh{j}{k}$, and this is expressed with the 
independent momenta as in \eqn{eq:Y2yctStCSirtSrt1}, after a $l\to k$ 
replacement. To evaluate the integral over $[\rd p^{(\ha{r})}_{1,m}]$ in 
\eqn{eq:StSrt-raw}, we use the partial fraction identity (\ref{eq:partialfrac}) 
for the eikonal factor $ \calS_{\ha{i}\ha{k}}(\ha{r})$ (with the substitutions 
$j \to i$ and $l \to k$). 
Further, we restrict our attention to the case when there are precisely three 
hard partons in the final state. As discussed around \eqn{eq:frameCktCSktr-3jet}, 
this leads to a constrained kinematics for the three momenta $\ti{p}_i^\mu$, 
$\ti{p}_k^\mu$ and $\ti{p}_j^\mu$, and we take this into account below.
It is convenient to introduce two different orientations of the frame
\eqn{eq:softframe} and integrate the two terms of the partial fraction in 
these different frames. In the first, we set
\beq
\bsp
\wti{p}_i^\mu &= E_{\wti{i}} (1,\dots,1)\,, \qquad
\wti{p}_k^\mu = E_{\wti{k}} (1,\ldots,\sin\chi_{ik},\cos\chi_{ik})\,,
\\[2mm]
\wti{p}_{j}^\mu &= E_{\wti{j}} (1,\ldots,-\sin\chi_{ij},\cos\chi_{ij})\,,
\label{eq:StSrt3j-framei}
\esp
\eeq
so
\bal
\frac{2y_{\wti{i}\ha{r}}}{y_{\wti{i}Q}} &= y_{\ha{r}Q} (1 - \cos\vth)\,,
\\[2mm]
\frac{2y_{\wti{k}\ha{r}}}{y_{\wti{k}Q}} &= y_{\ha{r}Q}
(1 - \sin\chi_{ik} \sin\vth \cos\vphi - \cos\chi_{ik} \cos\vth)\,,
\\[2mm]
\frac{2y_{\wti{j}\ha{r}}}{y_{\wti{j}Q}} &= y_{\ha{r}Q}
(1 + \sin\chi_{ij} \sin\vth \cos\vphi - \cos\chi_{ij} \cos\vth)\,,
\eal
where
\bal
&\cos\chi_{ik} = \cos\chi(\Yt{i}{k})\,,
&\cos\chi_{ij} = \cos\chi(\Yt{i}{j})\,,
\eal
with $\cos\chi(Y)$ given by \eqn{eq:coschi(Y)}.  In the second frame we
exchange $i$ and $k$, whose only effect on the integrand is to interchange
$\chi_{ij}$ and $\chi_{jk}$. Thus we find
\beq
\bsp
\cI^{(11)}_{\cS{}{};ik,jk}(Y_1,Y_2,Y_3;&\ep,y_0,d'_0) =
-4 Y_1 {\Gamma^{2}(1-\ep) \over 2\pi \Gamma(1-2\ep)}
\int_{0}^{y_0} \rd y\, y^{-1-2\ep} (1-y)^{d'_0-1+\ep} 
\\[2mm]&\times	
\int_{-1}^{1}\rd{(\cos\vth)}\, \rd{(\cos\vphi)}\,
(\sin\vth)^{-2\ep} (\sin\vphi)^{-1-2\ep}
(1-\cos\vth)^{-1}
\\[2mm]&\qquad\times	
\Big(
2-\cos\vth - \sin\chi(Y_1) \sin\vth \cos\vphi - \cos\chi(Y_1) \cos\vth\Big)^{-1}
\\[2mm]&\qquad\times	
\Big[\cJ\Big(\Yh{j}{k}(Y_1,Y_2, y,\vth,\vphi);\ep,y_0,d'_0\Big)
\\[2mm]&\qquad\quad	
+ \cJ\Big(\Yh{j}{k}(Y_1,Y_3, y,\vth,\vphi);\ep,y_0,d'_0\Big)\Big]
\,,
\label{eq:IS11}
\esp
\eeq 
where
\beq
\bsp
\Yh{j}{k}(Y_1,Y_2,y,\vth,\vphi) &= 4 Y_3 (1-y)
\\[2mm] &\times
\Big(2-y [1+\sin\chi(Y_1) \sin\vth \cos\vphi + \cos\chi(Y_1) \cos\vth]\Big)^{-1}
\\[2mm] &\times
\Big(2-y [1-\sin\chi(Y_2) \sin\vth \cos\vphi + \cos\chi(Y_2) \cos\vth]\Big)^{-1}
\,.
\esp
\eeq
Above, $Y_1 \equiv \Yt{i}{k}$, $Y_2 \equiv \Yt{i}{j}$ and $Y_3 \equiv \Yt{j}{k}$. 
The two terms in the square bracket in \eqn{eq:IS11} correspond to the two 
terms of the partial fraction in \eqn{eq:partialfrac}. Their integral representations 
are formally identical, only the kinematic invariants $Y_2$ and $Y_3$ are 
interchanged.

In terms of the integral $\cI^{(11)}_{\cS{}{}ik,jk}$
the integrated subtraction term defined in \eqn{eq:IcStIcSrtab} can be
expressed as in \eqn{eq:IcStIcSrtabresult}.

Finally, case (iii) is obtained trivially from the previous one. Setting $j=i$ implies 
\beq
\Yt{i}{j} \to \Yt{i}{i} = 0 \,,\qquad
\Yt{j}{k} \to \Yt{i}{k} \,,
\eeq
therefore, $\cI^{(11)}_{\cS{}{};ik,ik}$ depends only on $\Yt{i}{k}$. Then
integral (\ref{eq:IS11}) simplifies to
\beq
\bsp
\cI^{(11)}_{\cS{}{};ik,ik}(Y_1;\ep,y_0,d'_0) &=
-4 Y_1 {\Gamma^{2}(1-\ep) \over 2\pi \Gamma(1-2\ep)}
\int_{0}^{y_0} \rd y\, y^{-1-2\ep} (1-y)^{d'_0-1+\ep} 
\\[2mm]&\times	
\int_{-1}^{1}\rd{(\cos\vth)}\, \rd{(\cos\vphi)}\,
(\sin\vth)^{-2\ep} (\sin\vphi)^{-1-2\ep}
(1-\cos\vth)^{-1}
\\[2mm]&\qquad\times	
\Big(
2-\cos\vth - \sin\chi(Y_1) \sin\vth \cos\vphi - \cos\chi(Y_1) \cos\vth\Big)^{-1}
\\[2mm]&\qquad\times	
2 \cJ\Big(\Yh{i}{k}(Y_1,y,\vth,\vphi);\ep,y_0,d'_0\Big)
\,,
\esp
\eeq 
and it replaces $\cI^{(11)}_{\cS{}{};ik,jk}$ in \eqn{eq:IcStIcSrtabresult}. Above
\beq
\bsp
\Yh{i}{k}(Y_1,y,\vth,\vphi) &= 4 Y_1 (1-y)
\Big(2-y [1+ \cos\vth]\Big)^{-1}
\\[2mm] &\times
\Big(2-y [1+\sin\chi(Y_1) \sin\vth \cos\vphi + \cos\chi(Y_1) \cos\vth]\Big)^{-1}
\,.
\esp
\eeq

Turning now to the `non-abelian' contribution, \eqn{eq:IcStIcSrtnab}, we
see that we need to consider a single new integral,
\beq
\bsp
\cI^{(12)}_{\cS{}{};ik} =
\frac{16\pi^2}{S_\ep}Q^{2\ep}
\int &[\rd p^{(\ha{r})}_{1,m}] \frac{1}{2} \calS_{\ha{i}\ha{k}}(\ha{r})
f(y_0,y_{\ha{r}Q},d'(m,\ep))
\\[2mm] &\times
\cJ_1(\Yh{i}{r},\ep,y_0,d'_0)
\esp
\eeq
(obtained by performing the integral over $[\rd p^{(t)}_{1,m+1}]$ as
before). We must express $\Yh{i}{r}$ in terms of $\wti{p}_i^\mu$ 
(see \eqn{eq:hattotilde}),
\beq
\Yh{i}{r} = \frac{y_{\ha{i}\ha{r}}}{y_{\ha{i}Q} y_{\ha{r}Q}} 
= \frac{1}{y_{\ha{r}Q}}
\frac{2y_{\wti{i}\ha{r}}/y_{\wti{i}Q}}
{2(1-y_{\hr Q}) + 2y_{\wti{i}\hr}/y_{\wti{i}Q}}\,.
\eeq
We use the partial fraction identity (\ref{eq:partialfrac}) to write the
eikonal factor as a sum of two terms,
and choose two different orientations of the
frame (\ref{eq:softframe}) for each term. In the first one
we set 
\beq
\wti{p}_i^\mu = E_{\wti{i}} (1,\dots,1)\,, \qquad
\wti{p}_k^\mu = E_{\wti{k}} (1,\ldots,\sin\chi_{\wti{k}},\cos\chi_{\wti{k}})\,,
\label{eq:StSrt2-framei}
\eeq
thus
\beq
{2 y_{\wti{i}\ha{r}} \over y_{\wti{i}Q}} =  y_{\ha{r}Q} (1 - \cos\vth)\,,
\qquad
{2 y_{\wti{k}\ha{r}} \over y_{\wti{k}Q}} = y_{\ha{r}Q} 
(1 - \sin\chi_{\wti{k}} \sin\vth \cos\vphi - \cos\chi_{\wti{k}} \cos\vth)\,,
\eeq
where
\beq
\cos\chi_{\wti{k}} = \cos\chi(\Yt{i}{k})\,,
\eeq
with \eqn{eq:coschi(Y)} for $\cos\chi(Y)$.
The second frame is obtained by the interchange $i\leftrightarrow k$,
which again implies change only in the argument of the soft function.
Then the integral $\cI^{(12)}_{\cS{}{};ik}$ equals
\beq
\bsp
\cI^{(12)}_{\cS{}{};ik}(Y;&\ep,y_0,d'_0) =
	4Y {\Gamma^{2}(1-\ep) \over 2\pi \Gamma(1-2\ep)}
	\int_{0}^{y_0} \rd y\, y^{-1-2\ep} (1-y)^{d'_0-1+\ep} 
\\[2mm]&\times	
	\int_{-1}^{1}\rd{(\cos\vth)}\, \rd{(\cos\vphi)}\, 
	(\sin\vth)^{-2\ep} (\sin\vphi)^{-1-2\ep}(1-\cos\vth)^{-1}
\\[2mm]&\qquad\times	
\Big(2-\cos\vth
-\sin\chi(Y) \sin\vth \cos\vphi-\cos\chi(Y) \cos\vth\Big)^{-1}
\\[2mm]&\qquad\times
\Bigg[\cJ_1\Bigg(\frac{1-\cos\vth}{2-y(1+\cos\vth)};\ep,y_0,d'_0\Bigg)
\\[2mm]&\qquad\quad
+ \cJ_1\Bigg(
\frac{1 - \sin\chi(Y) \sin\vth \cos\vphi - \cos\chi(Y) \cos\vth}
{2-y (1 + \sin\chi(Y) \sin\vth \cos\vphi + \cos\chi(Y) \cos\vth)};
\ep,y_0,d'_0\Bigg)
\Bigg]
\,.
\label{eq:IS12}
\esp
\eeq 

In terms of the integrals $\cI^{(11)}_{\cS{}{}ik,ik}$ and
$\cI^{(12)}_{\cS{}{};ik}$, the integrated subtraction term
defined in \eqn{eq:IcStIcSrtnab} can be expressed as in
\eqn{eq:IcStIcSrtnabresult}.




\section{Integrating the soft-collinear-type terms}
\label{sec:IntsCktStA2}

In this appendix, we discuss the integration of the soft-collinear-type 
counterterms of \sect{sec:IntsCktStCTs}. In particular, we define and 
give an explicit integral representation of all $\cI^{(i)}_{\cSCS{}{}}$
functions ($i=1,\ldots,3$).

%
%

\subsection{Soft-collinear-triple collinear-type counterterms}
\label{sec:CktStCktr-type}

The soft-collinear-triple collinear-type counterterms involve a soft
momentum mapping followed by a collinear one, which leads to an exact
factorisation of the original $m+2$-particle phase space in the form
of \eqn{eq:psfact_CitStCirt-type}.  To evaluate the integrals of
\eqns{eq:IcCktIcStIcCkrt}{eq:IcCktIcStIcSCSirt} over the factorized
one-particle phase space measures in \eqn{eq:psfact_CitStCirt-type},
we first observe that the integral over the soft measure is a $\cK_1$
soft-collinear function, and the remaining the integral over the
collinear measure can be expressed as linear combination of the integrals
\beq
\cI^{(1)}_{\cSCS{}{}} =
\frac{16\pi^2}{S_\ep}Q^{2\ep}
\int [\rd p^{(\ha{i}\ha{r})}_{1,m}] 
\frac{1}{s_{\ha{i}\ha{r}}} \tzz{\ha{r}}{\ha{i}}^k
f(\al_0,\al_{\ha{i}\ha{r}},d(m,\ep))
\cK_1(\ep,y_0,d'_0)
\,,
\eeq
which is just twice the integral in \eqn{eq:IS1-def}, computed in \eqn{eq:IS1}, thus
\beq
\cI^{(1)}_{\cSCS{}{}}(x;\ep,\al_0,d_0,y_0,d'_0;k) =
2\cI^{(1)}_{\cS{}{}}(x;\ep,\al_0,d_0,y_0,d'_0;k)
\,.
\label{eq:ISCS1}
\eeq
%

%
%

\subsection{Soft-collinear-double soft-type counterterms}
\label{sec:CktStSrt-type}

The soft-collinear-double soft-type counterterms involve two successive
soft momentum mappings, which leads to an exact factorisation of the
original $m+2$-particle phase space in the form of
\eqn{eq:psfact_CktStSrt-type}. To evaluate the integrals of
eqs.~(\ref{eq:IcCktIcStIcCkrtIcSrt}), (\ref{eq:IcCktIcStIcSkt}) and
(\ref{eq:IcCktIcStIcSCSirtIcSrt}) over the factorized
one-particle phase space measures in \eqn{eq:psfact_CktStSrt-type},
we first observe that the integral over the soft measure 
$[\rd p^{(t)}_{1,m+1}]$ is again a $\cK_1$ soft-collinear function, and the
remaining integral over the second soft measure contains either an
eikonal factor or its collinear limit,
\beq
\bsp
\Big\{ \cI^{(2)}_{\cSCS{}{}},\, \cI^{(3)}_{\cSCS{}{}} \Big\} =
\frac{16\pi^2}{S_\ep}Q^{2\ep}
\int &[\rd p^{(\ha{r})}_{1,m}]
\Bigg\{
-\frac{1}{2} \calS_{\ha{j}\ha{l}}(\ha{r}),\,
\frac{2}{s_{\ha{i}\ha{r}}}
\frac{\tzz{\ha{i}}{\ha{r}}}{\tzz{\ha{r}}{\ha{i}}}
\Bigg\}
f(y_0,y_{\ha{r}Q},d'(m,\ep))
\\[2mm]&\times
\cK_1(\ep,y_0,d'_0)
\,,
\esp
\eeq
where in $\cI^{(3)}_{\cSCS{}{}}$ we recognise (twice) the function already
defined in \eqn{eq:IS6IS7-def}, hence
\beq
\cI^{(3)}_{\cSCS{}{}}(\ep,y_0,d'_0) = 2 \cI^{(6)}_{\cS{}{}}(\ep,y_0,d'_0)
\label{eq:ISCS3}
\,.
\eeq
The second integral decouples from the first one also for 
$\cI^{(2)}_{\cSCS{}{}}$ because $\cK_1$ is 
independent of the kinematics. Then, the final integral over 
$[\rd p^{(\ha{r})}_{1,m}]$ gives a soft function $\cJ_1$, so in this case,
\beq
\cI^{(2)}_{\cSCS{}{}}(Y;\ep,y_0,d'_0) =
\cJ_1(Y;\ep,y_0,d'_0-1+\ep) \cK_1(\ep,y_0,d'_0)
\,,
\label{eq:ISCS2}
\eeq
where $Y$ corresponds to $\Yt{j}{l}$.

In terms of the functions $\cI^{(i)}_{\cSCS{}{}}$, ($i = 1$, 2, 3) we
find the results given in \eqn{eq:softcollresults}.




\end{document}